\documentclass[acmtog,timestamp]{acmart}

\setcopyright{acmcopyright}

\usepackage{amsmath}
\usepackage{graphicx}
\usepackage{array}
\usepackage[titletoc,title]{appendix}
\usepackage{mathtools}
\usepackage{multirow}

\makeatletter
\newcommand\incircbin
{%
  \mathpalette\@incircbin
}
\newcommand\@incircbin[2]
{%
  \mathbin%
  {%
    \ooalign{\hidewidth$#1#2$\hidewidth\crcr$#1\bigcirc$}%
  }%
}
\newcommand{\oland}{\incircbin{\land}}
\makeatother

\DeclareMathOperator*{\argmin}{arg\,min}

\newcommand\addcomment[1]{\textcolor{black}{#1}}
\usepackage[normalem]{ulem}
\newcommand\addnote[1]{\textcolor{black}{#1}}
\usepackage{todonotes}
\newcommand\rev[1]{\textcolor{black}{#1}}
\newcommand\new[1]{\textcolor{black}{#1}}
\newcommand\newa[1]{\textcolor{black}{#1}}
\DeclareMathOperator*{\cumsum}{\textrm{cumsum}}

\newcommand\revb[1]{\textcolor{black}{#1}}

\usepackage{stfloats}

\usepackage{csvsimple}
\usepackage{xstring}

\newcommand{\ourmethod}{ALIGNet}

\RequirePackage[normalem]{ulem} 
\RequirePackage{color}\definecolor{RED}{rgb}{1,0,0}\definecolor{BLUE}{rgb}{0,0,1} 
\providecommand{\DIFadd}[1]{{\protect\color{blue}\uwave{#1}}} 
\providecommand{\DIFdel}[1]{{\protect\color{red}\sout{#1}}}                      
\providecommand{\DIFaddbegin}{} 
\providecommand{\DIFaddend}{} 
\providecommand{\DIFdelbegin}{} 
\providecommand{\DIFdelend}{} 

\usepackage[export]{adjustbox}
\usepackage{hyperref}

\begin{document}
\title{\rev{ALIGNet}: Partial-Shape Agnostic Alignment via Unsupervised Learning}

\author{Rana Hanocka}
\affiliation{\institution{Tel Aviv University}}

\author{Noa Fish}
\affiliation{\institution{Tel Aviv University}}

\author{Zhenhua Wang}\affiliation{\institution{Hebrew University}}

\author{Raja Giryes}
\affiliation{\institution{Tel Aviv University}}

\author{Shachar Fleishman}
\affiliation{\institution{Intel Corporation}}

\author{Daniel Cohen-Or}
\affiliation{\institution{Tel Aviv University}}

\begin{abstract}
The process of aligning a pair of shapes is a fundamental operation in computer graphics. Traditional approaches rely heavily on matching corresponding points or features to guide the alignment, a paradigm that falters when significant shape portions are missing. 
\rev{These techniques generally do not incorporate prior knowledge about expected shape characteristics, which can help compensate for any misleading cues left by inaccuracies exhibited in the input shapes.}
We present an approach based on a deep neural network, leveraging shape datasets 
\addcomment{to learn a \emph{shape-aware} prior for source-to-target alignment}
that is robust to shape incompleteness.
In the absence of ground truth alignments for supervision, we train a network on the task of shape alignment using incomplete shapes generated from full shapes for self-supervision.
Our network, called \emph{\ourmethod{}}, is trained to warp complete source shapes to incomplete targets, as if the target shapes were complete, thus essentially rendering the alignment \emph{partial-shape agnostic}.
We aim for the network to develop specialized expertise over the common characteristics of the shapes in each dataset, thereby achieving a higher-level understanding of the expected shape space to which a local approach would be oblivious.
We constrain \emph{\ourmethod{}} through an anisotropic total variation identity regularization to promote piecewise smooth deformation fields, facilitating both partial-shape agnosticism and post-deformation applications. We demonstrate that \emph{\ourmethod{}} learns to align geometrically distinct shapes, and is able to infer plausible mappings even when the target shape is significantly incomplete. We show that our network learns the common expected characteristics of shape collections, without over-fitting or memorization, enabling it to produce plausible deformations on unseen data during test time.

\end{abstract}

\citestyle{acmauthoryear}
\setcitestyle{square}

\maketitle

\section{Introduction}

Shape registration is a fundamental problem in computer graphics and computer vision, with diverse applications ranging from object recognition and scene understanding to texture or attribute transfer and synthesis. Classic approaches draw from a predefined family of transformations to obtain a warp that optimizes the registration between two shapes. While successful in straight-forward scenarios, the limitations of this paradigm are revealed as we seek to match increasingly distinct shapes differing in geometry and topology. Given such differences, restricting the pool of allowed transformations (\emph{e.g.,} affine transformations) is often insufficient, requiring higher-order deformations (\emph{e.g.,} free-form deformation). The problem becomes even more challenging when the shape is partial, for example due to missing data or occlusions.

Typically, the desired transformation between two shapes is obtained by solving a set of constraints defined by sparsely matching corresponding points or features on the shapes in question. Point matching necessitates a modicum of geometric resemblance between point neighborhoods, which is not always at hand under moderate to large differences. To alleviate this difficulty, features or descriptors are computed and matched instead. However, designing descriptors to be sensitive to interesting aspects of the shape yet robust to less significant variations, as well as to a wide range of shape types, is a daunting, and at times even intractable, task. 
As a further complication, when the shape is incomplete,
the process of feature matching becomes even more ambiguous and prone to errors. 
\rev{Moreover, traditional methods locally optimize the alignment between a pair of shapes without incorporating prior knowledge about their geometric and semantic identity. This information may provide invaluable cues as to the expected outcome, and guide the alignment both generally, as well as in and around the missing regions, such that geometric features typical to the relevant class of objects are preserved.}

\begin{figure}[t!]
\newcommand{\sfig}{3}
\setlength\tabcolsep{2pt}
\begin{tabular}{ c c c }

\includegraphics[width=\sfig cm]{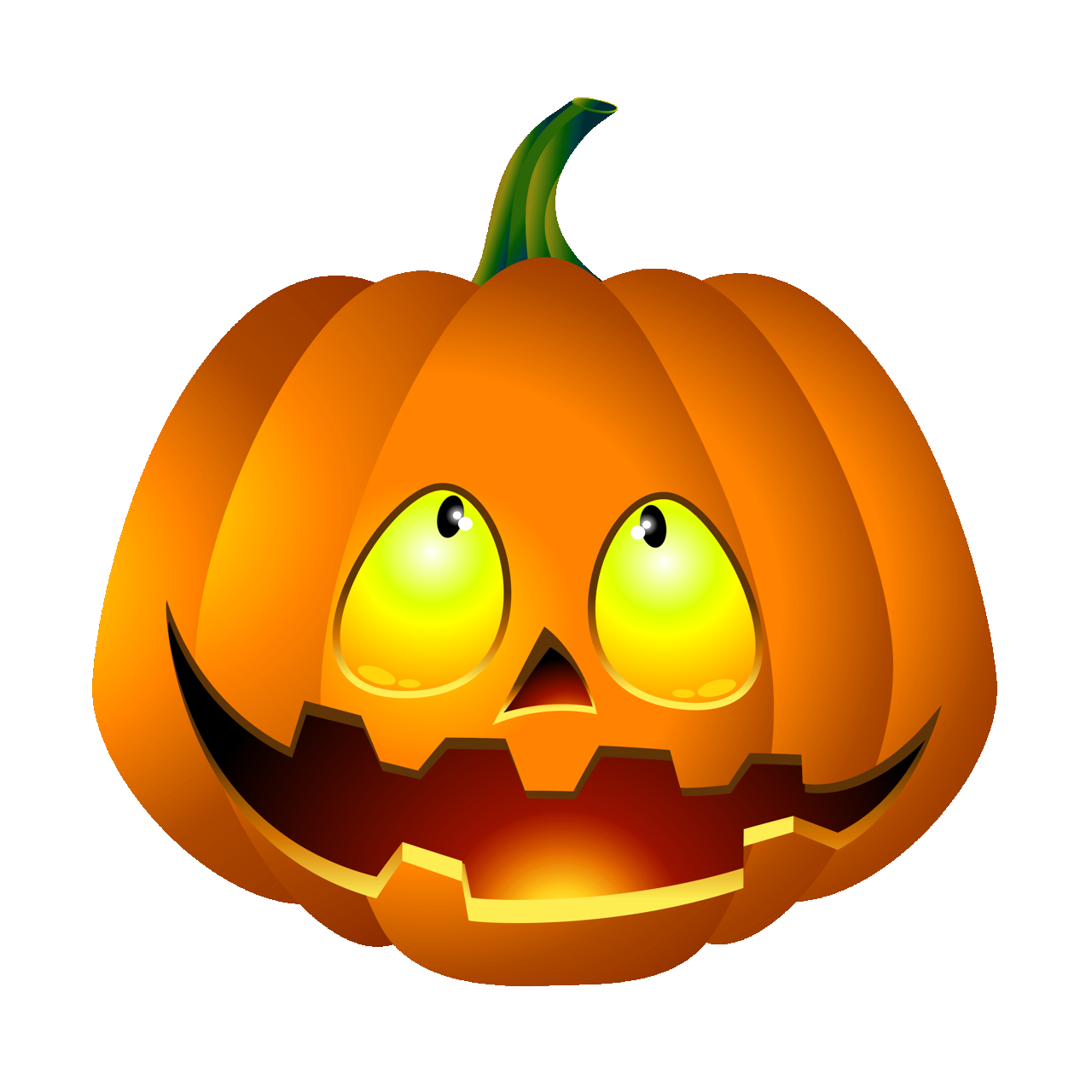} &
\includegraphics[width=\sfig cm]{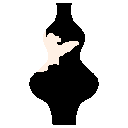} &
\includegraphics[width=\sfig cm]{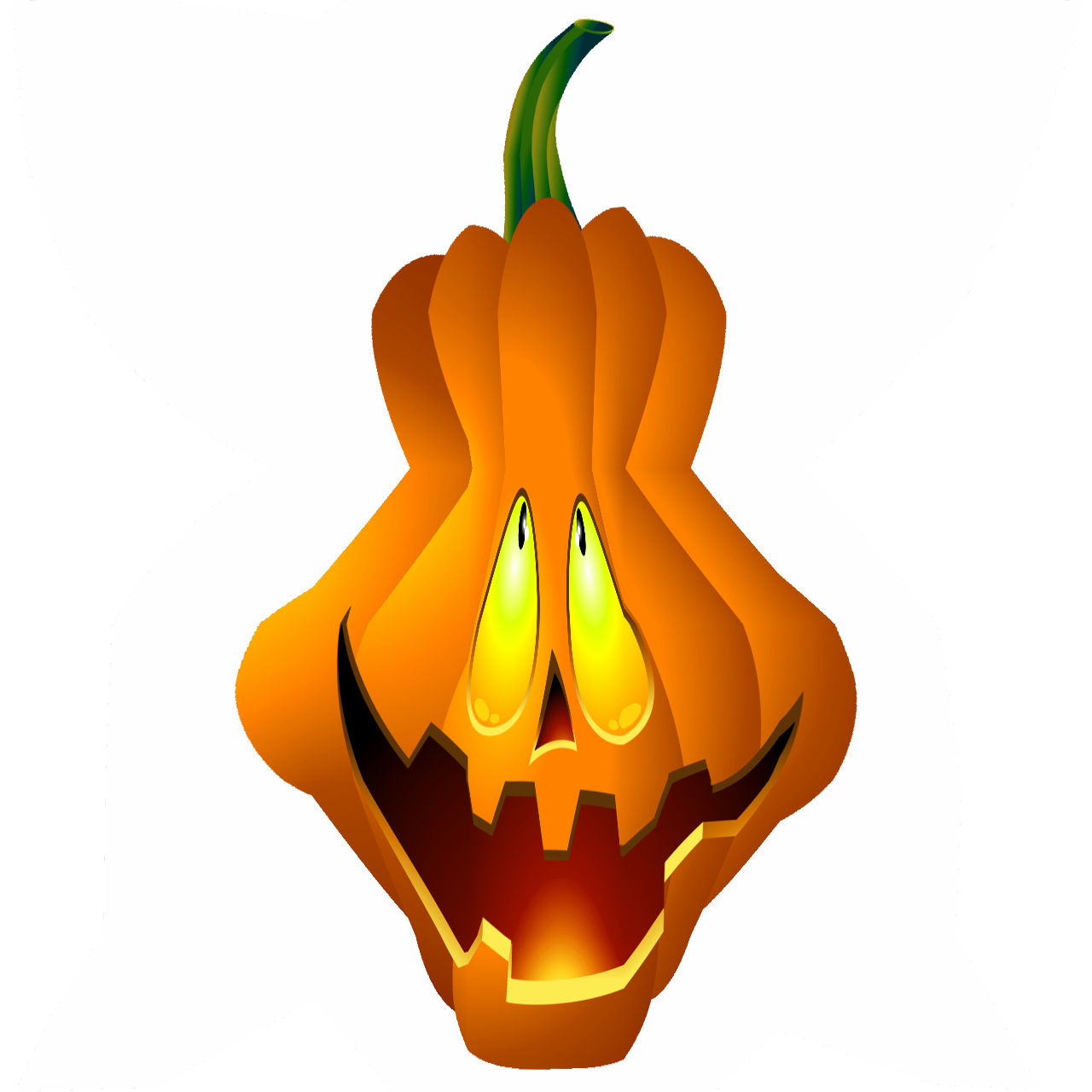} \\

source &
target &
aligned \\

\end{tabular}    
\caption{\rev{\ourmethod{} aligns a \emph{pumpkin} to a partial \emph{vase} (missing part visualized in pink), resulting in a \emph{vase}-like \emph{pumpkin} (aligned).}}
\label{fig:teaser2}
\end{figure}

\begin{figure}

\newcommand{\tfig}{1.6}
\setlength\tabcolsep{1pt}
\centering
\begin{tabular}{c c c c c}

 &
\multicolumn{2}{c}{target A} &
\multicolumn{2}{c}{target B} \\

 &
\multicolumn{2}{c}{$\overbrace{\rule{2.6cm}{0pt}}^\text{}$} &
\multicolumn{2}{c}{$\overbrace{\rule{2.6cm}{0pt}}^\text{}$} \\
 
 &
full &
partial &
full &
partial \\

\includegraphics[height=\tfig cm]{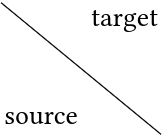} &
\includegraphics[height=\tfig cm]{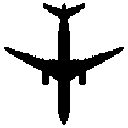} &
\includegraphics[height=\tfig cm]{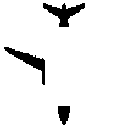} &
\includegraphics[height=\tfig cm]{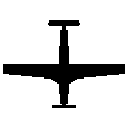} &
\includegraphics[height=\tfig cm]{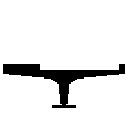} \\

\includegraphics[height=\tfig cm]{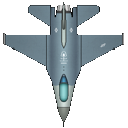} &
\includegraphics[height=\tfig cm]{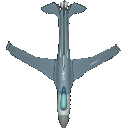} &
\includegraphics[height=\tfig cm]{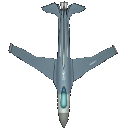} &
\includegraphics[height=\tfig cm]{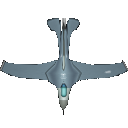} &
\includegraphics[height=\tfig cm]{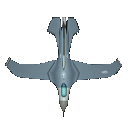} \\

\includegraphics[height=\tfig cm]{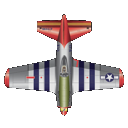} &
\includegraphics[height=\tfig cm]{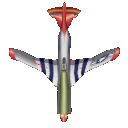} &
\includegraphics[height=\tfig cm]{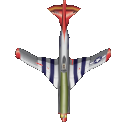} &
\includegraphics[height=\tfig cm]{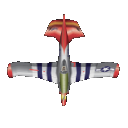} &
\includegraphics[height=\tfig cm]{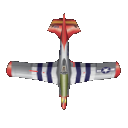} \\

\includegraphics[height=\tfig cm]{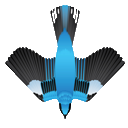} &
\includegraphics[height=\tfig cm]{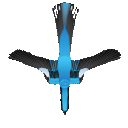} &
\includegraphics[height=\tfig cm]{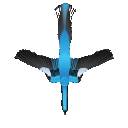} &
\includegraphics[height=\tfig cm]{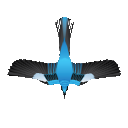} &
\includegraphics[height=\tfig cm]{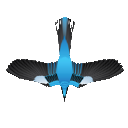} \\

\includegraphics[height=\tfig cm]{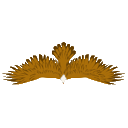} &
\includegraphics[height=\tfig cm]{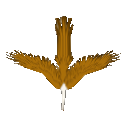} &
\includegraphics[height=\tfig cm]{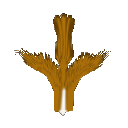} &
\includegraphics[height=\tfig cm]{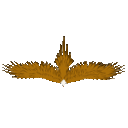} &
\includegraphics[height=\tfig cm]{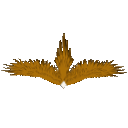} \\


\end{tabular}    
\caption{Examples of partial-shape agnostic deformations (test set). \ourmethod{} computes a deformation between source (left column) and full target (A/B) that is nearly identical to the deformation between source and a significantly partial target (A/B). The symmetry of airplane wings is preserved when one wing is missing in target A. Despite the ambiguous length of partial target B, the vertical scale is correctly estimated. 
\ourmethod{}, trained on the airplane class,
generalizes well to birds (bottom rows), a completely novel class. 
Note: RGB textures are applied post-deformation for visualization purposes only (not considered by network).
}
\label{fig:teaser}
\end{figure}

Inspired by the recent irrefutable success of deep learning methods applied to various computer vision problems, we propose a data-driven learning approach to address the problem of \addcomment{partial} shape alignment. Instead of providing a local solution, we opt to leverage the ever-growing availability of large collections of shapes to learn a data-aware global warping model. Many of these shape collections are consistently pre-oriented (or can be oriented \emph{e.g.,} by employing hierarchical alignment~\cite{shapenet2015}), and contain a rich variety of examples that collectively form a comprehensive guide capturing the characteristic attributes of the set, such as common geometry, topology, and even semantic high-level features such as part existence.
We arrange this data into random pairs of source and target shapes, and train a convolutional neural network (CNN) to learn the mapping between input source and target shapes.
We employ grid-based free-form deformation (FFD\new{~\cite{sederberg1986free}}), a new type of high order spatial transformer \addcomment{(STN)~\cite{spatialtransformer}}, which is simple yet expressive enough to support a wide range of alignments.
\rev{Unlike STN, we deliberately learn to align a source shape to an incomplete target, as though it were complete (\emph{e.g.,} Figure~\ref{fig:teaser2}).}
As shown in Figure~\ref{fig:teaser}, we are able to warp a model's texture to different partial target shapes, even when the target is missing more than 50\% of the original shape (all shapes are from the test set and therefore unseen during training). Furthermore, we are able to faithfully predict the scale (even in ambiguous scenarios such as target B) and correctly reconstruct symmetries of missing parts (target A).


Our approach is completely unsupervised, in that we do not supply any ground-truth warps that the network is expected to reproduce. Instead, the network is trained with a shape alignment loss, comparing the overlap between the warped source and the expected (complete) target, acting as a proxy for learning the desired underlying FFD parameters. As the network is trained, it essentially familiarizes itself with the common characteristics of the shapes in each dataset, allowing it to form a higher-level understanding of the expected shape space 
to which a local approach would be oblivious.
As such, our network is able to deal with incomplete, partial shapes, inferring plausible mappings in the missing regions by computing a \emph{class-aware} alignment. By training to warp complete source shapes to incomplete targets, as if the target shapes were complete, 
our network learns to be \emph{partial-shape agnostic}: unaffected by missing shape parts.

Due to the unsupervised nature of our training, in its simplest form, the model is given free reign in terms of the warp fields it is allowed to output, as long as the end result is deemed appropriate. 
Specifically, there are many vastly different FFDs that, when applied on \rev{\emph{binary shapes}}, yield the same shape alignment cost.
Such a behavior is undesirable when the warp field is needed for further processing; thus, to ensure the quality and dependability of our warps, we introduce a key addition to regulate the space of possible transformations.
This addition encourages piecewise smooth warp fields by penalizing the total variation of the deviation of the warp field from the identity.
By tuning the weighted contribution of this regularization component to the overall score, we are better able to control the degree of freedom given to the network for warp computation. We show that this step benefits not only post-deformation processes, but also partial agnosticism.

\section{Related Works}
Common shape registration methods estimate a transformation from a pre-determined class of transformations, by evaluating the overall shape-to-shape alignment. The classic Iterative Closest Point (ICP) method~\cite{ICP} uses nearest neighbor correspondences to refine the transformation, which minimizes the mismatch between the source and target points. Follow-up works propose numerous variants of ICP involving modifying the constraints defined to compute the transformation~\cite{ICPVariants_Levoy2001}. Other methods assume a global probabilistic approach~\cite{CPD,Jian2011,Tsin2004}, or more recently incorporate local structure to further improve results~\cite{ma2014robust,ma2016non}. To handle partial shapes, these approaches must explicitly incorporate outlier rejection and intelligently select a subset of appropriate points, a challenging task in and of itself.

To handle partial shapes, approaches such as RANSAC~\cite{RANSAC}, sample a small number of points to form candidate transformation models, and determine the best one by voting. This approach works well when the two shapes share similar geometries~\cite{amo_fpcs_sig_08}, and the transformation model can be expressed with a rather small number of variables, e.g., rigid or affine transformation. However, their search space becomes prohibitively large and even intractable if their geometries significantly differ. Later, we show (Figure~\ref{fig:comparisons_qual}) a restrictive alignment between shapes using affine transformations. The shapes that we successfully align in our work require higher order deformations, for example FFDs with 128 degrees of freedom.

The use of local descriptors compensates for significant geometry differences~\cite{Belongie2002,Zheng2006,Mori2003,Thayananthan2003,ling2007shape}. \rev{Works which use functional maps for correspondence rely on good local descriptors to drive the matching process~\cite{litany2017fully}}. These descriptors are typically hand-crafted and often tailored to a specific subset of shapes. When the transformation model is a high-order deformation, it is challenging to find good correspondences between the local descriptors, which are often ambiguous. 

Recently, many works have taken a deep learning approach to directly compute correspondence or learn an invariant feature descriptor, e.g.,~\cite{choy2016universal,simo2015discriminative, tian2017l2}. WarpNet~\cite{kanazawa2016warpnet} uses exemplar warps for training a network to compute visual correspondences. Zbontar et al.~\shortcite{zbontar2016stereo} train a CNN to accurately match correspondences across stereo image pairs (later improved by~\cite{shaked2016improved}). 
\revb{Fischer et. al.~\shortcite{FischerDIHHGSCB15} proposed using supervised CNNs to solve the optical flow estimation problem.} 
In 3D, Zeng et al.~\shortcite{zeng20163dmatch} used 3D-CNNs to learn a noise invariant feature representation. 
\revb{Yumer \& Mitra~\shortcite{yumer2016learning} pioneered the idea of 3D shape deformation using deformation fields. To achieve their goal, they incorporated semantic words in a supervised learning setting (\emph{e.g.,} deforming a car to become more \emph{sporty}).}
\rev{Tewari et al.~\shortcite{tewari2017mofa} estimated facial parameters (such as expression and lighting) from images via an end-to-end autoencoder. Garg et al.~\shortcite{garg2016unsupervised} used self-supervision for estimating depth from a single image. }

The Spatial Transformer Network (STN)~\cite{spatialtransformer} showed that incorporating a transformation module into the network enables the task at hand to be transformation invariant (further expanded in~\cite{li2017dense}), demonstrating inspiring results on digit classification. \ourmethod{} can be considered of as a new class of high-order spatial transformers, designed for the task of partial shape non-rigid alignment. In our presented work, \ourmethod{} learns an FFD, however it can similarly learn other classes of transformations, such as affine, projective or TPS~\cite{TPS}, as used in STN. Note that unlike any of the aforementioned works, we deliberately learn to align a source shape to an incomplete target shape, as though the target were complete.

\section{Overview}

\begin{figure*}[ht]
\includegraphics[width=16cm]{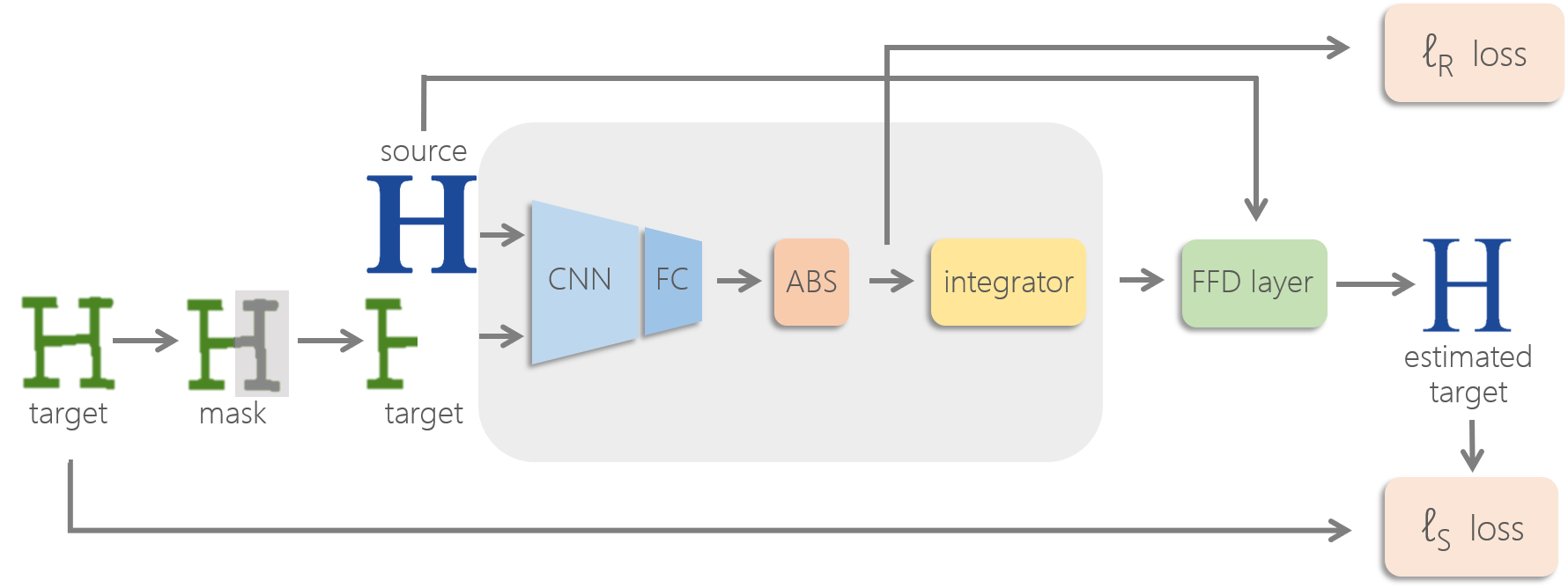}
\caption{Illustration of \ourmethod{} architecture. During training, \ourmethod{} learns to compute FFD grids without labeled ground-truth FFDs via the proposed unsupervised training scheme.
Given input source and target shapes, a random mask is applied to the target shape to form its partial version. A feature extraction CNN and a fully connected (FC) plus absolute value layer are then employed to output the differential of the warp field. The integrator layer, implemented via cumulative sum, results in the warp field. To compute the shape loss $\ell_s$, the warp field is applied on the source shape through the FFD warping layer, and the regularization loss $\ell_r$ is computed on the differential warp field. At test time, after the network weights have been learned, only the sub-portion of \ourmethod{} that computes the FFD grid deformation is required (bounded inside the gray rectangle).}
\label{fig:FFD_arch}
\end{figure*}

\ourmethod{} is a neural network that aligns a source shape to a target shape via FFD. The FFD controllers are defined on a uniform grid, whose resolution trades off between deformation expressiveness and warping smoothness. Defining the FFD on a uniform grid is a judicious technique for providing an expressive, yet smooth, deformation, with a relatively small number of parameters. The power of \ourmethod{} lies is its ability to learn to align a source shape to an incomplete target shape as though the target were complete.

Training a network when ground-truth information is unavailable is a major challenge.
Our approach employs unsupervised learning, \emph{i.e.,} the true FFD that best aligns the source shape to the target is not utilized during training. The network scheme is illustrated in Figure~\ref{fig:FFD_arch}. \ourmethod{} is trained by drawing two instances from a pool of training data of objects from the same class, deeming one instance as the source shape and the other as the target shape. To mimic missing data, we remove a random part of the target shape, and feed the partial target as input to the network, holding out the full shape for loss computation. The input shapes are then fed forward through a series of neural network layers yielding an FFD, which when applied on the source shape yields a deformed source, called the estimated target. The training loss measures the difference between the estimated target and the complete target. Consequently, \ourmethod{} is \emph{partial shape agnostic}, that is, it learns to estimate the deformation that aligns the source shape to the partial target as if it were the complete target.

Without any restrictions on the FFD, \ourmethod{} can conjure up deformations which do not preserve internal structure. 
To this end, we introduce a total variation penalty that is key to 
\addcomment{producing} 
smooth deformations.
Rather than directly computing the \emph{actual displacement} values, we compute \emph{relative displacements}, where each displacement is relative to its preceding displacement, such that the cumulative summed result (in each axis) yields the actual grid displacement field.
Since these relative displacements are essentially displacement gradients, we utilize them to employ anisotropic total variation (TV) on the deviation from the identity warp field, which encourages uniformly spaced piecewise smooth deformation fields.
To further encourage smoothness, we maintain axial monotonicity by enforcing positive relative grid displacements which ensure consecutively increasing values in each axis of the displacement grid.
We present and evaluate \ourmethod{} on 2D shapes, and demonstrate its extension to 3D in section~\ref{sec:threed}.



\section{Method}

The input to our system is made up of pairs of source and target shapes, where a target shape may be partial at various locations and to varying extents. The baseline system consists of a neural network component $f$ that computes an FFD in the form of a grid-based warp field $W_c$.
The input source shape $S$ and the potentially partial target $T^p$ are stacked as a two-channel image and passed to $f$ to obtain the low resolution FFD warp field: $f(S,T^p) = W_c$. The architecture of $f$ is given by a series of max pooling, convolutions and rectified linear activation units (repeated four times), followed by two fully-connected layers. Similar to~\cite{spatialtransformer}, we train the network from scratch by setting the convolutional layer weights randomly, and initializing the last FC to yield the identity displacement field (weights set to zero and biases to the identity displacement field). This is followed by the differentiable FFD sampling layer, which warps the source to the target by upsampling the low resolution FFD.

The network is trained in an unsupervised manner, such that at no point during training is it shown the ground-truth warp field that correctly aligns the source to the target shape.
Instead, the pairs of shapes given as input during training are in fact triplets, where the target is passed twice - once in its complete form and once with missing parts. 
This scheme facilitates training by bestowing a self-supervised shape similarity loss function $(\ell_s)$: comparing the warped source to the full target (see Figure~\ref{fig:FFD_arch}), aimed at teaching the network to become oblivious to possible missing data. 

In addition to the shape similarity loss function, we add a regularization loss $(\ell_r)$ which encourages the network to produce a warp field that is varying uniformly vis-\`a-vis the identity warp, limiting the likelihood of fold-overs. Our loss function thus amounts to: $loss = \ell_s + \lambda \ell_r$, where $\lambda$ is a user defined parameter that trades off between shape alignment and smoothness of the warp field.

\begin{figure}[h]
\newcommand{\sfig}{2.7}
\setlength\tabcolsep{2pt}
\begin{tabular}{ c c }

\includegraphics[height=\sfig cm]{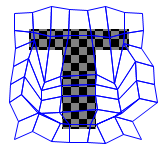} &
\includegraphics[height=\sfig cm]{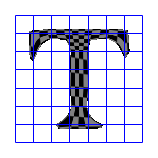} \\

source &
warped \\
\end{tabular}   
\caption{
\addcomment{2D FFD grid deformation. The uniform 8x8 grid in the warped domain (right) shown in blue, applied in a backwards manner: each location in the warped domain grid corresponds to a look-up location in the source (left).}}
\label{fig:FFD_visualization}
\end{figure}

\subsection{FFD Representation}
\begin{figure}[h!]
\includegraphics[height=5cm]{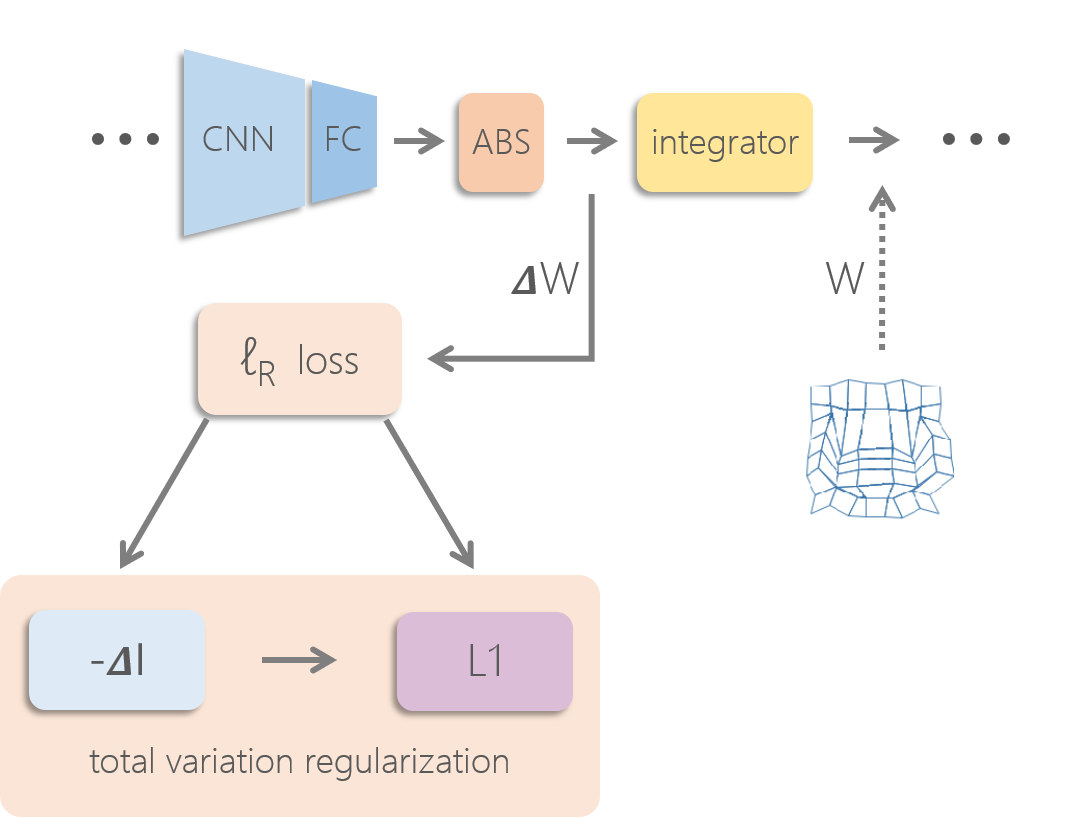}
\caption{Regularization component commissioned for training \ourmethod{} to compute plausible FFD grids. The network weights predict a differential grid, passed through an absolute value operator to enforce axial monotonicity. Followed by $\ell_1$ regularization on $\Delta W - \Delta I$ (equivalent to anisotropic total variation regularization on $W - I$). Lastly, the differential $\cumsum$ layer (Integrator) restores the absolute displacement grid.}
\label{fig:FFD_regularization}
\end{figure}

We constrain the space of possible warp fields by incorporating regularization that encourages piecewise smooth deformation fields and simultaneously enforces axial monotonicity (illustrated in Figure~\ref{fig:FFD_regularization}).
%
%
We facilitate these warp field constraints by representing the FFD as a \emph{differential} displacement grid, where each value is relative to the preceding value in the grid. Instead of directly computing the absolute FFD values after the FC layer, we compute the differential FFD values, which when cumulatively summed result in an absolute FFD grid. The differential displacement grids encourage shift-invariance, since the displacement at each grid cell is relative to its neighborhood. 

To illustrate differential displacement fields, consider a 1D absolute sequence $a$ of length n, and the corresponding differential sequence $\Delta a$ as
\begin{subequations}
\begin{align}
a  & = \Big\{ a_k \Big\}_{k=1}^n  = ( a_1, a_2, a_3, \cdots, a_n ) \label{eq:oned_seq_abs}\\
\begin{split}
\Delta a & = \Big\{a_k - a_{k-1} \Big\}_{k=1}^n = ( \Delta a_1, \Delta a_2, \Delta a_3, \cdots, \Delta a_n )
\\
& = (a_1 - a_0, a_2 - a_1, a_3 - a_2,  \cdots, a_n - a_{n-1} ),
\end{split}
\label{eq:oned_seq_rel}
\end{align}
\end{subequations}
where $a_0$ is a scalar offset (here $a_0 = 0$) that can be learned during training. Given the differential 1D $\Delta a$, the absolute sequence can be exactly recovered using the cumulative sum operator
\begin{equation}
a = \cumsum( \Delta a ,a_0) = \Big\{ a_0 + \sum_{i=1}^k \Delta a_i \Big\}_{k=1}^n.
\label{eq:cumsum}
\end{equation}
The proposed $\cumsum$ layer is differentiable and can be added into a network for training. It is apparent in the 1D example that the derivative is given by
\begin{equation}
\partial \Delta a=\Big\{\sum_{i=k}^n \partial a_i  \Big\}_{k=1}^n.
\label{eq:cumsum_deriv}
\end{equation}

The deformation field in \ourmethod{} is a 2D grid with vertical and horizontal displacement components, where the $\cumsum$ operates along each dimension, respectively. The learned \ourmethod{} weights map the input shapes to the differential displacement field $\Delta W_c$, subsequently passing through the absolute and $\cumsum$-layer, resulting in the final displacement field $W_c = \cumsum(\Delta W_c, W_0)$ (where $W_0$ is a learned scalar).

\paragraph{\textbf{Axial Monotonicity.}}
Observe that for any strictly positive differential 1D sequence ($\Delta a_k \geq 0$ for all $k$), the recovered absolute 1D sequence $a$ is monotonic
\begin{equation}
a   = \Big\{ a_k \Big\}_{k=1}^n \quad \text{\textit{s.t.}  } \; a_k \leq a_{k+1}
\label{eq:cumsum_mono}
\end{equation}
%
In \ourmethod{}, we enforce the differential sequence at the end of the network to be positive. In the 2D case, enforcing strictly positive relative displacement values $\Delta W_c \geq 0$, we ensure axial monotonicity in the displacement fields. Axial monotonicity implies that each axis increases monotonically independently of the opposing axis.

\paragraph{\textbf{FFD Layer.}}
The displacement map used to express the FFD is a low resolution 2D displacement grid, defining the inverse warp field from target to source (see Figure~\ref{fig:FFD_visualization}). To obtain the full resolution target image, pixels in the target image are copied from their corresponding source image locations as indicated by a bilinear interpolation upsampling of the low resolution warp field. This is accomplished by a 2-step bilinear sampling operation. First, use the bilinear sampler $\oland$ to upsample the lower resolution grid $W_c$ to the full size of the source image ($mxn$) by $W = W_c \oland \mathcal{G}^{mxn}$,
where $\mathcal{G}^{mxn}$ is a regular grid of coordinates for every pixel in the full resolution $mxn$. 
The second sampling applies the upsampled warp field on the source shape $\hat{T} = S \oland W$,
resulting in an estimate of the target shape. For brevity, we represent the two-step sampling as
\begin{equation}
\hat{T} = S \oland W_c \leftarrow S \oland  (W_c \oland \mathcal{G}^{mxn}).
\label{eq:sampler}
\end{equation}

\subsection{Loss Functions}
\label{sec:loss}
Estimating FFDs without direct supervision is ill-posed, since the large degree of freedom (DoF) in the FFD results in wildly different deformations - all with very similar cost. Specifically, \ourmethod{} operates solely on binary silhouettes, since knowing the true corresponding RGB textures between source and target shapes requires a form of ground-truth annotation. Binary silhouettes are problematic since any FFD that deforms source foreground pixels to target foreground pixels, even if parts are completely rearranged, results in the same shape similarity loss. Absent regularization, \ourmethod{} is entitled to falsely conjecture about the possible deformation field.

Our overall loss function is composed of two components. The first is the shape alignment loss, encouraging a faithful deformation from source to target. The second, which measures anisotropic total variation with regards to deviation from the identity warp, serves as a strong regularization component that helps produce smoother warp fields. 

\subsubsection{Shape Alignment Loss}
The self-supervised training loss evaluates the difference between the complete target $T$ and the source that has been aligned by the FFD (\emph{i.e.,} estimated target $\hat{T}$) to train \ourmethod{}. For every example $n$ ($S^n$, $T^n$) in the training set $\mathcal{N}$, we train the network by minimizing the loss given by
\begin{equation}
f = \argmin_{f} 
\sum_{n \in \mathcal{N}}
|| S^{n} \oland f(S^{n},{T^{p}}^n) - T^{n} || 
= \sum_{n \in \mathcal{N}}
|| S^{n} \oland W_c - T^{n} ||,
\label{eq:FFDnet_loss}
\end{equation}
where $||\cdot||$ can be any error measure. We use the $\ell_2$-norm, which is given by
%
%
\begin{equation}
 \ell_s(T,\hat{T}) = \sum_{i \in \forall (\hat{T},T)} \frac{1}{2}(T_i-\hat{T_i})^2,
\label{eq:huber}
\end{equation}
for every pixel $i$ in the target and estimated target.

The proposed approach enables training without ground-truth FFDs while simultaneously learning to be partial-shape agnostic. At test time, given a source shape $S$ and a target shape with different parts missing $T^{p_1}, T^{p_2}, \cdots T^{p_n} \in T$ (from test set), \ourmethod{} demonstrates partiality-agnostic behavior by computing very similar FFDs for different partialities of the same underlying target (see Figure~\ref{fig:sweeping_partialT}). We attribute this generalization to the unsupervised nature of the training process.

\subsubsection{Anisotropic Total Variation Identity Regularization}
We facilitate minimal distortion deformation fields by imposing an additional penalty in the form of TV regularization. Observe that the differential sequence as defined in Equation~\ref{eq:oned_seq_rel}, is actually the discrete gradient of the absolute sequence in Equation~\ref{eq:oned_seq_abs}. 
Similarly for the $2$D counterpart, the differential grid $\Delta W$ is made up of two channels, $\Delta W_x$,$\Delta W_y$, which are the horizontal and vertical derivatives of $W$ respectively. Recall that the anisotropic total variation regularization for a 2D matrix is given by the $\ell_1$-norm of the sum of the horizontal and vertical gradients
\begin{equation}
\ell_{TV}(W) = ||\Delta W_x ||_1 + ||\Delta W_y ||_1.
\label{eq:TV_reg}
\end{equation}
TV regularization encourages deformations that are piecewise smooth, favoring smoothness but simultaneously allowing for large discontinuities due to the $\ell_1$-norm. Piecewise smoothness is the essence of a favorable deformation field: large discontinuities at the shape boundaries with a smooth shape interior. 

Yet, the displacement field $W$ that would lead to the lowest cost in the standard TV regularization in Equation~\ref{eq:TV_reg} would be simply a zero valued differential displacement field, which is an undesirable result. Instead, we encourage a non-zero-valued uniform spacing by subtracting the differential identity warp field from the differential warp, favoring a uniformly spaced grid over the size of the image.

For a 1D analogy, consider an identity sequence $I$ of length $n$ normalized to the interval $[-1, 1]$, which is uniformly spaced by a scalar $\delta_k = \frac{2}{n-1} $. Then we can write $I$ and the corresponding differential $\Delta I$ as
\begin{subequations}
\begin{align}
\begin{split}
\label{eq:id_seq}
I & = \Big\{ -1 + (k - 1) \cdot \delta_k  \Big\}_{k=1}^n \\ 
& = (-1, -1 +\delta_k  , -1 +2 \delta_k , \cdots, 1)
\end{split}
\\
\label{eq:id_seq_rel}
\Delta I & = \Big\{ \delta_k  \Big\}_{k=1}^n,
\end{align}
\end{subequations}
where the differential identity sequence $\Delta I$ contains a single unique value: the uniform spacing constant $\delta_k$. In $2$D, it follows that the proposed anisotropic TV identity regularization becomes 
\begin{equation}
\ell_{r} = \ell_{TV}(W - I) = ||\Delta W_x - \Delta I_x||_1 + ||\Delta W_y  - \Delta I_y||_1.
\label{eq:TV_reg_id}
\end{equation}
The overall contribution of the regularization term is controlled by a weighting factor $\lambda$.

\section{Results and Evaluation}

In this section we examine and test \ourmethod{} through various qualitative and quantitative evaluations (code and data are available at \href{https://github.com/ranahanocka/ALIGNet}{\textbf{https://github.com/ranahanocka/ALIGNet}}). All experiments are performed on the reserved \emph{test set data}, meaning that these shapes were never trained on. All networks are trained on strictly binary silhouettes, but we overlay a checkerboard texture on the source shape to clearly visualize the smoothness of the estimated mappings and display the texture transfer abilities facilitated by our system. Note that the missing components of the target are illustrated in light pink (for visualization purposes only), but the actual network input is binary. For all results and classes, we use the same parameters and architecture described in Table~\ref{table:params}, unless stated otherwise (\emph{e.g.,} in the ablation study).

\begin{table}
\begin{center}
 \begin{tabular}{|c || c|} 
 \hline
 \multicolumn{2}{|c|}{Architecture} \\
 \hline
 conv1 &  $5 \times 5 \times 20$  \\ 
 conv2 &  $5 \times 5 \times 20$  \\ 
 conv3 &  $2 \times 2 \times 20$  \\ 
 conv4 &  $4 \times 4 \times 20$  \\ 
 fc1 &  $20$  \\ 
 fc2 &  $2 \cdot m \cdot n$  \\ 
 \hline
\end{tabular}
\quad
 \begin{tabular}{|c || c|} 
 \hline
 \multicolumn{2}{|c|}{Parameters} \\
 \hline
 n & 8 \\ 
 m & 8 \\ 
 $\lambda$ & $1e-5$ \\ 
solver & ADAM \\ 
LR & $1e-3$ \\ 
 \hline
\end{tabular}
\end{center}
\caption{\ourmethod{} architecture and parameters used for all results (except where otherwise specified), \addcomment{where $m$ and $n$ represent the resolution of the grid.}}
\label{table:params}
\end{table}

\subsection{Training Data}
\label{sec:un_data}
In our experiments, we train \ourmethod{} on a class of shapes where pairs of source and target instances are randomly drawn from the pre-defined training set. Specifically, we have experimented with sets containing renders of lower and upper case letters for 330 fonts (resulting in 56 classes), as well as 2D projections of 3D objects from ShapeNet~\shortcite{shapenet2015} and COSEG~\cite{wang2012active}. 

\paragraph{Train and Test Sets}
For each class dataset, we reserve 30 models for testing and use the rest for training. For example, a class with a set of 330 shapes yields $\sim$90k distinct training pairs and 870 test pairs. We augment the number of training set examples by generating an arbitrarily large amount of partial data, and by applying small vertical and horizontal scaling on-the-fly during training. We generate rectangular partial data masks by setting a random foreground pixel as the center, with width and height sampled from a uniform distribution. 

During testing, we evaluate performance on rectangular masks and verify generalization on non-rectangular partial data masks created from the MPEG-7~\cite{latecki2000shape} dataset. 

\begin{figure}
\newcommand{\conv}{3}
\setlength\tabcolsep{5pt} 
\begin{tabular}{cc}
\hspace*{4mm} Train &
\hspace*{4mm} Test \\
\includegraphics[width=\conv cm]{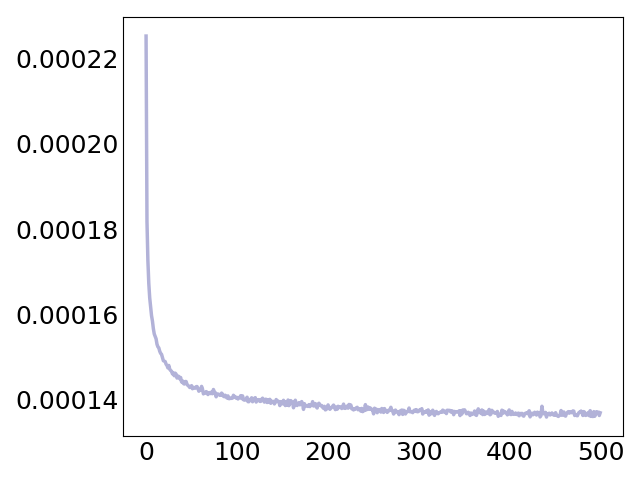} &
\includegraphics[width=\conv cm]{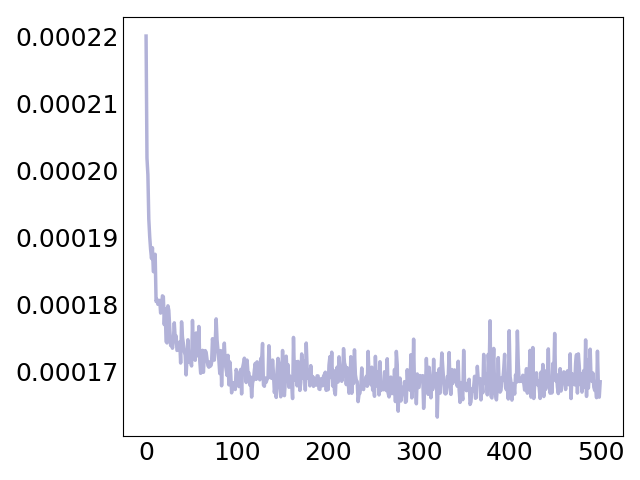} \\

\hspace*{5mm}Iterations &
\hspace*{5mm}Iterations \\

\end{tabular}
\caption{\revb{Train (left) and test (right) error convergence plots for the \emph{vase} class.}}
\label{fig:conv_plot}
\end{figure}

\subsection{Ablation Study}
We analyze the qualitative and quantitative effects of grid regularization by training \ourmethod{} in three different settings: without any regularization (none), total variation regularization (TV) and total variation regularization with axial monotonicity (TV \& M). Regularization controls the trade-off between: fidelity, in our case the silhouette alignment accuracy, and the prior, which is the smoothness of the warp field. We quantitatively measure accuracy via IOU between the source and true underlying target, and measure smoothness by the difference between the warp field gradient and the identity field gradient (TV loss in Equation~\ref{eq:TV_reg_id}). 

\begin{figure}[h]
\newcommand{\ablfig}{1.6}
\newcommand{\ablfigsp}{0.2}
\setlength\tabcolsep{1pt} 
\centering
\begin{tabular}{c c c c c}

\includegraphics[height=\ablfig cm]{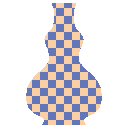} &
\includegraphics[height=\ablfig cm]{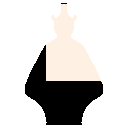} &
\includegraphics[height=\ablfig cm]{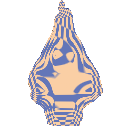} &
\includegraphics[height=\ablfig cm]{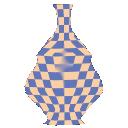} &
\includegraphics[height=\ablfig cm]{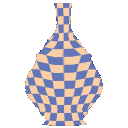} \\

\includegraphics[height=\ablfig cm]{figures/ablation/vase/sample530_source.png} &
\includegraphics[height=\ablfig cm]{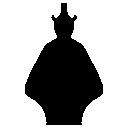}  &
\includegraphics[height=\ablfig cm]{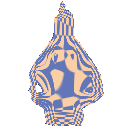} &
\includegraphics[height=\ablfig cm]{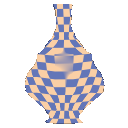} &
\includegraphics[height=\ablfig cm]{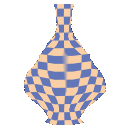} \\[\ablfigsp cm]
\hline
\vspace{\ablfigsp cm}

\includegraphics[height=\ablfig cm]{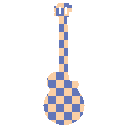} &
\includegraphics[height=\ablfig cm]{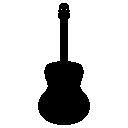}  &
\includegraphics[height=\ablfig cm]{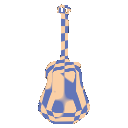} &
\includegraphics[height=\ablfig cm]{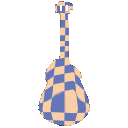} &
\includegraphics[height=\ablfig cm]{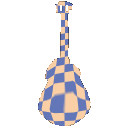} \\[\ablfigsp cm]
\hline
\vspace{\ablfigsp cm}


\includegraphics[height=\ablfig cm]{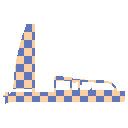} &
\includegraphics[height=\ablfig cm]{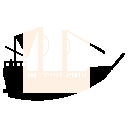} &
\includegraphics[height=\ablfig cm]{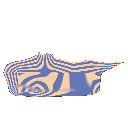} &
\includegraphics[height=\ablfig cm]{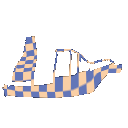} &
\includegraphics[height=\ablfig cm]{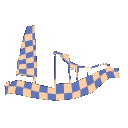} \\

\includegraphics[height=\ablfig cm]{figures/ablation/vessel2/sample466_source.png} &
\includegraphics[height=\ablfig cm]{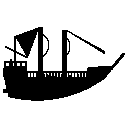} &
\includegraphics[height=\ablfig cm]{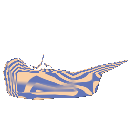} &
\includegraphics[height=\ablfig cm]{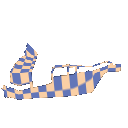} &
\includegraphics[height=\ablfig cm]{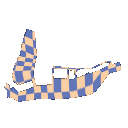} \\
 
source &
target &
none &
TV &
TV \& M \\
\end{tabular}
\caption{Ablation study visual examples for increasing levels of FFD regularization: no regularization, TV regularization and TV regularization with axial monotonicity. Without regularization, \ourmethod{} is not partial agnostic: the deformation is disparate for the full and partial version of the same target.  }
\label{fig:visual_reg_ablation}
\end{figure}

\begin{figure}

\newcommand{\sfig}{3.1}
\setlength\tabcolsep{2pt}
\begin{tabular}{ c c }
\includegraphics[height=\sfig cm]{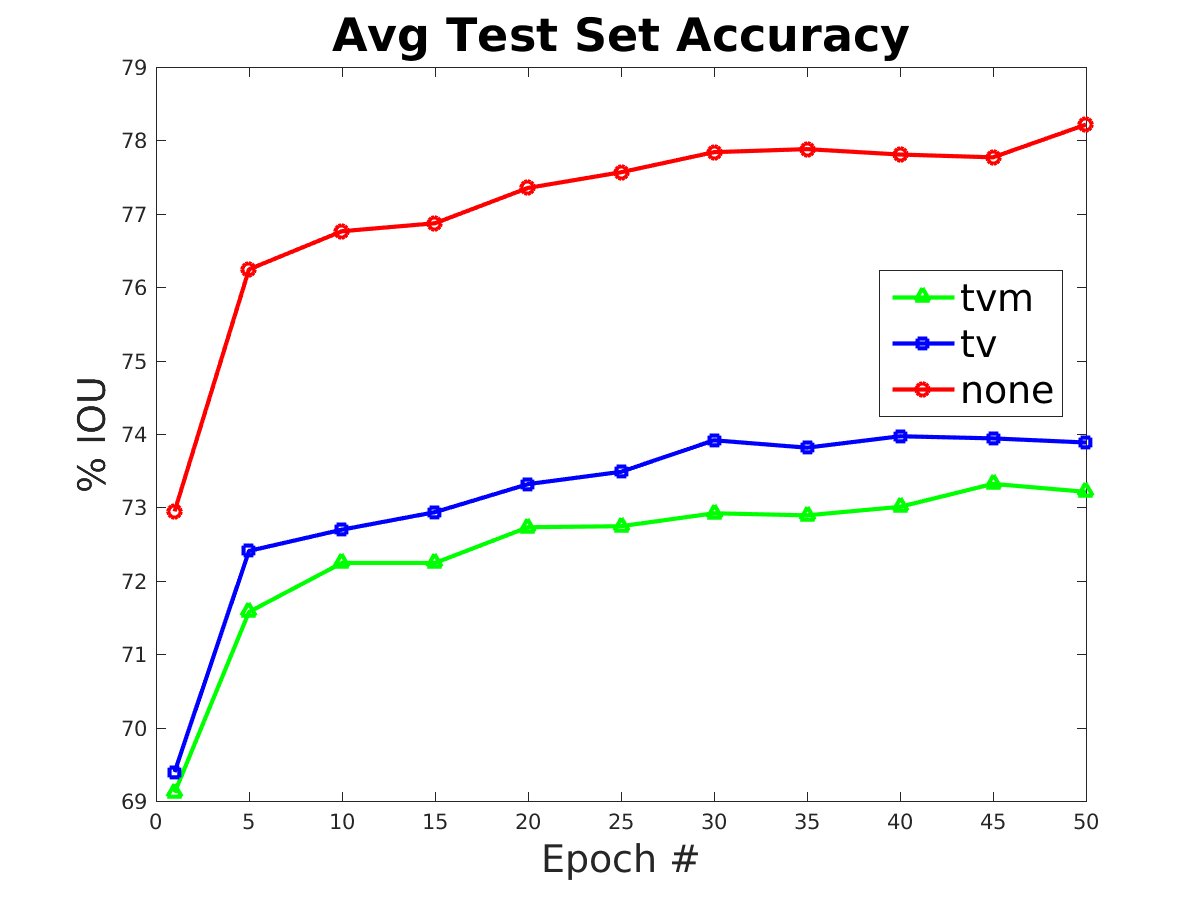} &
\includegraphics[height=\sfig cm]{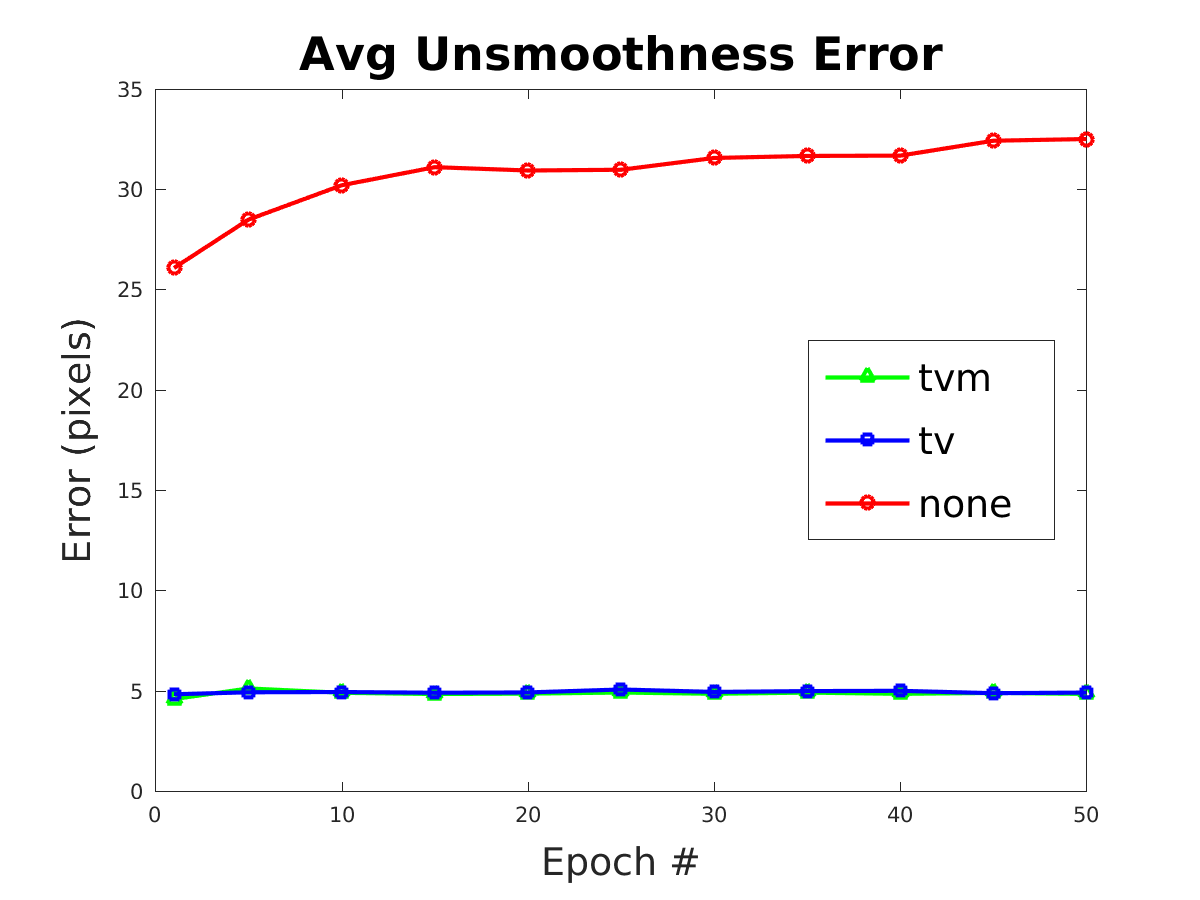} \\
\end{tabular} 
\caption{Quantitative ablation study for no regularization (none: red), total variation regularization (tv: blue) and total variation regularization plus axial monotonicity (tvm: green). Left: increasing gains in accuracy do not necessarily correspond to plausible warp fields (visuals in Figure~\ref{fig:visual_reg_ablation}). Right: observe that without regularization the unsmoothness error metric (Equation~\ref{eq:TV_reg_id}) increases as training advances, however the smoothness remains constant with regularization present.}
\label{fig:itr_reg_plot}
\end{figure}

The fidelity term in \ourmethod{} (Equation~\ref{eq:FFDnet_loss}) without additional regularization is a greedy criteria, aspiring to improve the alignment accuracy (synonymous to IOU) at any cost. Without any regularization, advanced training iterations yield increasingly unsmooth warp fields. This continued training results in a deceitful improvement in alignment accuracy, since the resulting deformation is inconsistent with plausible FFDs (visuals in Figure~\ref{fig:visual_reg_ablation}). Importantly, training with TV regularization preserves smooth fields for increasing training iterations, stipulating faithful improvements in alignment accuracy (plots in Figure~\ref{fig:itr_reg_plot}). Naturally, incorporating an additional regularization term leads to slightly inferior IOU accuracy, but gains significantly smoother, and therefore more conceivable, warp fields.

\begin{figure*}
\newcommand{\gridresfig}{2.1}

\begin{tabular}{c c c c c c c}

\includegraphics[height=\gridresfig cm]{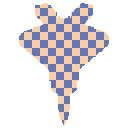} &
\includegraphics[height=\gridresfig cm]{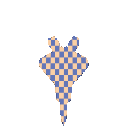} &
\includegraphics[height=\gridresfig cm]{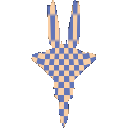} &
\includegraphics[height=\gridresfig cm]{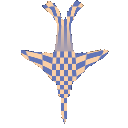} &
\includegraphics[height=\gridresfig cm]{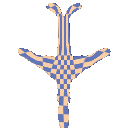} &
\includegraphics[height=\gridresfig cm]{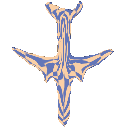} &
\includegraphics[height=\gridresfig cm]{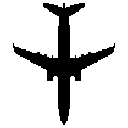} \\

source &
$2\times2$ &
$4\times4$ &
$6\times6$ &
$8\times8$ &
\new{$23\times23$} &
target \\

\end{tabular}
\caption{Grid resolution expressiveness. We experiment with increasing grid resolutions to explore the change in warping flexibility. Results were obtained using \ourmethod{} under the exact same conditions, apart from grid size. \new{Observe that a smaller grid resolution facilitates a smooth but not expressive enough deformation, while a higher resolution yields  a deformation that is more expressive \newa{but contains redundant deformations.}}}
\label{fig:grid_res}

\end{figure*}

\new{\newa{Furthermore, we explore the effects of varying grid resolutions on deformation smoothness, expressivity and high-frequency oscillations of the shape boundary and internal structure (Figure~\ref{fig:grid_res})}. A grid of very low resolution (\emph{e.g.,} 2$\times$2 or 4$\times$4) produces an inherently low-frequency deformation. Yet, the alignments in these resolutions are noticeably inadequate for the disparate geometries between the airplanes. Conversely, a grid of very high resolution (\emph{e.g.,} 23$\times$23) produces an expressive deformation, which tends to contain redundant oscillations. Note in some cases it may be possible to achieve finer results using a higher grid resolution, this would necessitate more training data, deeper networks, and longer training time. For this reason, we opt for a moderate grid resolution (\emph{e.g.,} 8$\times$8), to produce smooth and expressive deformations.}

\begin{figure}
\newcommand{\rfig}{1.6}
\setlength\tabcolsep{2pt}
\begin{tabular}{ c c c c }
\includegraphics[width=\rfig cm]{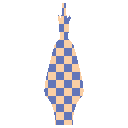} &
\includegraphics[width=\rfig cm]{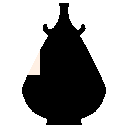} &
\includegraphics[width=\rfig cm]{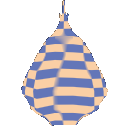} &
\includegraphics[width=\rfig cm]{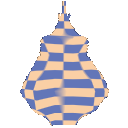} \\
source &
target & 
isotropic &
anisotropic \\
\end{tabular}
\caption{\revb{\ourmethod{} is modular, with the ability to employ alternative forms of regularization depending on the application. Applying isotropic vs. anisotropic TV regularization can lead to slight differences in the outcome.}}
\label{fig:isotv}
\end{figure}

\revb{Additionally, we investigated using an alternative loss function (results in Figure~\ref{fig:isotv}), namely isotropic total variation.
Recall Equation~\ref{eq:TV_reg} which defined anisotropic TV, isotropic TV is given by}
\begin{equation}
\revb{\ell_{TV}(W) = ||\sqrt{\Delta W_x^2 + \Delta W_y^2} ||_1,}
\end{equation}
\revb{where the square and square-root operations are applied elementwise.}

\subsection{\ourmethod{} Generalization}
\begin{figure*}[h]
\newcommand{\nnfigc}{1.7}
\setlength\tabcolsep{1pt} 
\begin{tabular}{ c c c c c c c c c c}
 & 
\includegraphics[width=\nnfigc cm]{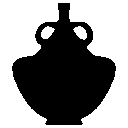} &
\includegraphics[width=\nnfigc cm]{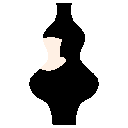} &
\includegraphics[width=\nnfigc cm]{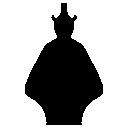} &
\includegraphics[width=\nnfigc cm]{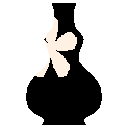} &
 & 
\includegraphics[width=\nnfigc cm]{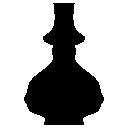} &
\includegraphics[width=\nnfigc cm]{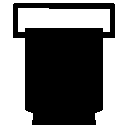} &
\includegraphics[width=\nnfigc cm]{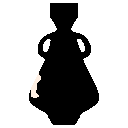} &
\includegraphics[width=\nnfigc cm]{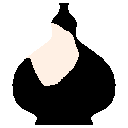} \\
\includegraphics[width=\nnfigc cm]{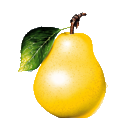} &
\includegraphics[width=\nnfigc cm]{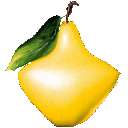} &
\includegraphics[width=\nnfigc cm]{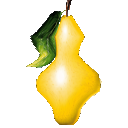} &
\includegraphics[width=\nnfigc cm]{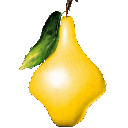} &
\includegraphics[width=\nnfigc cm]{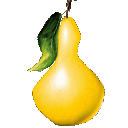} &
\includegraphics[width=\nnfigc cm]{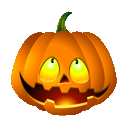} &
\includegraphics[width=\nnfigc cm]{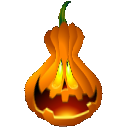} &
\includegraphics[width=\nnfigc cm]{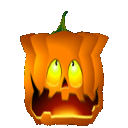} &
\includegraphics[width=\nnfigc cm]{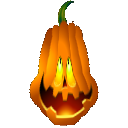} &
\includegraphics[width=\nnfigc cm]{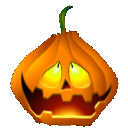} \\
\hline
 & 
\includegraphics[width=\nnfigc cm]{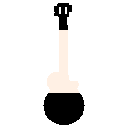} &
\includegraphics[width=\nnfigc cm]{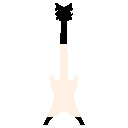} &
\includegraphics[width=\nnfigc cm]{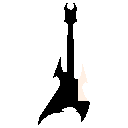} &
\includegraphics[width=\nnfigc cm]{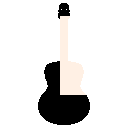} &
 & 
\includegraphics[width=\nnfigc cm]{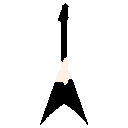} &
\includegraphics[width=\nnfigc cm]{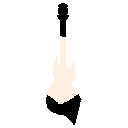} &
\includegraphics[width=\nnfigc cm]{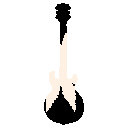} &
\includegraphics[width=\nnfigc cm]{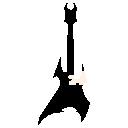} \\
\includegraphics[width=\nnfigc cm]{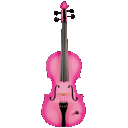} &
\includegraphics[width=\nnfigc cm]{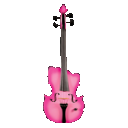} &
\includegraphics[width=\nnfigc cm]{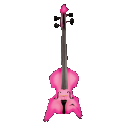} &
\includegraphics[width=\nnfigc cm]{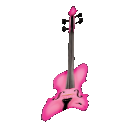} &
\includegraphics[width=\nnfigc cm]{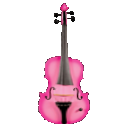} &
\includegraphics[width=\nnfigc cm]{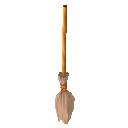} &
\includegraphics[width=\nnfigc cm]{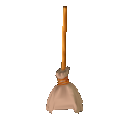} &
\includegraphics[width=\nnfigc cm]{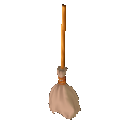} &
\includegraphics[width=\nnfigc cm]{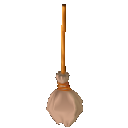} &
\includegraphics[width=\nnfigc cm]{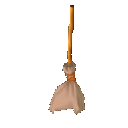} \\
\hline
 & 
\includegraphics[width=\nnfigc cm]{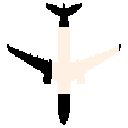} &
\includegraphics[width=\nnfigc cm]{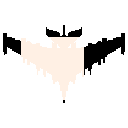} &
\includegraphics[width=\nnfigc cm]{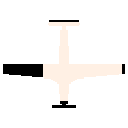} &
\includegraphics[width=\nnfigc cm]{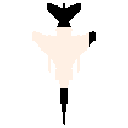} &
 & 
\includegraphics[width=\nnfigc cm]{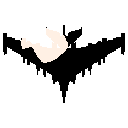} &
\includegraphics[width=\nnfigc cm]{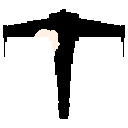} &
\includegraphics[width=\nnfigc cm]{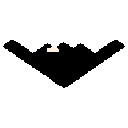} &
\includegraphics[width=\nnfigc cm]{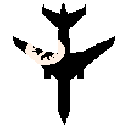} \\
\includegraphics[width=\nnfigc cm]{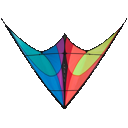} &
\includegraphics[width=\nnfigc cm]{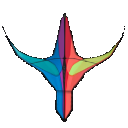} &
\includegraphics[width=\nnfigc cm]{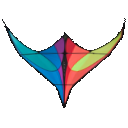} &
\includegraphics[width=\nnfigc cm]{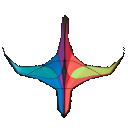} &
\includegraphics[width=\nnfigc cm]{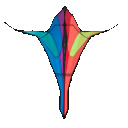} &
\includegraphics[width=\nnfigc cm]{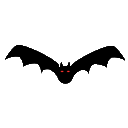} &
\includegraphics[width=\nnfigc cm]{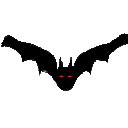} &
\includegraphics[width=\nnfigc cm]{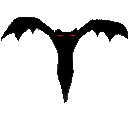} &
\includegraphics[width=\nnfigc cm]{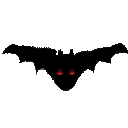} &
\includegraphics[width=\nnfigc cm]{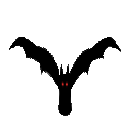} \\
\end{tabular}
\caption{
Qualitative results of applying \ourmethod{} to untrained classes. We utilize \ourmethod{} trained on three different classes to deform source shapes from novel classes (untrained) that share geometric commonalities with one of the trained classes. In these examples, we deform a \emph{pear} and a \emph{pumpkin} into \emph{vases} (top), a \emph{violin} and a \emph{broom} into \emph{guitars} (middle) and a \emph{kite} and a \emph{bat} into \emph{airplanes} (bottom).}
\label{fig:novel_source_class}
\end{figure*}

\begin{figure}[h]
\newcommand{\nnfig}{2}
\setlength\tabcolsep{1pt}
\begin{tabular}{c c c}

\includegraphics[width=\nnfig cm]{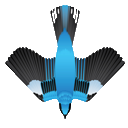} &
\includegraphics[width=\nnfig cm]{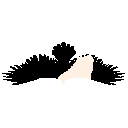} &
\includegraphics[width=\nnfig cm]{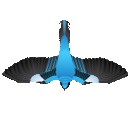} \\

\hline

\includegraphics[width=\nnfig cm]{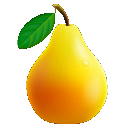} &
\includegraphics[width=\nnfig cm]{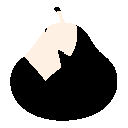} &
\includegraphics[width=\nnfig cm]{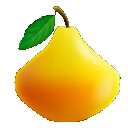} \\

\includegraphics[width=\nnfig cm]{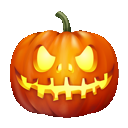} &
\includegraphics[width=\nnfig cm]{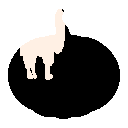} &
\includegraphics[width=\nnfig cm]{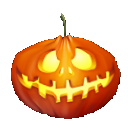} \\

novel source &
novel target &
warped \\

\end{tabular}

\caption{
Qualitative results of applying \ourmethod{} to untrained classes. 
Both source and target shapes belong to novel classes that were not trained on, yet they share geometric commonalities with the trained class.
The top row features a \emph{bird}-to-\emph{bird} deformation, performed using \ourmethod{} trained on \emph{airplanes}, while in the bottom two rows, \emph{pear}-to-\emph{pear} and \emph{pumpkin}-to-\emph{pumpkin} deformations are carried out using a \emph{vase}-trained network.}
\label{fig:novel_novel_class}
\end{figure}

\begin{figure}
\newcommand{\sfig}{1.6}
\setlength\tabcolsep{1pt}
\begin{tabular}{ c c c c c}

 &
\includegraphics[width=\sfig cm]{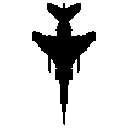} &
\includegraphics[width=\sfig cm]{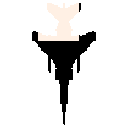} &
\includegraphics[width=\sfig cm]{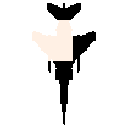} &
\includegraphics[width=\sfig cm]{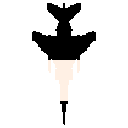} \\

\includegraphics[width=\sfig cm]{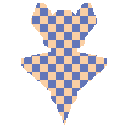} &
\includegraphics[width=\sfig cm]{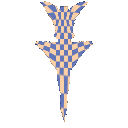} &
\includegraphics[width=\sfig cm]{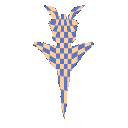} &
\includegraphics[width=\sfig cm]{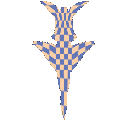} &
\includegraphics[width=\sfig cm]{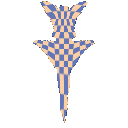} \\

 &
\includegraphics[width=\sfig cm]{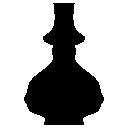} &
\includegraphics[width=\sfig cm]{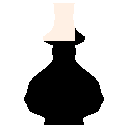} &
\includegraphics[width=\sfig cm]{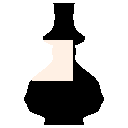} &
\includegraphics[width=\sfig cm]{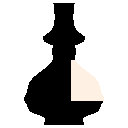} \\

\includegraphics[width=\sfig cm]{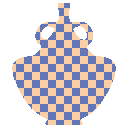} &
\includegraphics[width=\sfig cm]{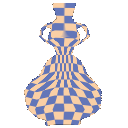} &
\includegraphics[width=\sfig cm]{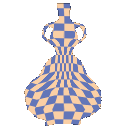} &
\includegraphics[width=\sfig cm]{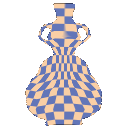} &
\includegraphics[width=\sfig cm]{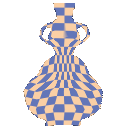} \\

 &
\includegraphics[width=\sfig cm]{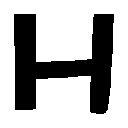} &
\includegraphics[width=\sfig cm]{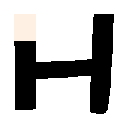} &
\includegraphics[width=\sfig cm]{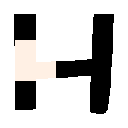} &
\includegraphics[width=\sfig cm]{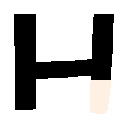} \\

\includegraphics[width=\sfig cm]{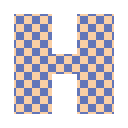} &
\includegraphics[width=\sfig cm]{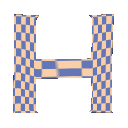} &
\includegraphics[width=\sfig cm]{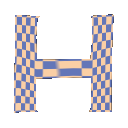} &
\includegraphics[width=\sfig cm]{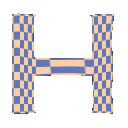} &
\includegraphics[width=\sfig cm]{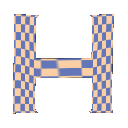} \\

\end{tabular}
\caption{Qualitative demonstration of network generalization and partial-agnostic behavior. Displayed are examples from the test set for three different classes (top: \emph{airplane}, middle: \emph{vase}, bottom: \emph{uppercase H}). The warp computed by \ourmethod{} on a particular source and different partialities of the same target results in a visually consistent deformed shape. }
\label{fig:sweeping_partialT}
\end{figure}

\begin{figure}
\newcommand{\sfig}{1.6}
\setlength\tabcolsep{1pt}
\begin{tabular}{ c c c c c}


 &
\includegraphics[width=\sfig cm]{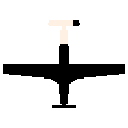} &
\includegraphics[width=\sfig cm]{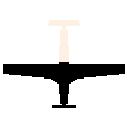} &
\includegraphics[width=\sfig cm]{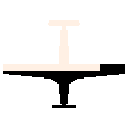} &
\includegraphics[width=\sfig cm]{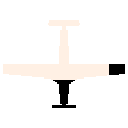} \\

\includegraphics[width=\sfig cm]{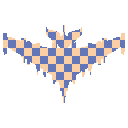} &
\includegraphics[width=\sfig cm]{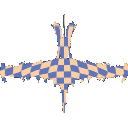} &
\includegraphics[width=\sfig cm]{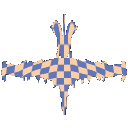} &
\includegraphics[width=\sfig cm]{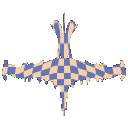} &
\includegraphics[width=\sfig cm]{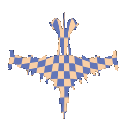} \\

 &
\includegraphics[width=\sfig cm]{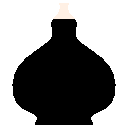} &
\includegraphics[width=\sfig cm]{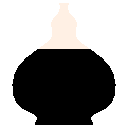} &
\includegraphics[width=\sfig cm]{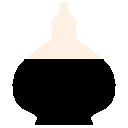} &
\includegraphics[width=\sfig cm]{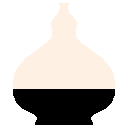} \\

\includegraphics[width=\sfig cm]{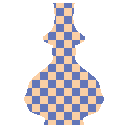} &
\includegraphics[width=\sfig cm]{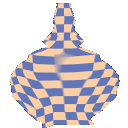} &
\includegraphics[width=\sfig cm]{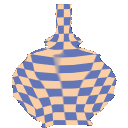} &
\includegraphics[width=\sfig cm]{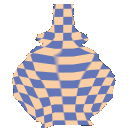} &
\includegraphics[width=\sfig cm]{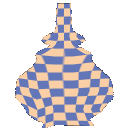} \\

\end{tabular}
\caption{Qualitative stress test for FFD computation under increasing percentage of missing shape data.}
\label{fig:stress_test}
\end{figure}

In this section we demonstrate the generalization capabilities of \ourmethod{}. We show test results on novel classes with similar characteristics, demonstrating \ourmethod{} does not over-fit or memorize the training data. We deform a novel class source to a target from the class trained on in Figure~\ref{fig:novel_source_class}. For example, consider the top section of Figure~\ref{fig:novel_source_class}, where \ourmethod{} trained on vases is used to align a pear and a pumpkin source shapes (novel yet geometrically similar classes) to vases from the test set, resulting in \emph{pear-esque} and \emph{pumpkin-esque} vases. 
We demonstrate further generalization abilities in Figure~\ref{fig:novel_novel_class}, where we utilize a network trained on a certain class to perform alignments from source to target such that \emph{both} belong to a novel class. 
Figure~\ref{fig:sweeping_partialT} features similarly predicted \ourmethod{} alignments for the same source and underlying target, with differing missing element locations. This missing part location indifference simultaneously demonstrates missing part generalization and partial agnosticism of \ourmethod{}. Similarly, in Figure~\ref{fig:stress_test}, we present a \emph{stress test} on the same source and underlying target by computing alignments for an increasing amount of missing data in the target. Observe that the estimated mappings are initially predicted consistently, and even when significant portions of the data are missing, the estimated mappings remain plausible.

\subsection{\rev{Applications in 3D}}
\label{sec:threed}
\begin{figure}
\newcommand{\tfig}{4}
\setlength\tabcolsep{2pt}
\begin{tabular}{c c}


\adjincludegraphics[width=\tfig cm,trim={{.2\width} {.1\height} {.2\width} {.1\height}},clip]{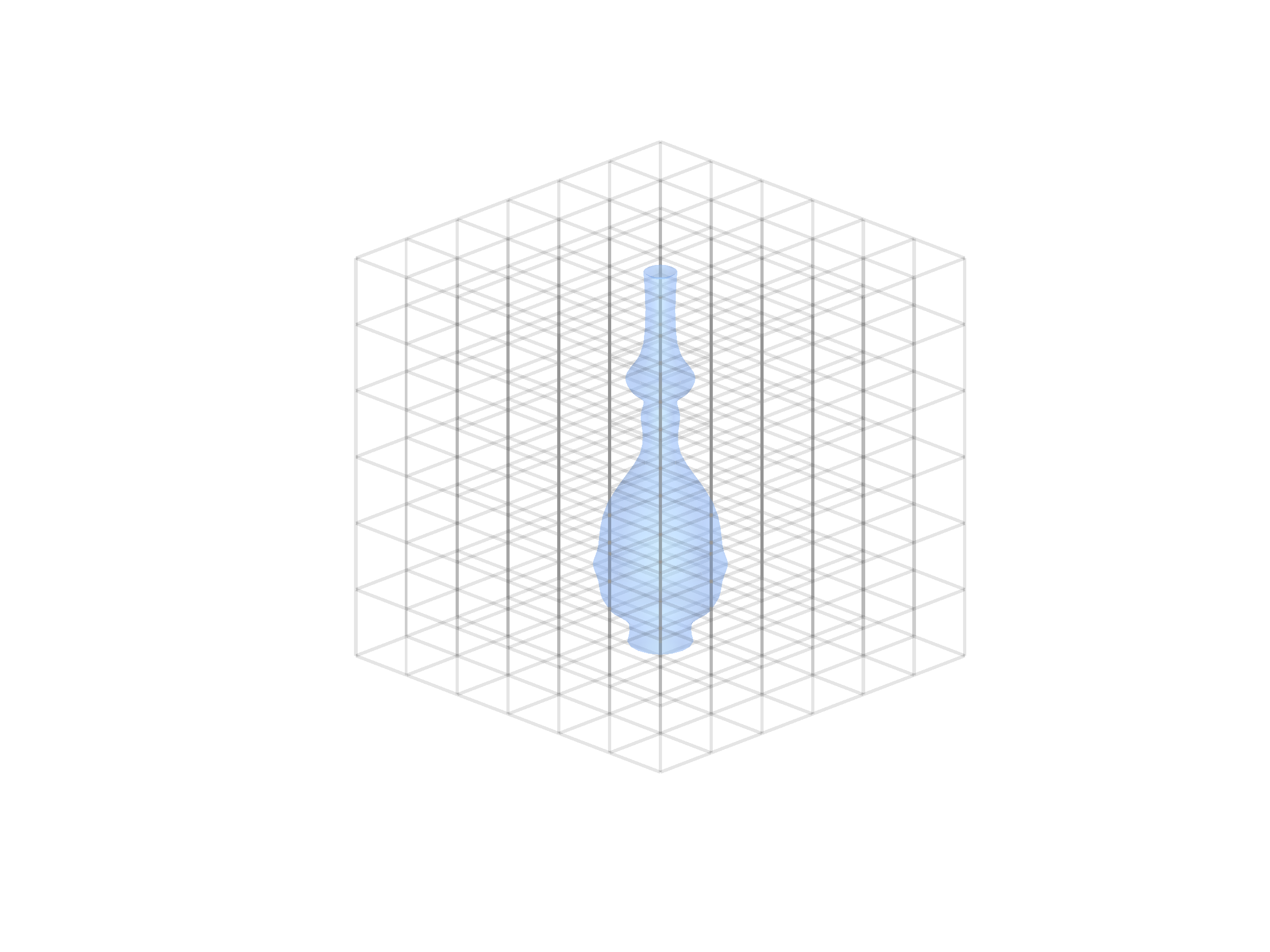} &
\adjincludegraphics[width=\tfig cm,trim={{.2\width} {.1\height} {.2\width} {.1\height}},clip]{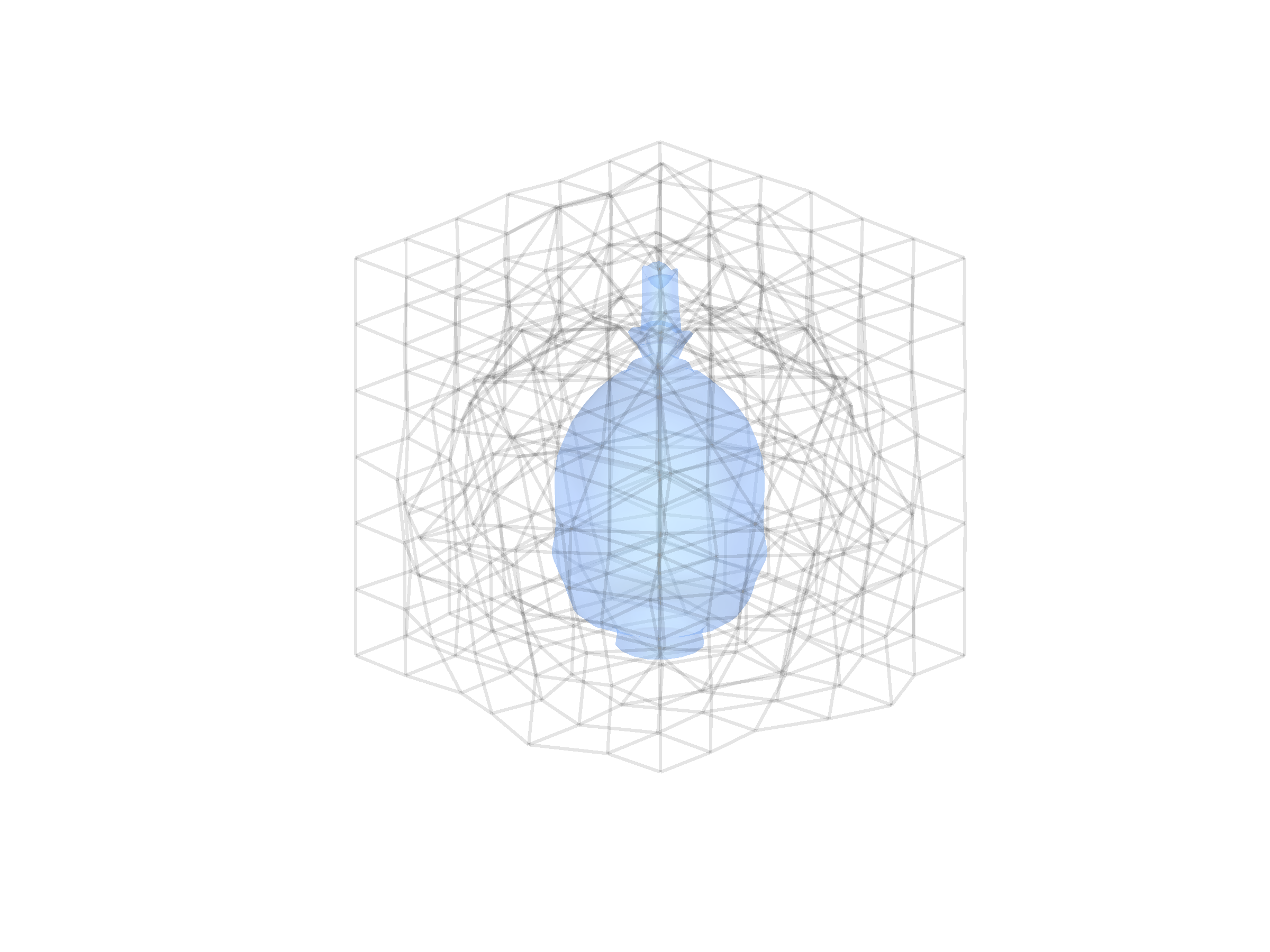} \\

\end{tabular}    
\caption{
\rev{3D deformation grid. The source shape is placed within the uniform 7x7x7 grid (left). A deformation on the grid induces the shape to warp as shown on the right.}}
\label{fig:threeD_grid}
\end{figure}

\rev{
Throughout the paper we have shown alignment results obtained by applying \ourmethod{} on shapes within the 2D domain, and demonstrated the integrity of the estimated warps and their readiness to support alternative post-process 2D applications.
In this section, we shift our focus to the 3D domain, and present additional applications following a simple extension of \ourmethod{} to 3D. We train on volumetric data (of resolution 32x32x32) and replace the convolution, pooling, integrator and FFD layer modules with their 3D counterparts. We learn to regress an FFD grid of resolution 7x7x7 (1029 DoF), which is applied directly on the source mesh vertices. \new{Note that neither the grid nor voxelization resolution affect the quality of the output mesh, since the deformation is applied on the input high-resolution mesh.} See Figure~\ref{fig:threeD_grid} for a visualization of our 3D grid.
}

\rev{
Volumetric CNNs typically operate on voxelized shape representations where not only is the surface of the shape marked as occupied, but also the inner volume. 
Generating a voxelized representation of surface-based shapes represented, \emph{e.g.}, as meshes, is straight-forward.
The same, however, cannot be said for shapes represented, \emph{e.g.}, as point clouds, which sparsely sample the surface of an object.
To this end, we train \ourmethod{} on (hollow) volumetric surfaces automatically generated from 3D meshes.}
\rev{
During training, we draw pairs of source and target shapes from the pool of voxelized surfaces, and remove random portions from the target surface. As in 2D, we train \ourmethod{} by comparing the overlap between the warped source surface and the expected (complete) target surface. While the 3D grid is estimated for the voxelized surface, we can simply apply it on the original mesh.
Below we show applications of 3D warping on point-cloud registration and segmentation transfer.}

\begin{figure}
\newcommand{\sfig}{2.7}
\newcommand{\vfig}{2}
\setlength\tabcolsep{0pt}
\begin{tabular}{ c c c}

\adjincludegraphics[width=\vfig cm,trim={{.28\width} {.15\height} {.28\width} {.15\height}},clip]{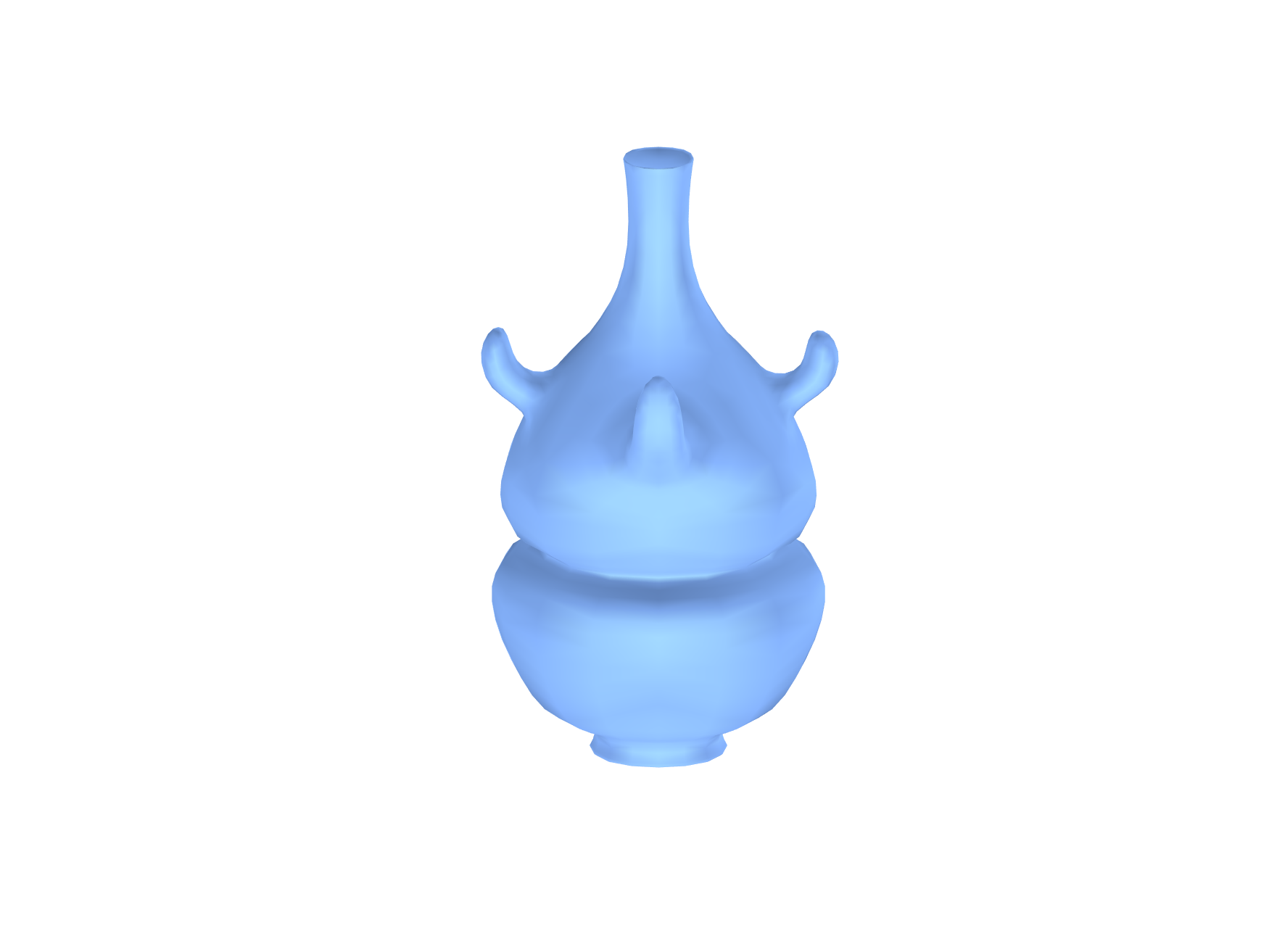} &
\adjincludegraphics[width=\vfig cm,trim={{.28\width} {.15\height} {.28\width} {.15\height}},clip]{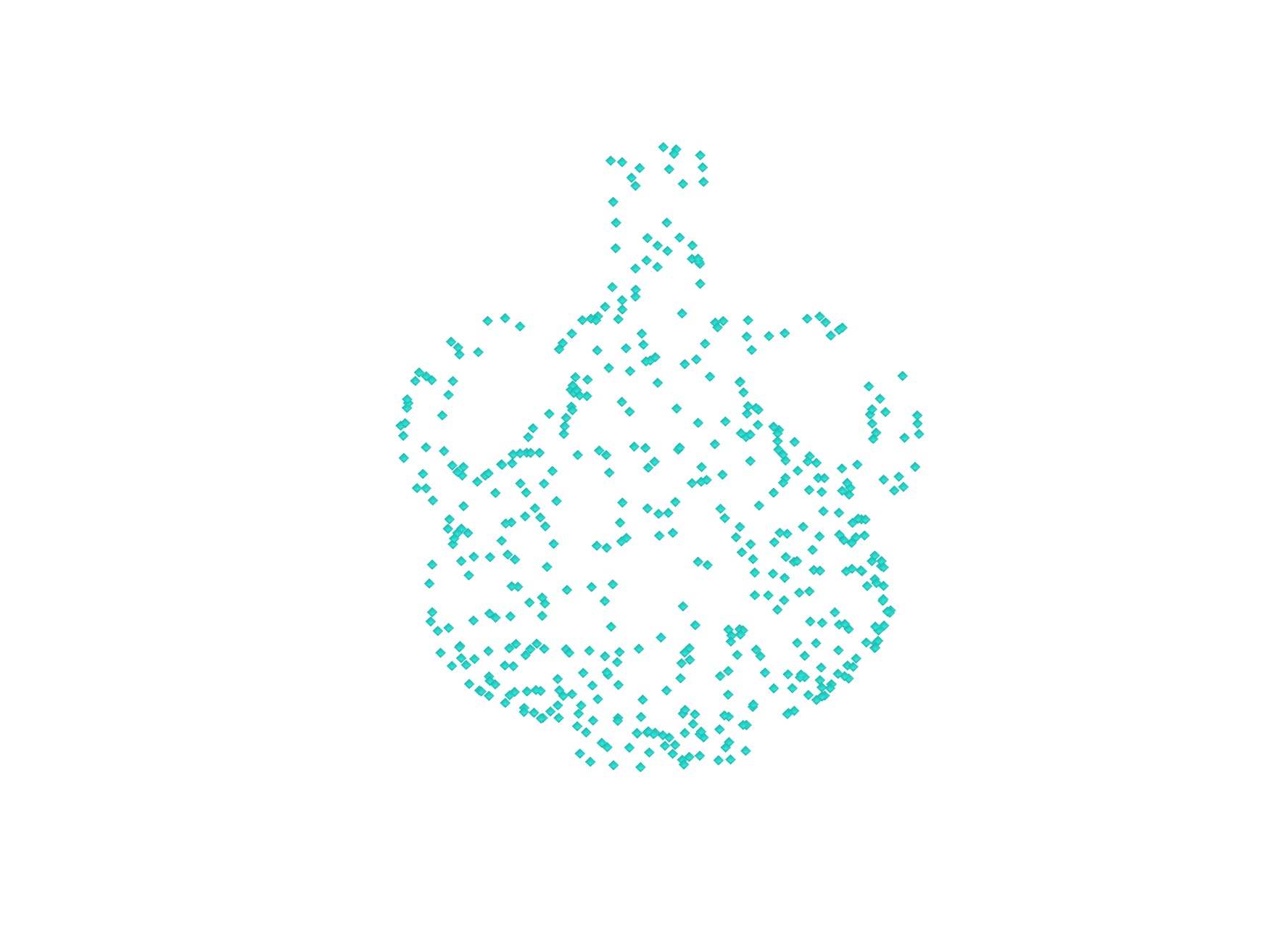} &
\adjincludegraphics[width=\vfig cm,trim={{.28\width} {.15\height} {.28\width} {.15\height}},clip]{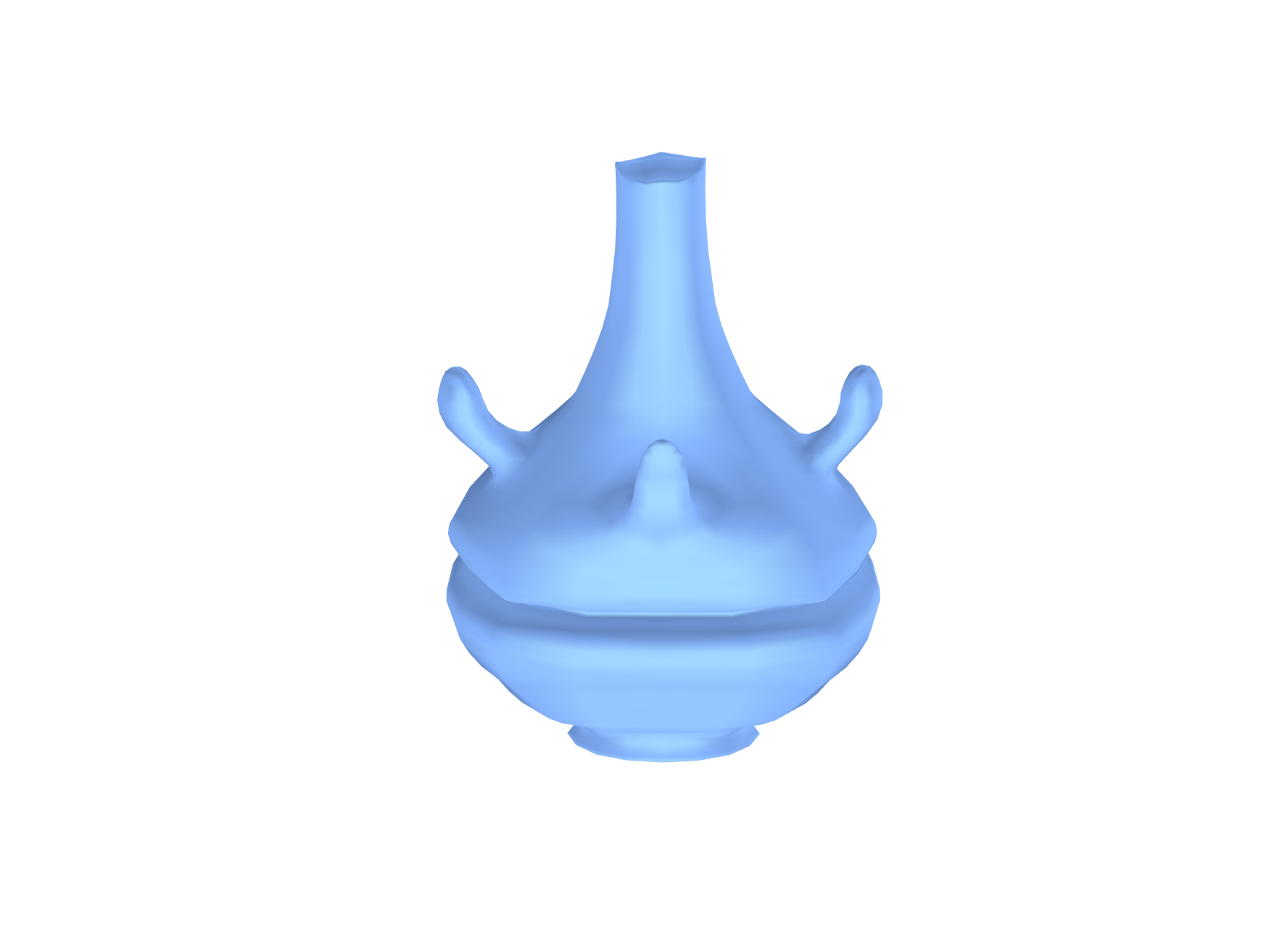} \\

\adjincludegraphics[width=\vfig cm,trim={{.28\width} {.15\height} {.28\width} {.15\height}},clip]{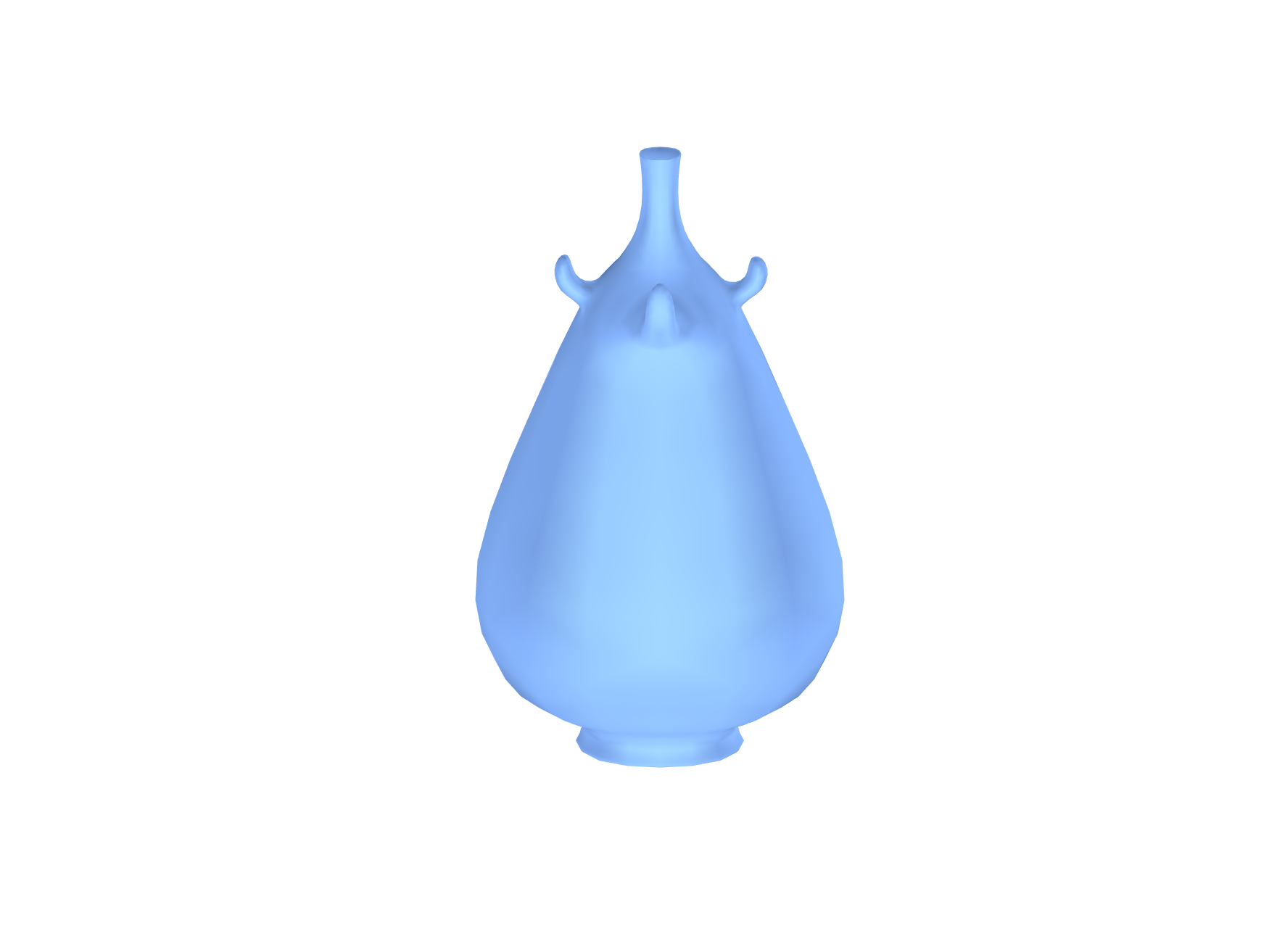} &
\adjincludegraphics[width=\vfig cm,trim={{.28\width} {.15\height} {.28\width} {.15\height}},clip]{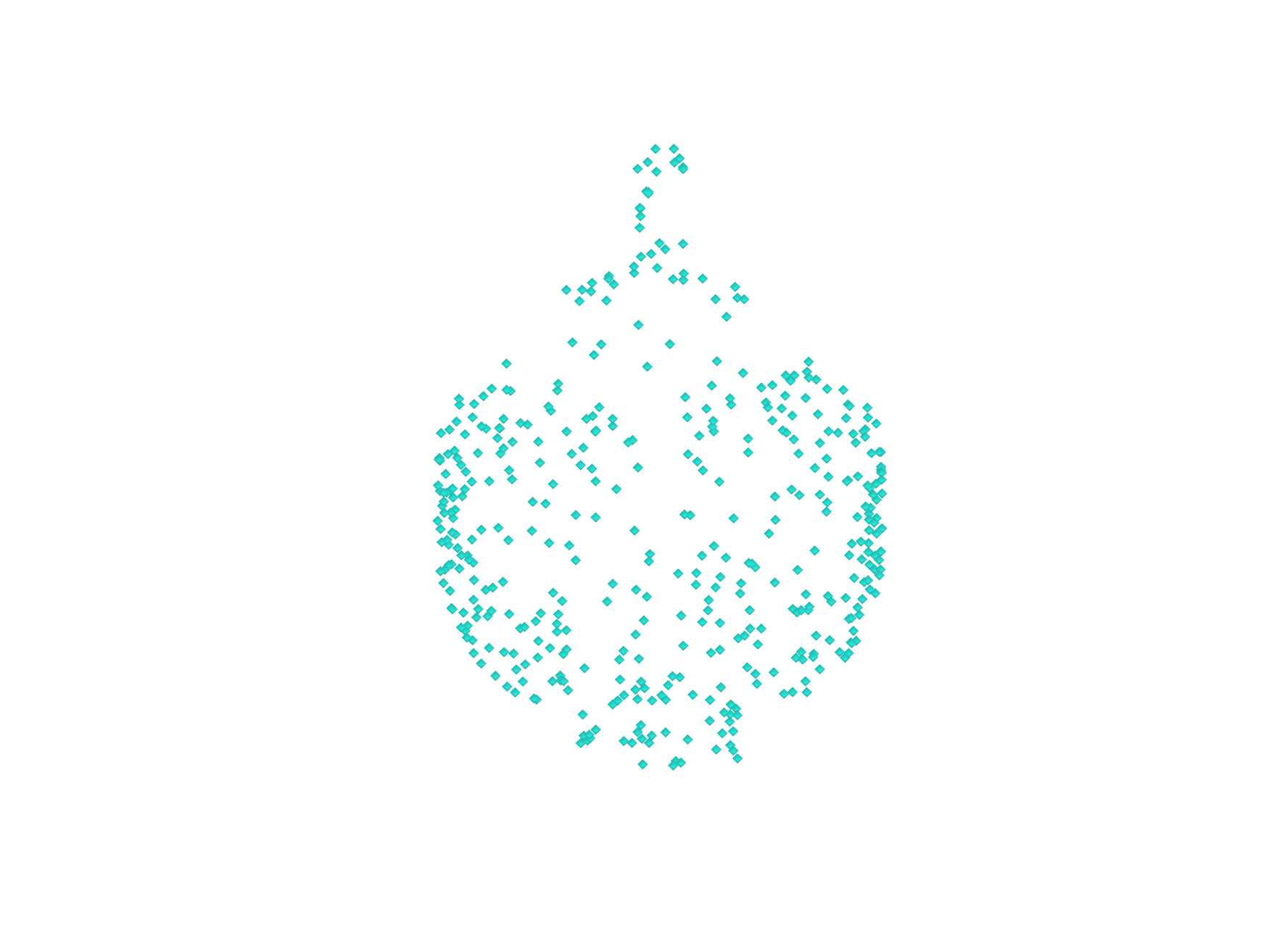} &
\adjincludegraphics[width=\vfig cm,trim={{.28\width} {.15\height} {.28\width} {.15\height}},clip]{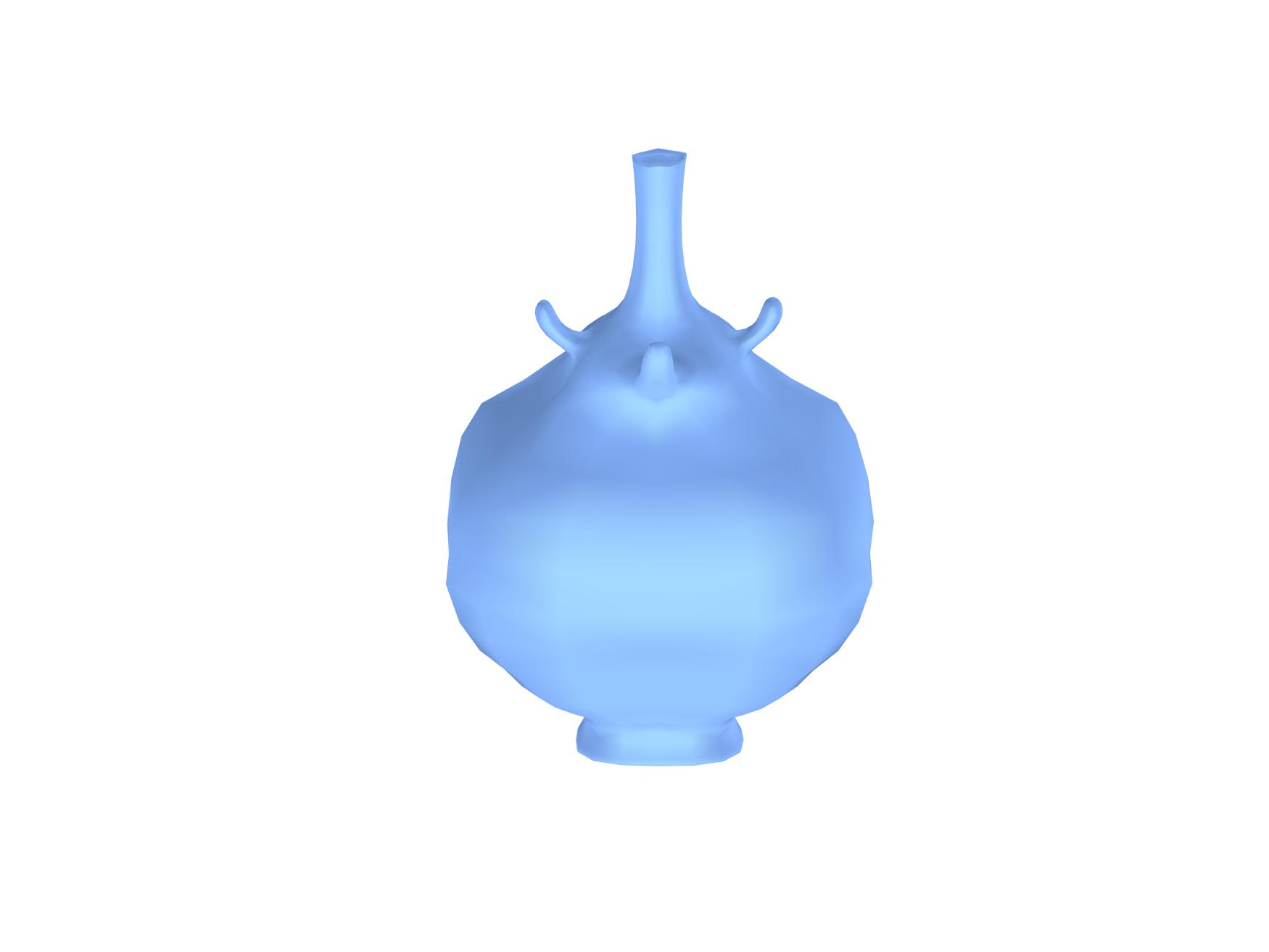} \\

\adjincludegraphics[width=\sfig cm,trim={{.33\width} {.33\height} {.25\width} {.3\height}},clip]{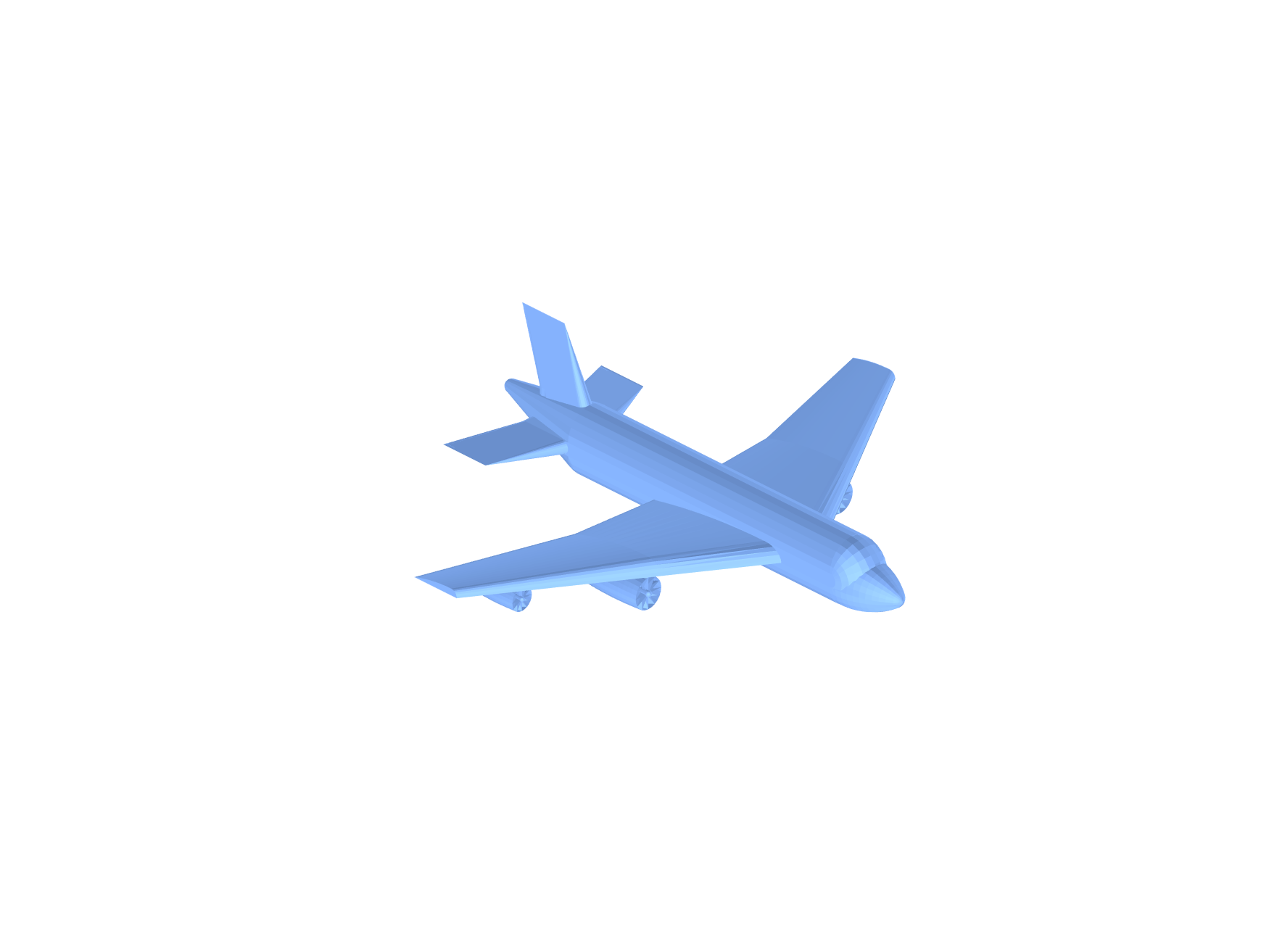} &
\adjincludegraphics[width=\sfig cm,trim={{.33\width} {.33\height} {.25\width} {.3\height}},clip]{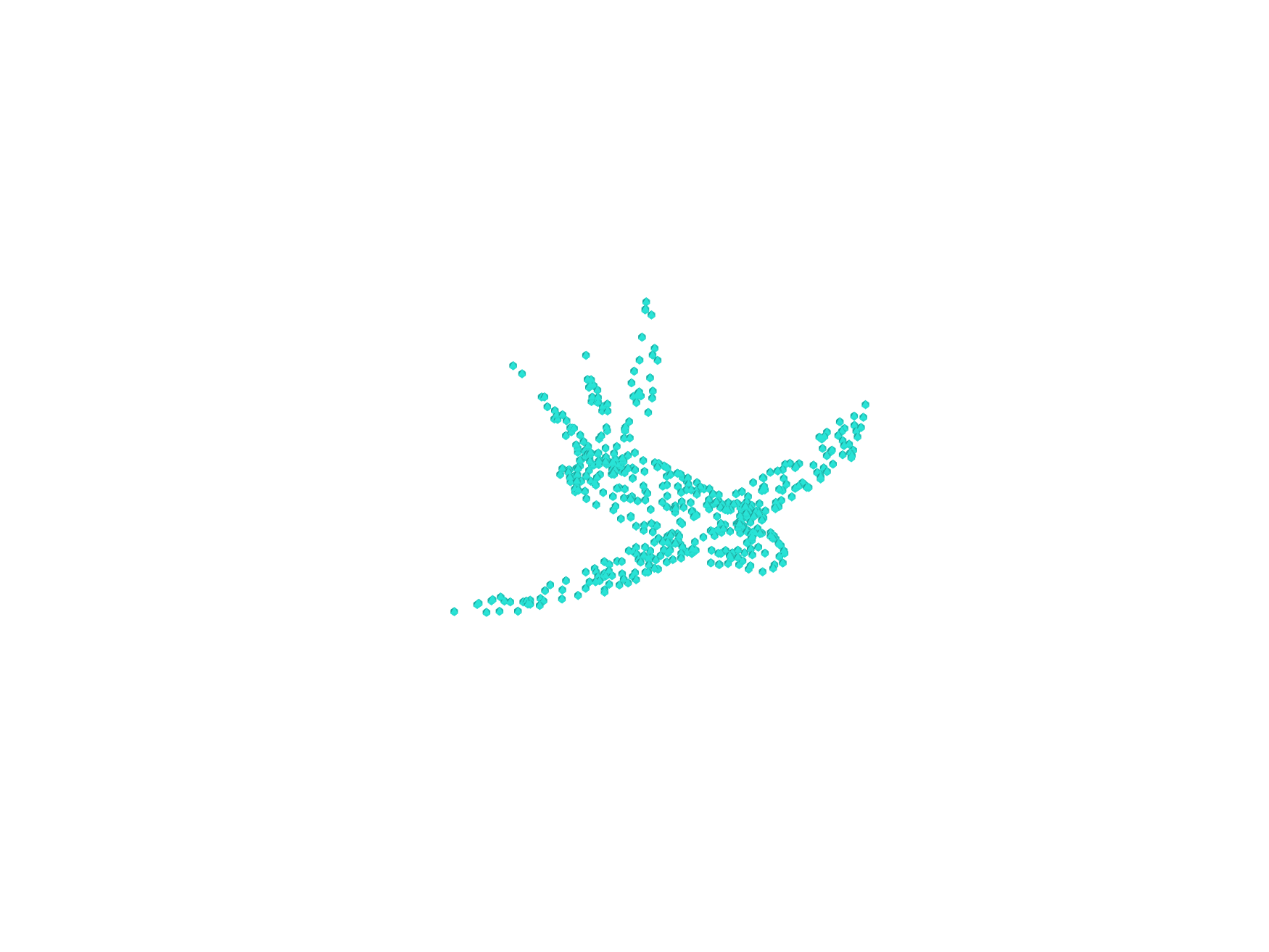} &
\adjincludegraphics[width=\sfig cm,trim={{.33\width} {.33\height} {.25\width} {.3\height}},clip]{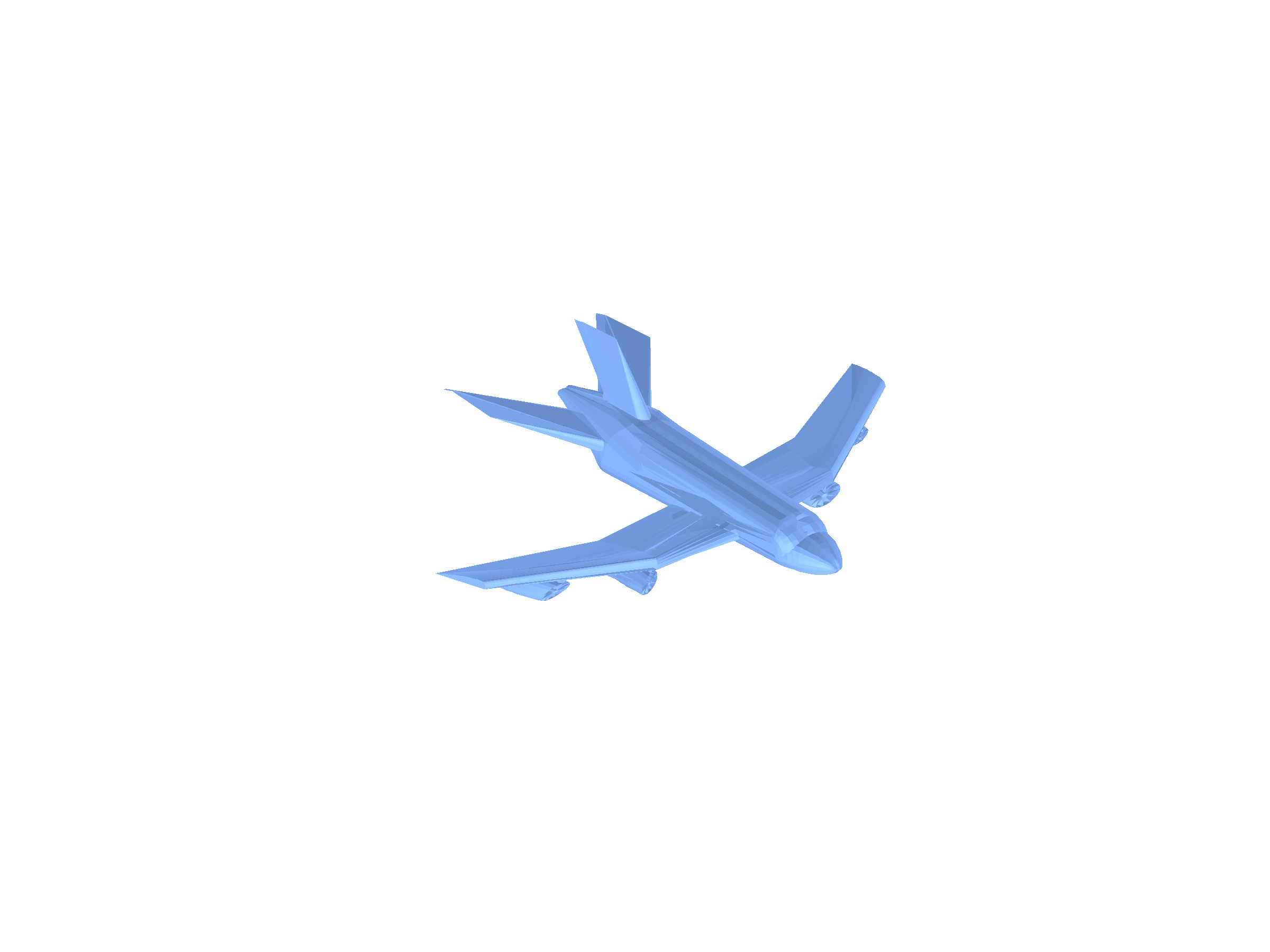} \\

\adjincludegraphics[width=\sfig cm,trim={{.33\width} {.33\height} {.25\width} {.3\height}},clip]{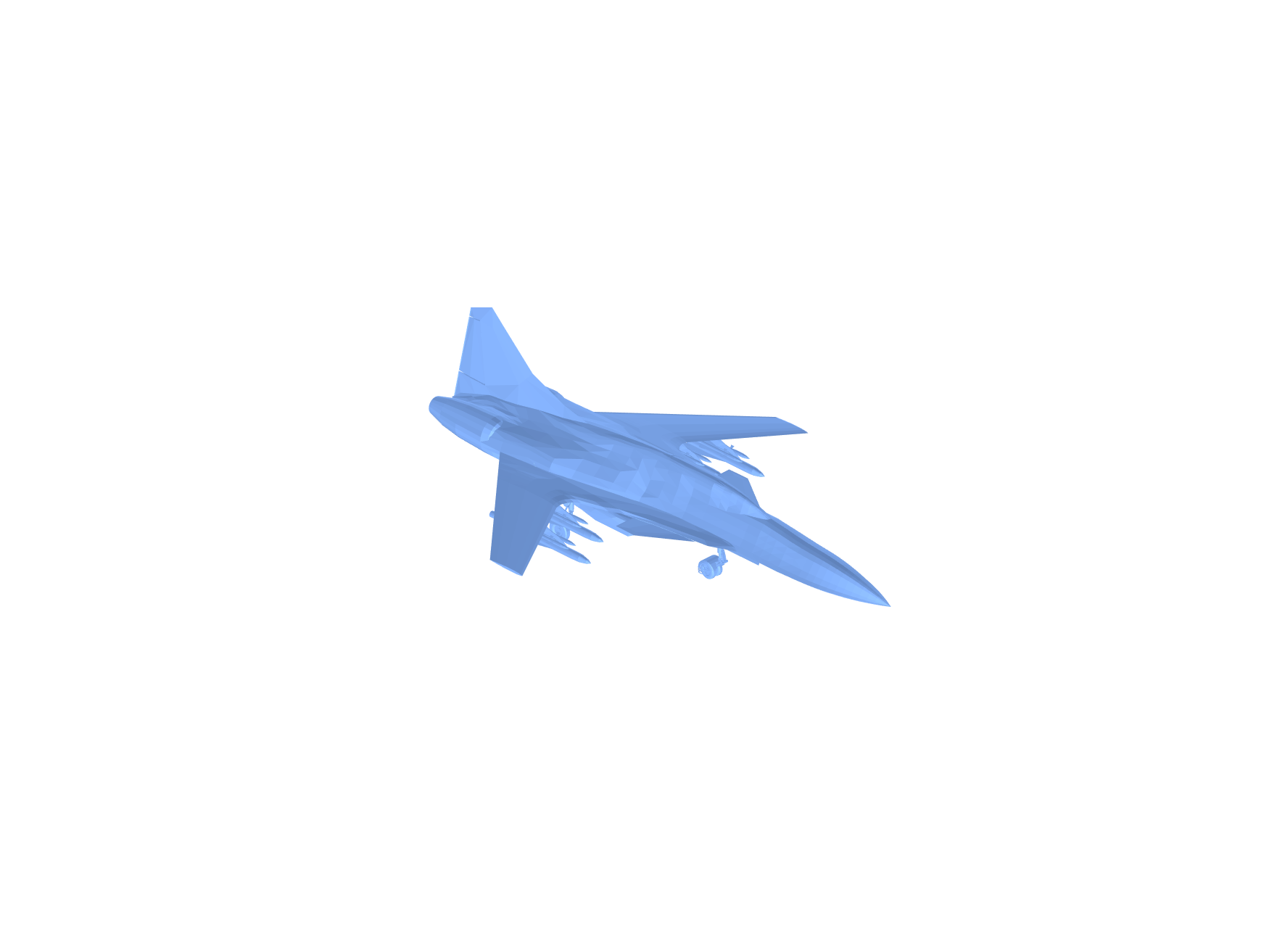} &
\adjincludegraphics[width=\sfig cm,trim={{.33\width} {.33\height} {.25\width} {.3\height}},clip]{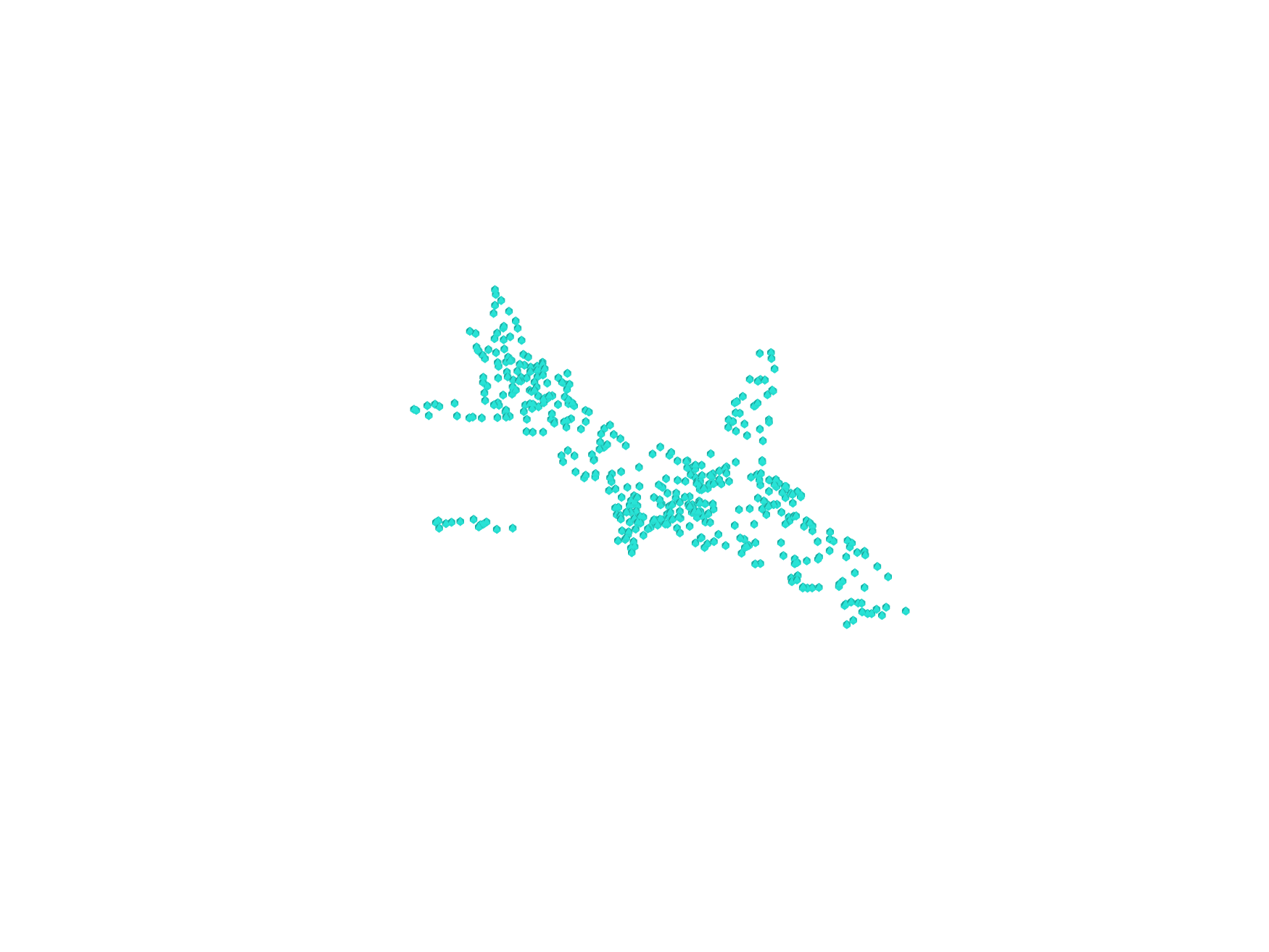} &
\adjincludegraphics[width=\sfig cm,trim={{.33\width} {.33\height} {.25\width} {.3\height}},clip]{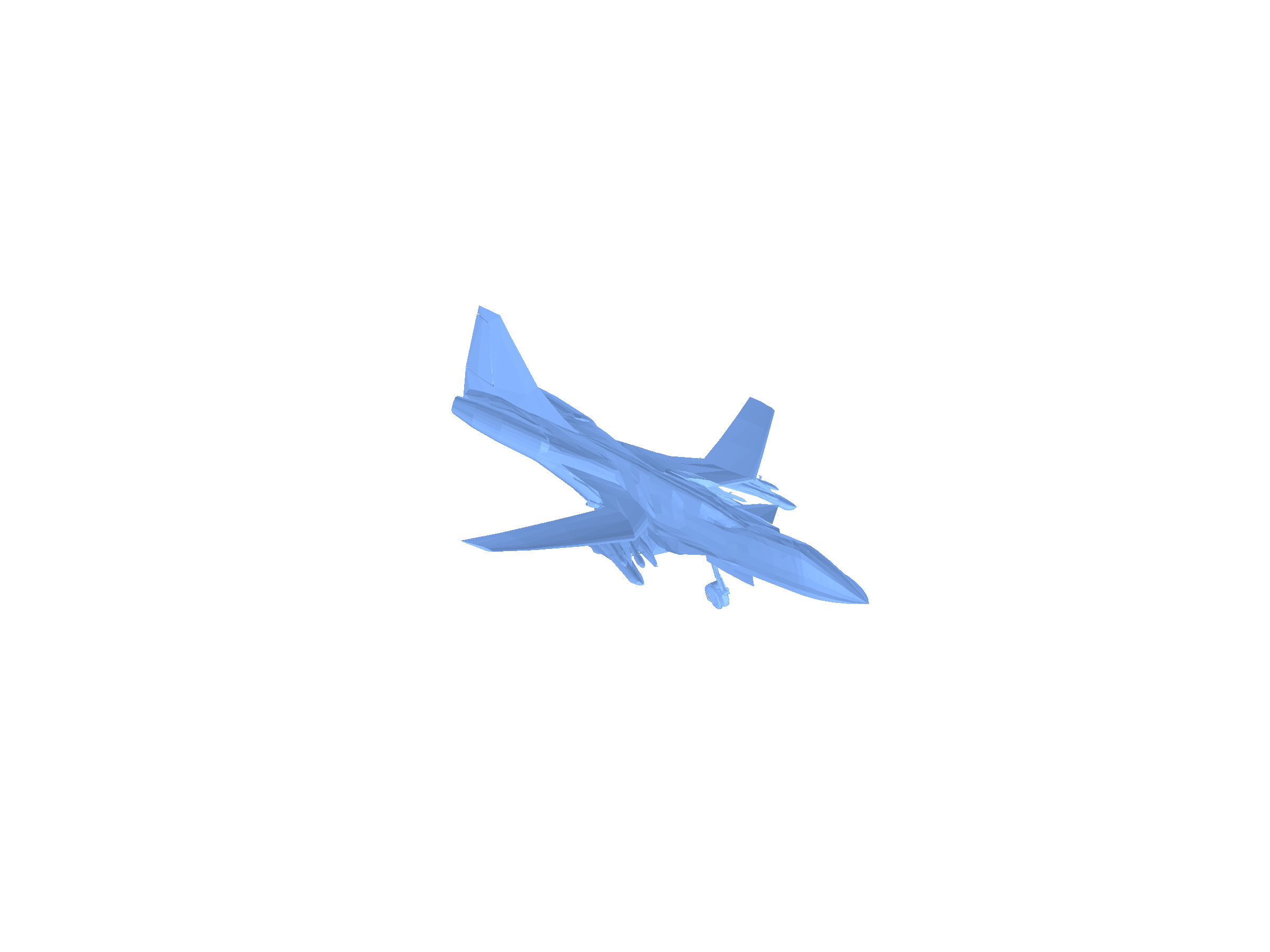} \\

\adjincludegraphics[width=\sfig cm,trim={{.33\width} {.33\height} {.25\width} {.3\height}},clip]{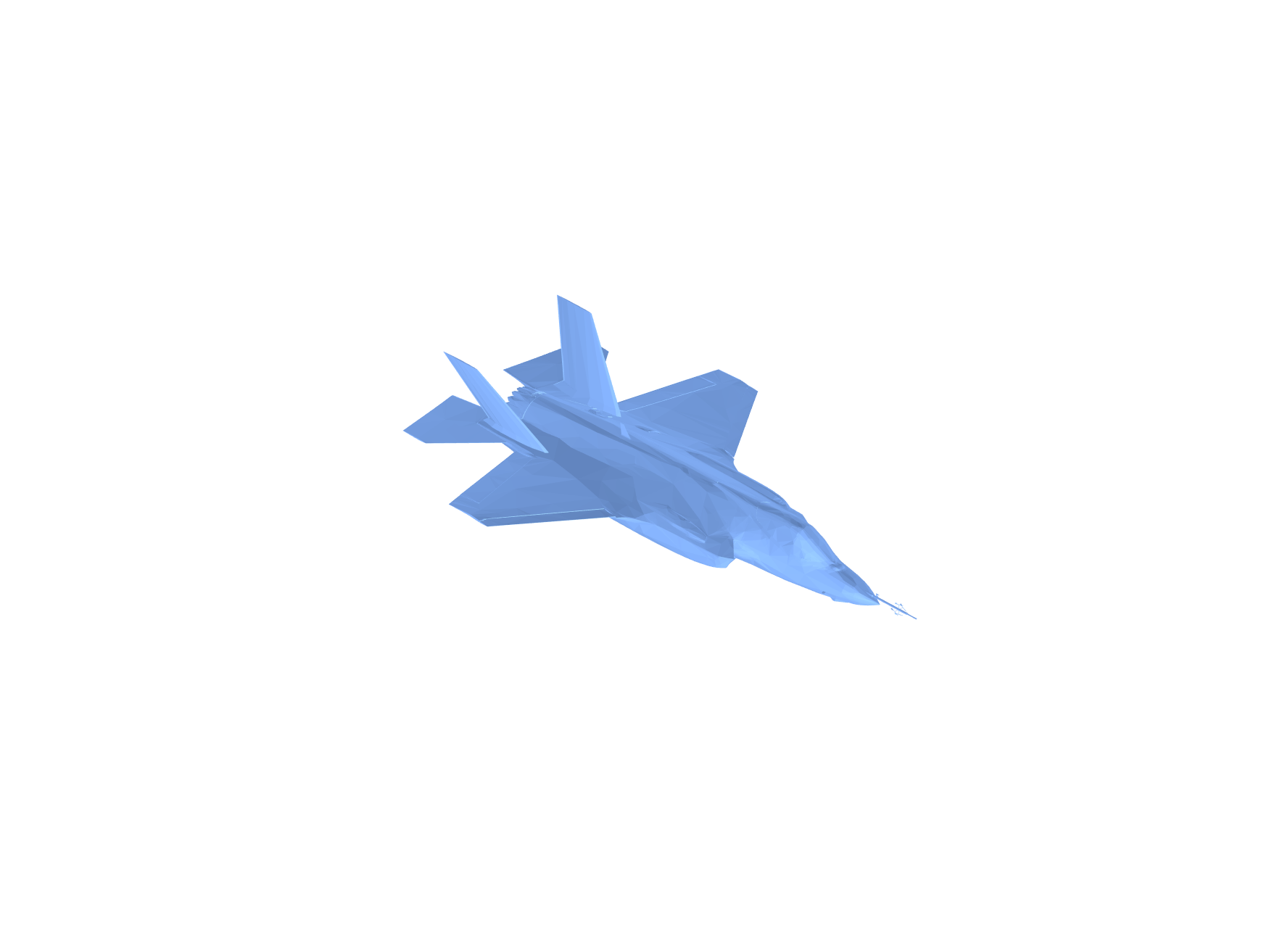} &
\adjincludegraphics[width=\sfig cm,trim={{.33\width} {.33\height} {.25\width} {.3\height}},clip]{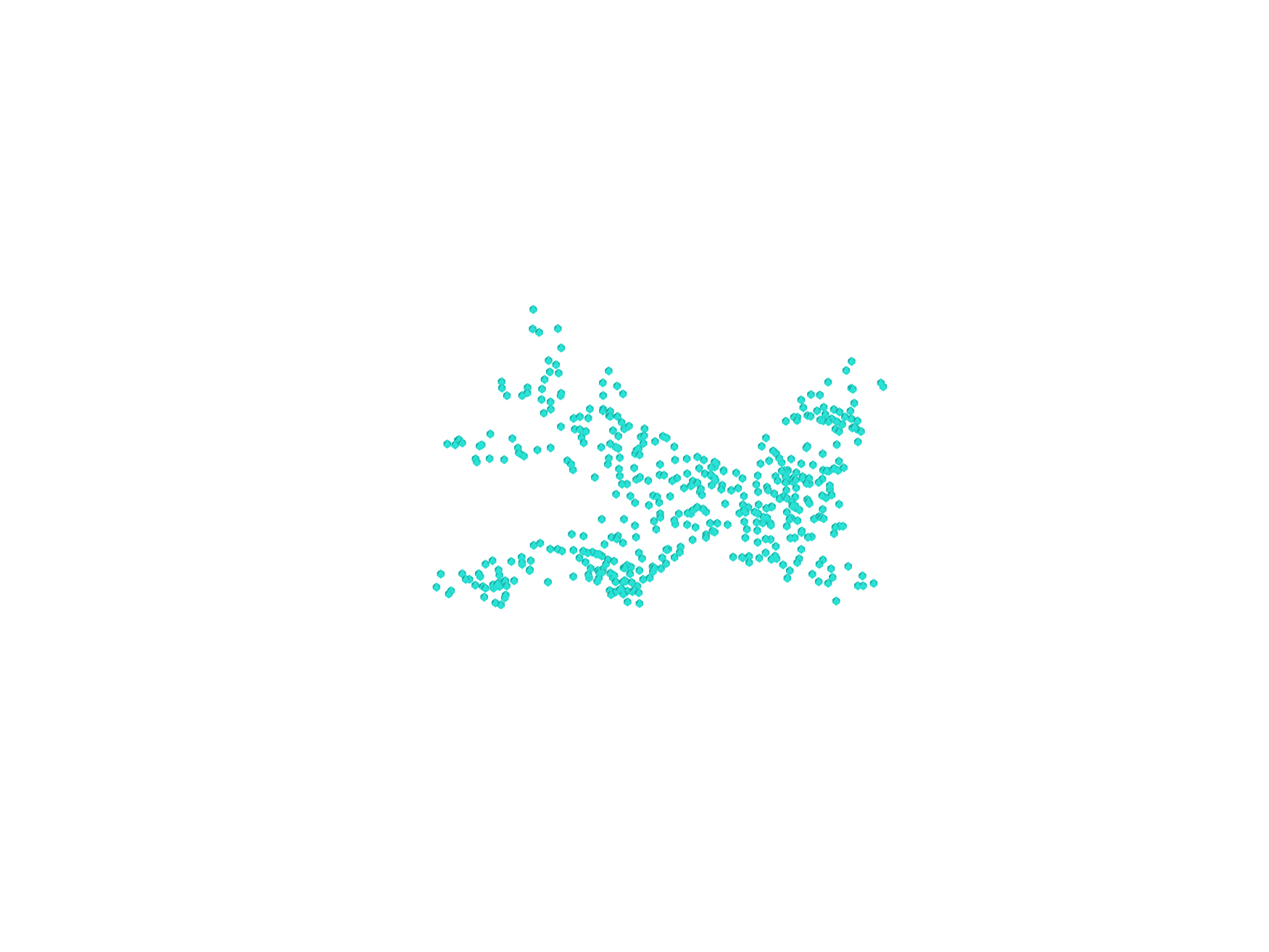} &
\adjincludegraphics[width=\sfig cm,trim={{.33\width} {.33\height} {.25\width} {.3\height}},clip]{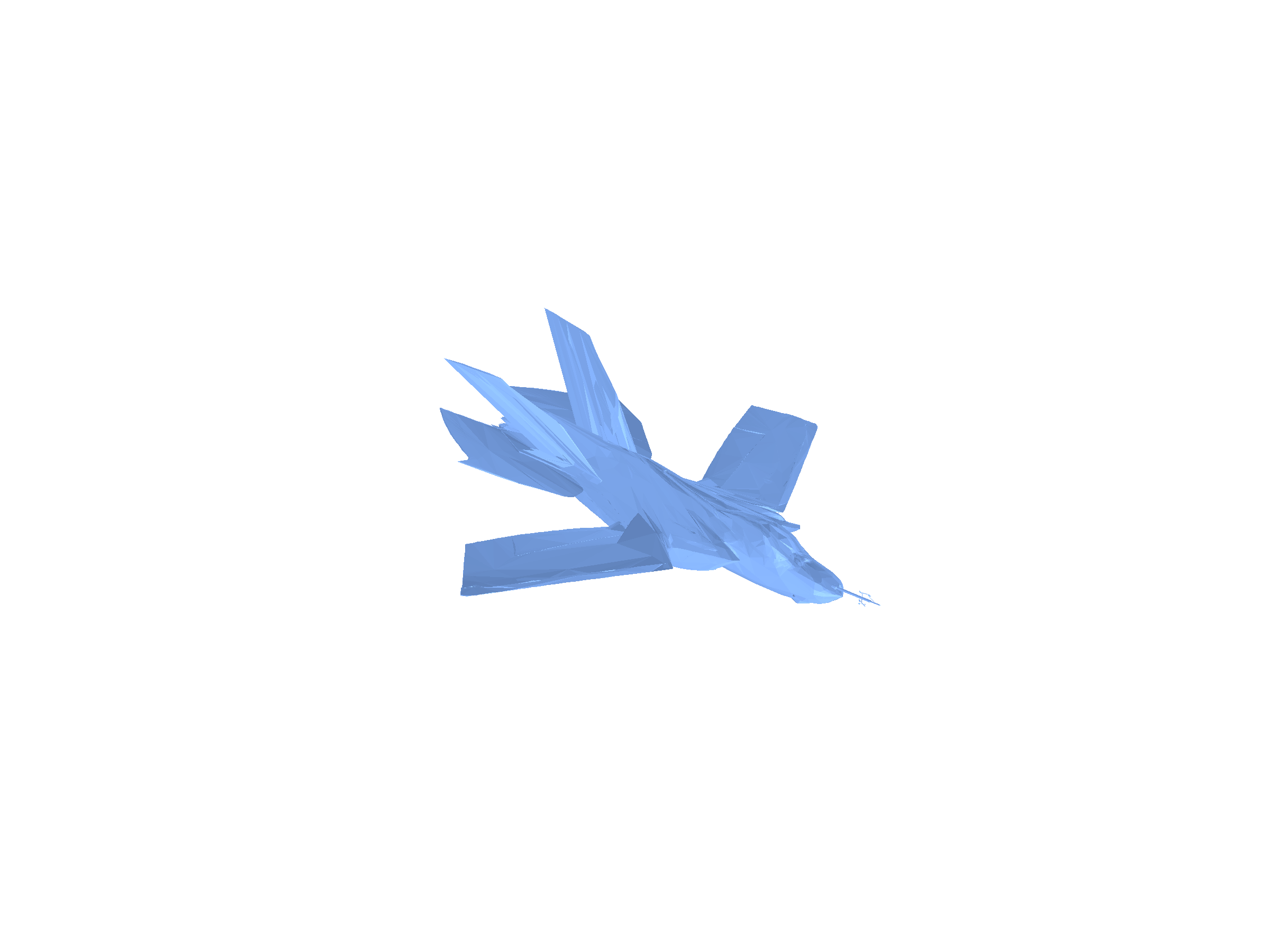} \\

\adjincludegraphics[width=\vfig cm,trim={{.28\width} {.14\height} {.28\width} {.09\height}},clip]{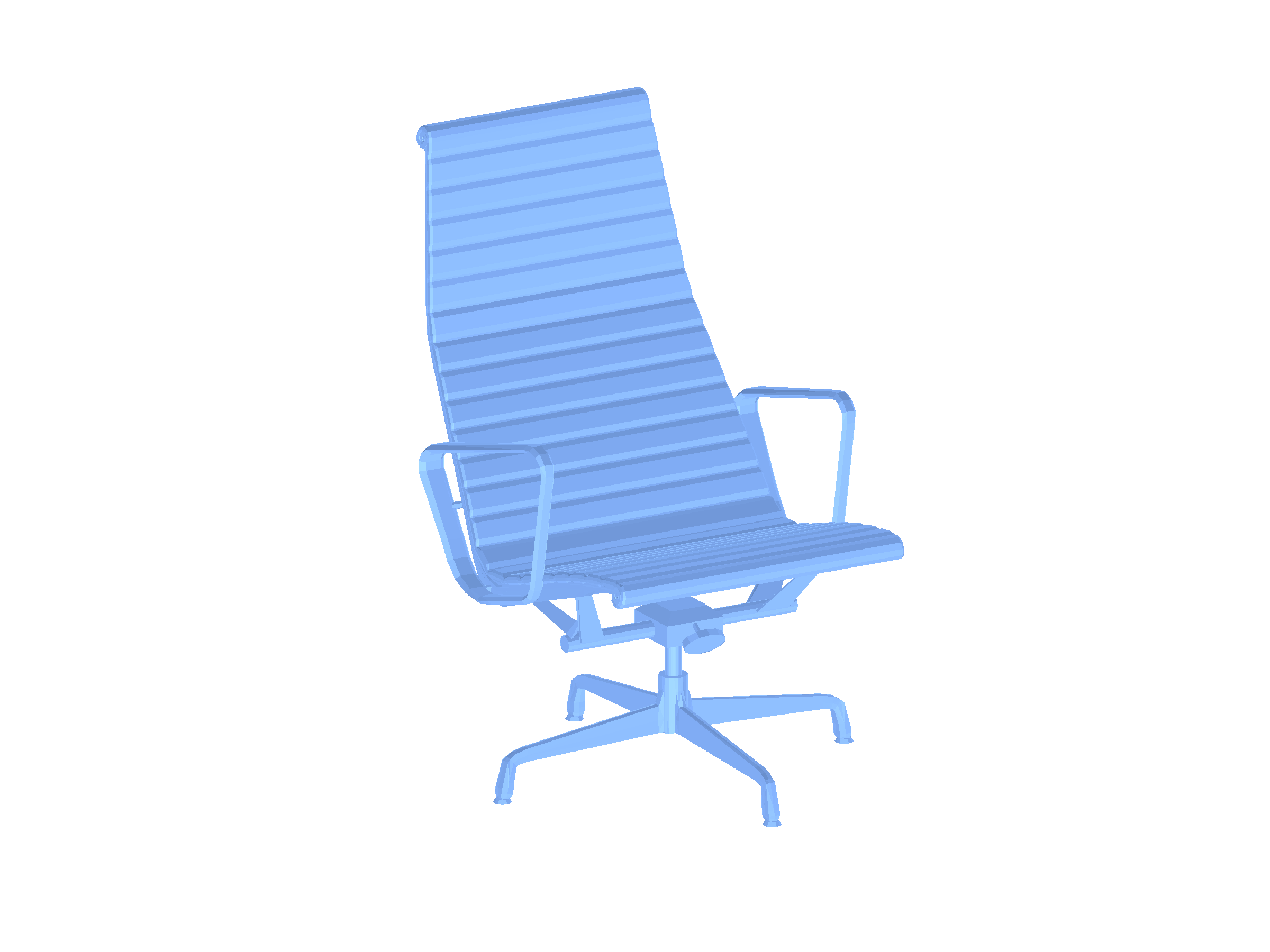} &
\adjincludegraphics[width=\vfig cm,trim={{.28\width} {.14\height} {.28\width} {.09\height}},clip]{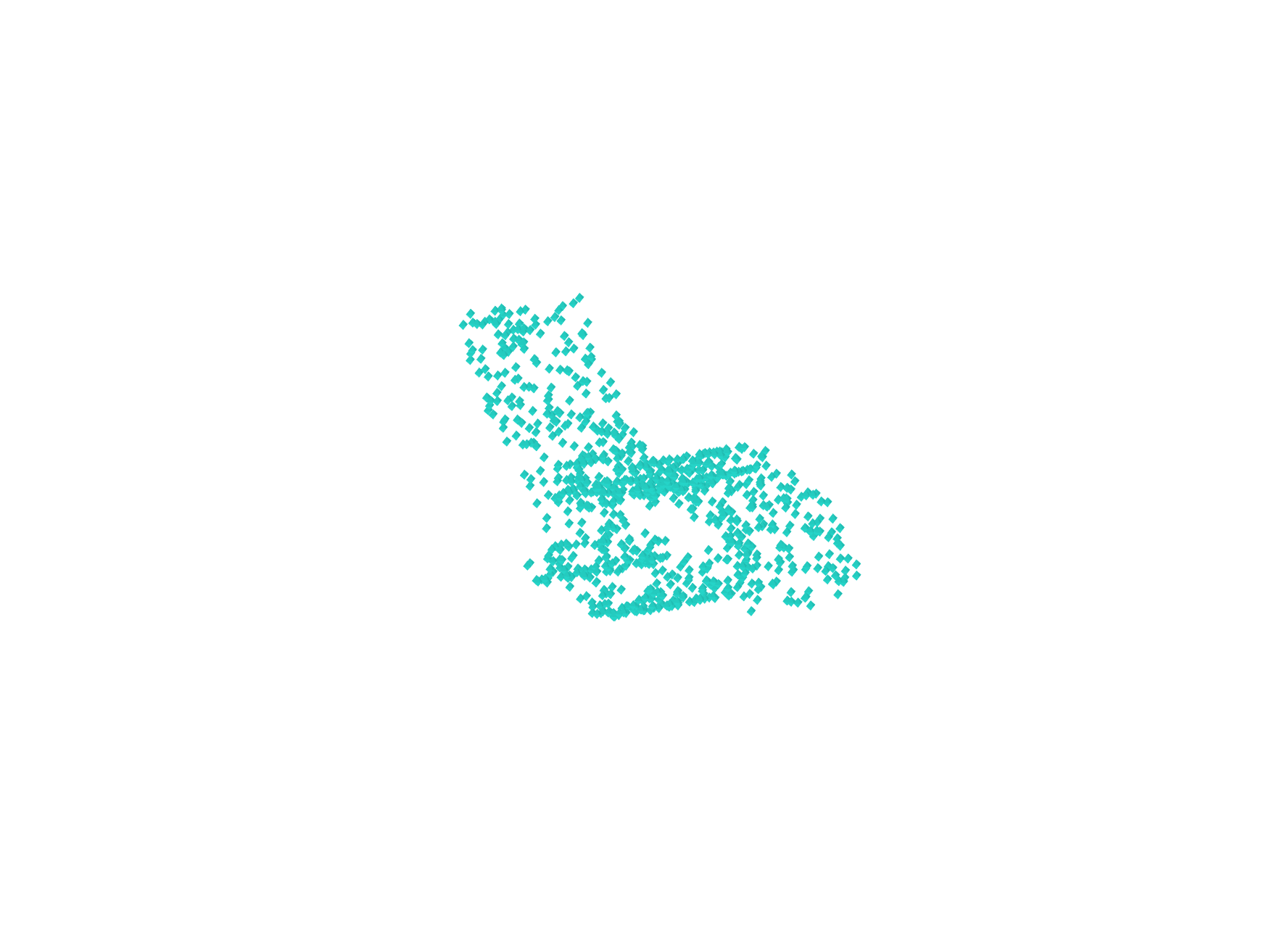} &
\adjincludegraphics[width=\vfig cm,trim={{.28\width} {.14\height} {.28\width} {.09\height}},clip]{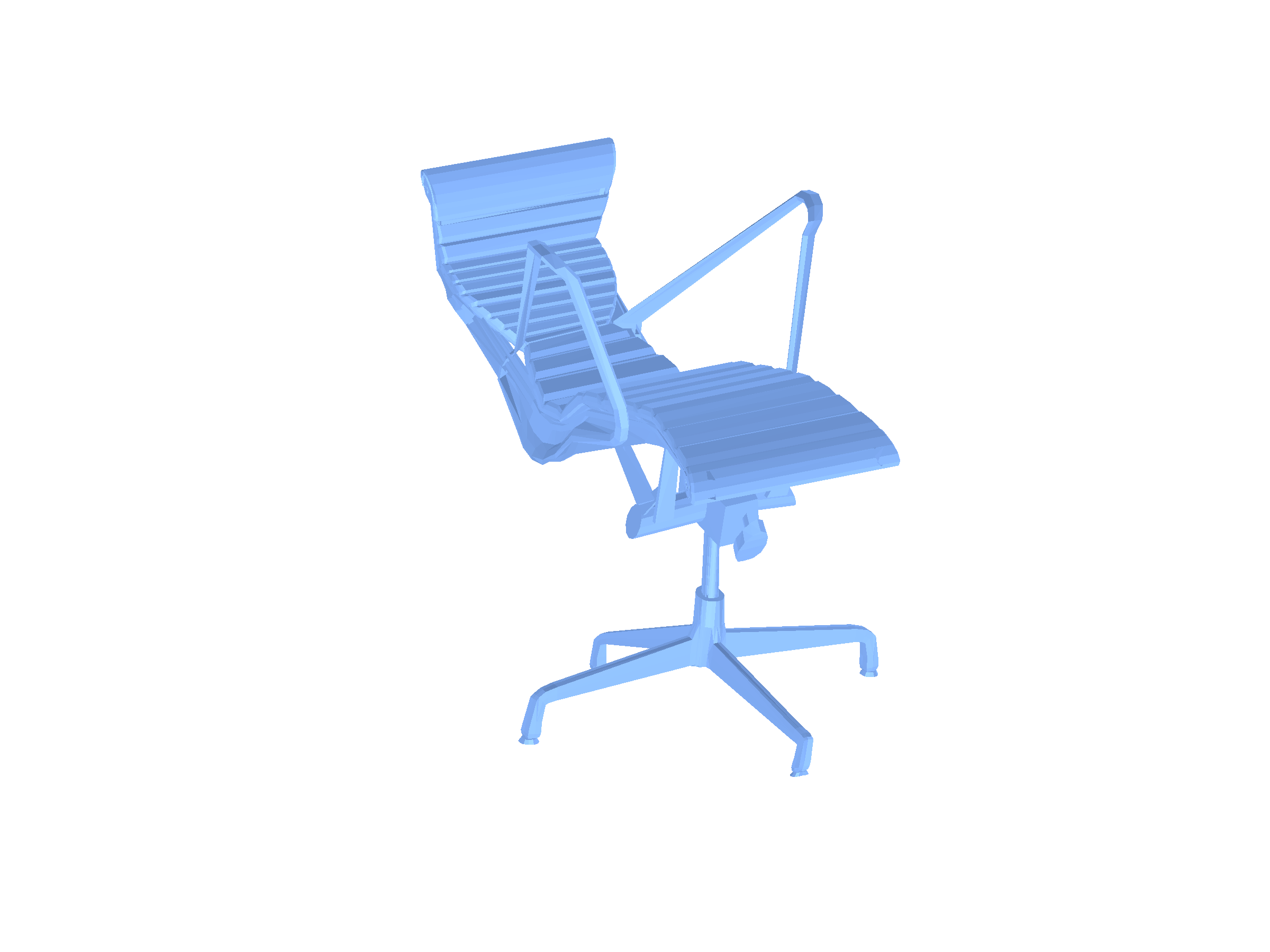} \\

source &
partial target &
aligned \\

\end{tabular}    
\caption{
\rev{Aligning a high quality 3D mesh to a point cloud. \ourmethod{} estimates the 7x7x7 grid deformation between a 3D source mesh (represented as a complete voxelized surface) and a point cloud (sparse voxelized surface), which we apply on the vertices of the source mesh to obtain the aligned version (last column). 
\ourmethod{}, trained on complete surfaces with segments removed at random, generalizes to sparse surfaces at test time.}}
\label{fig:threeD}
\end{figure}

\paragraph{\textbf{\rev{Point Cloud Registration.}}}
\rev{
In many applications ranging from VR and AR to scene understanding, it is desirable to align a mesh (containing semantic information) to a partial point cloud obtained by a scanning device.}
\rev{
Figure~\ref{fig:threeD} features several examples of alignments performed between 3D meshes and point clouds using our method \revb{(quantitative results in Table~\ref{tab:quant3d})}.
\rev{Recall that our 3D network processes pairs of voxelized surfaces as input, such that the source is complete, and the target is generated from a complete surface by removing parts at random. At test time, however, we are given a target shape represented as a point cloud, requiring conversion into our network format.  
To this end, we simply discretize the continuous coordinates of the point cloud and form a sparse surface occupancy grid. 
Despite not having trained on such sparsely sampled surfaces, \ourmethod{} generalizes well and is able to produce plausible deformations.}} 

\begin{table}
\begin{center}
\begin{tabular}{c||c || c} 
\revb{samples} & \revb{500} & \revb{3000} \\
\hline
\revb{IOU} & \revb{53} & \revb{57} \\ [0.2ex]
\end{tabular}
\end{center}
\caption{
\revb{Quantitative results for 3D point cloud registration on the \emph{airplane} class. We compute the average IOU between deformed source and full target given target in point cloud form for both 500 and 3000 point samples. Note that the IOU is computed on the full voxelized representation of the shapes generated from their mesh forms, where the applied warp field is computed using the point cloud representation of the target.}}
\label{tab:quant3d}
\end{table}


\paragraph{\textbf{\rev{Segmentation Transfer.}}}
\rev{Semantically segmenting a shape into its constituent parts is an important first step for a wide range of applications in shape analysis and synthesis. A popular approach to solve this task is the transfer of an existing high-quality segmentation from one shape to another. Naturally, some form of correspondence between the two shapes must be given or computed in order to support the transference.}

\begin{figure}[h!]
\newcommand{\tfig}{2.2}
\newcommand{\vfig}{1.7}
\setlength\tabcolsep{10pt}
\begin{tabular}{c c c}


\adjincludegraphics[width=\vfig cm,trim={{.3\width} {.18\height} {.3\width} {.15\height}},clip]{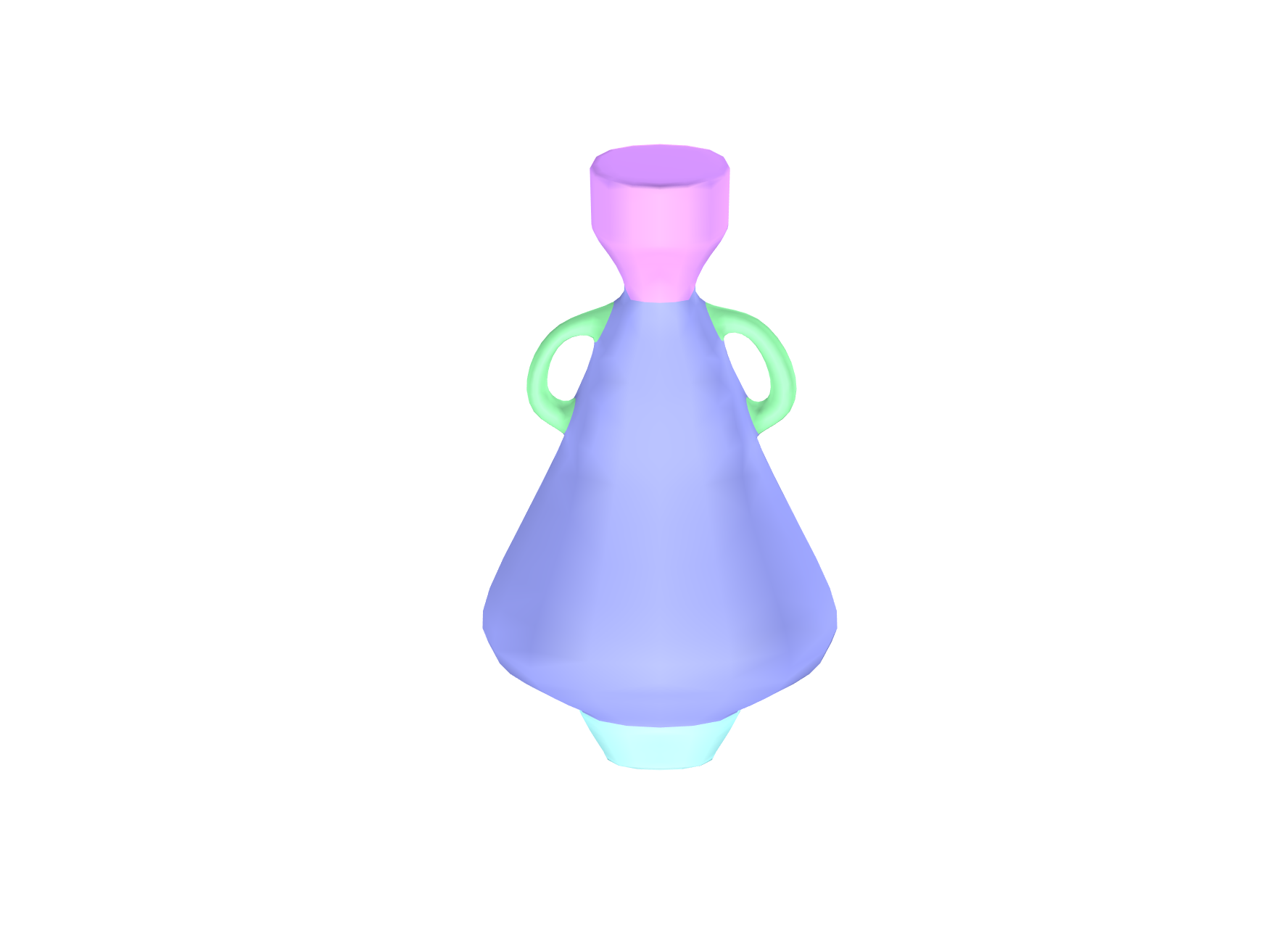} &
\adjincludegraphics[width=\vfig cm,trim={{.3\width} {.18\height} {.3\width} {.15\height}},clip]{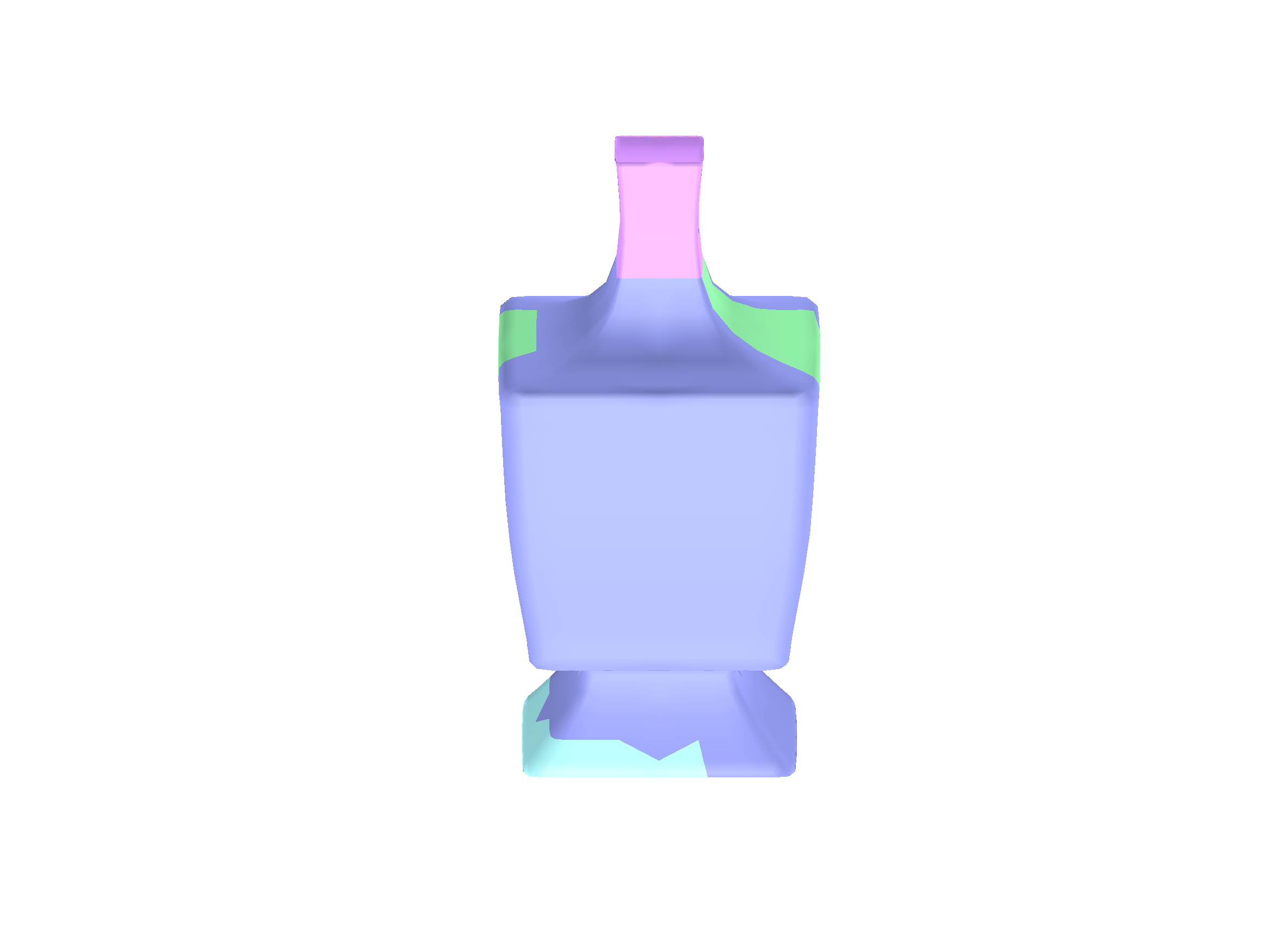} &
\adjincludegraphics[width=\vfig cm,trim={{.3\width} {.18\height} {.3\width} {.15\height}},clip]{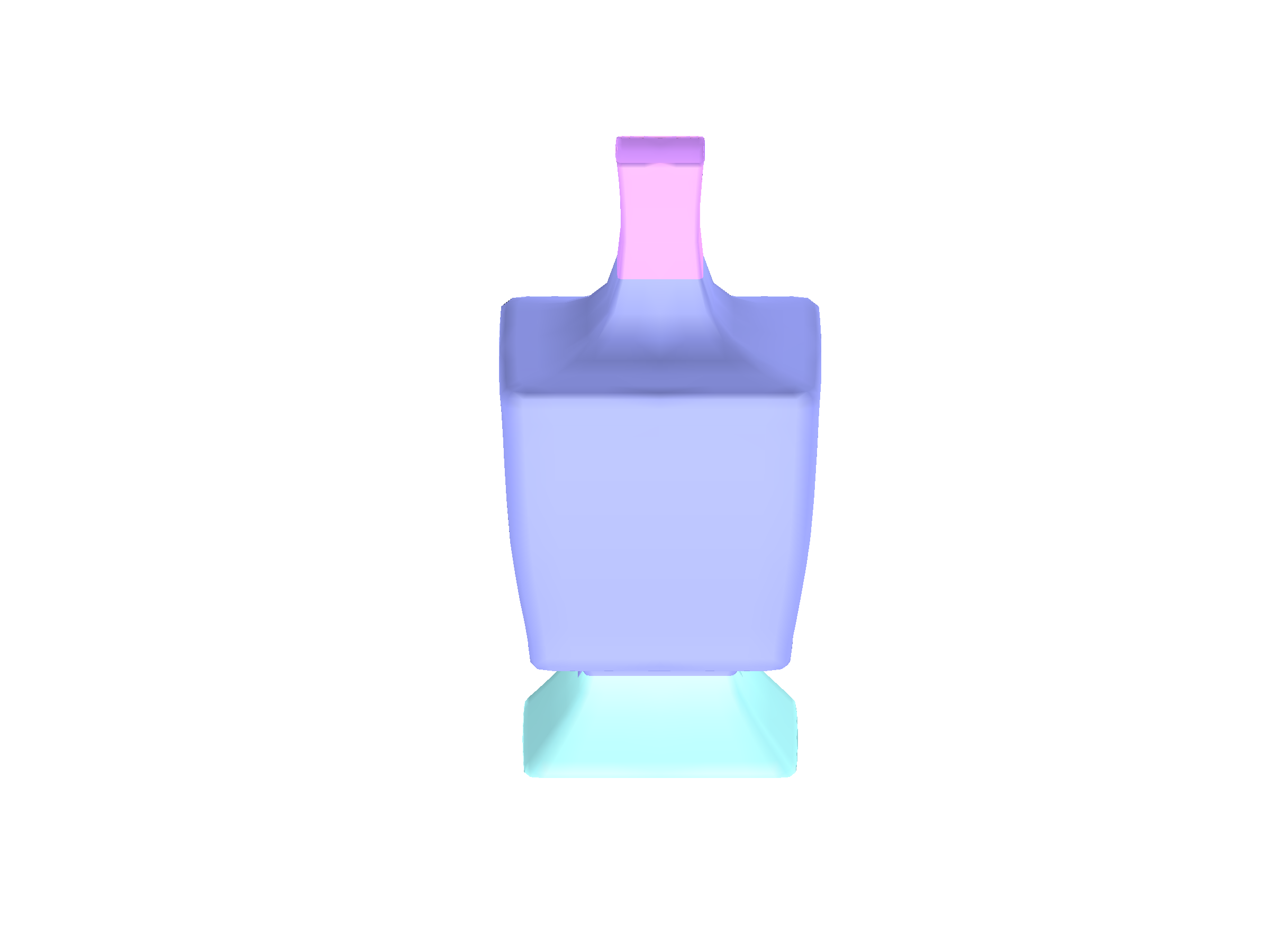} \\






\adjincludegraphics[width=\tfig cm,trim={{.3\width} {.3\height} {.25\width} {.3\height}},clip]{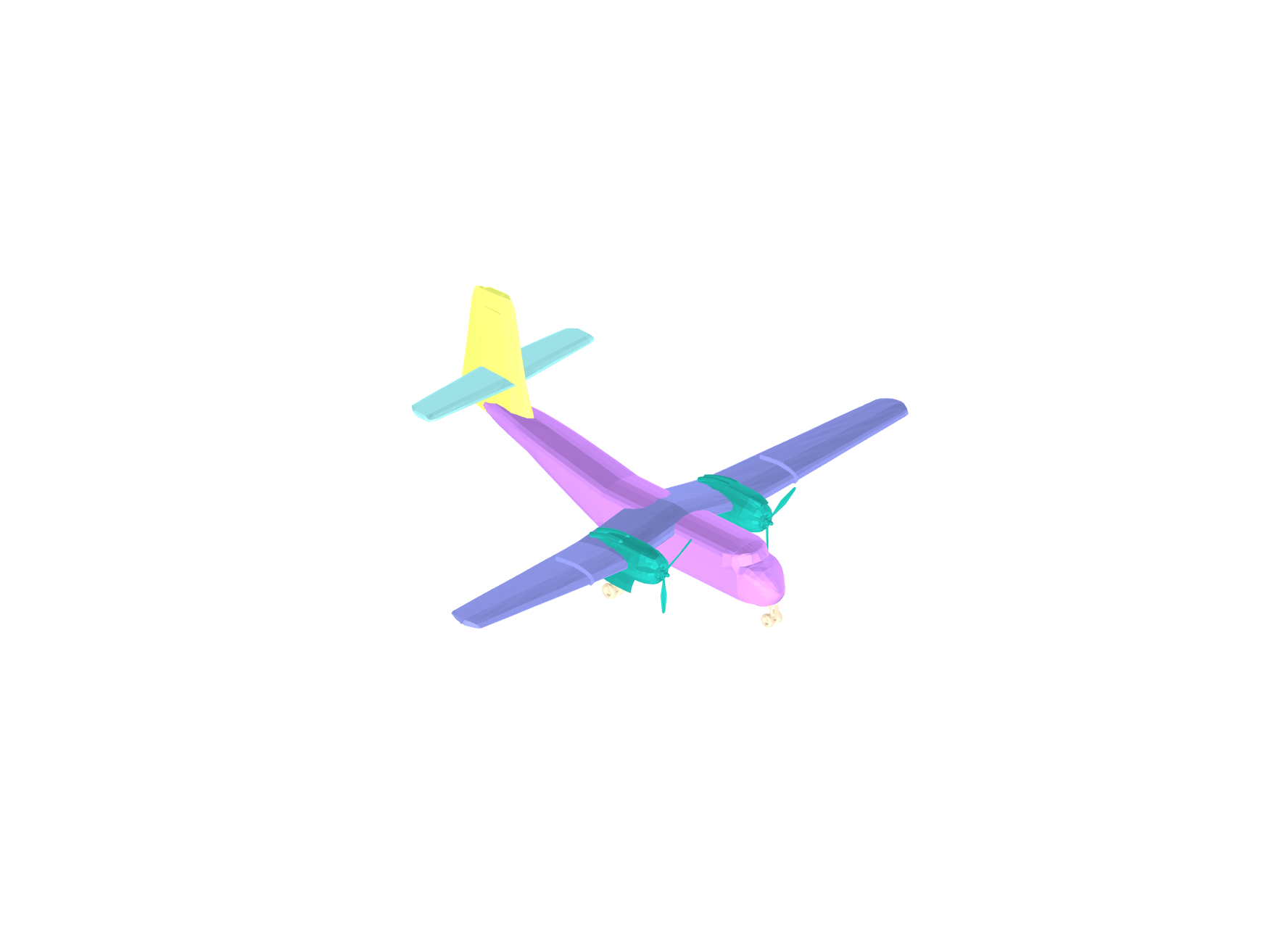} &
\adjincludegraphics[width=\tfig cm,trim={{.3\width} {.3\height} {.25\width} {.3\height}},clip]{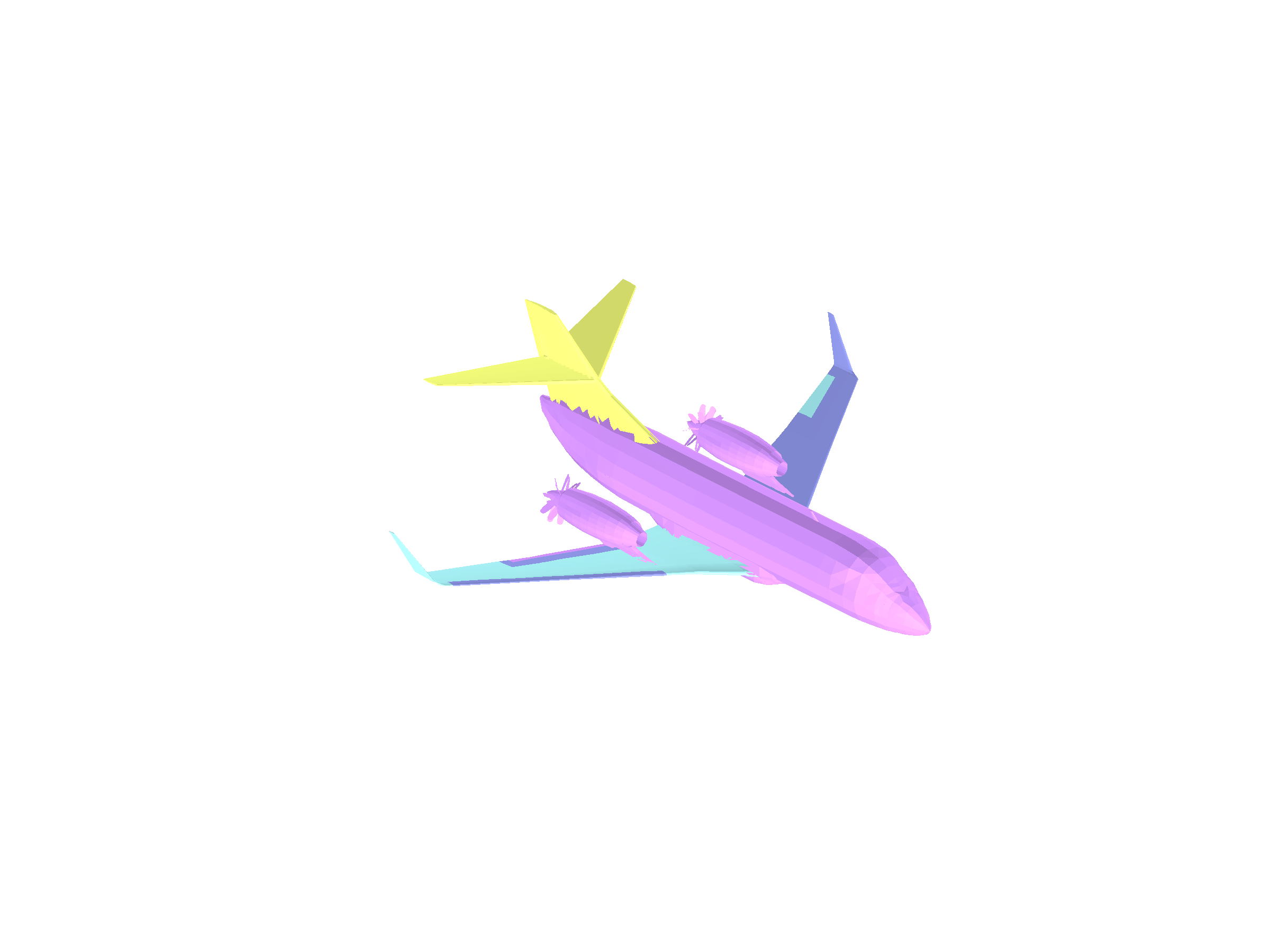} &
\adjincludegraphics[width=\tfig cm,trim={{.3\width} {.3\height} {.25\width} {.3\height}},clip]{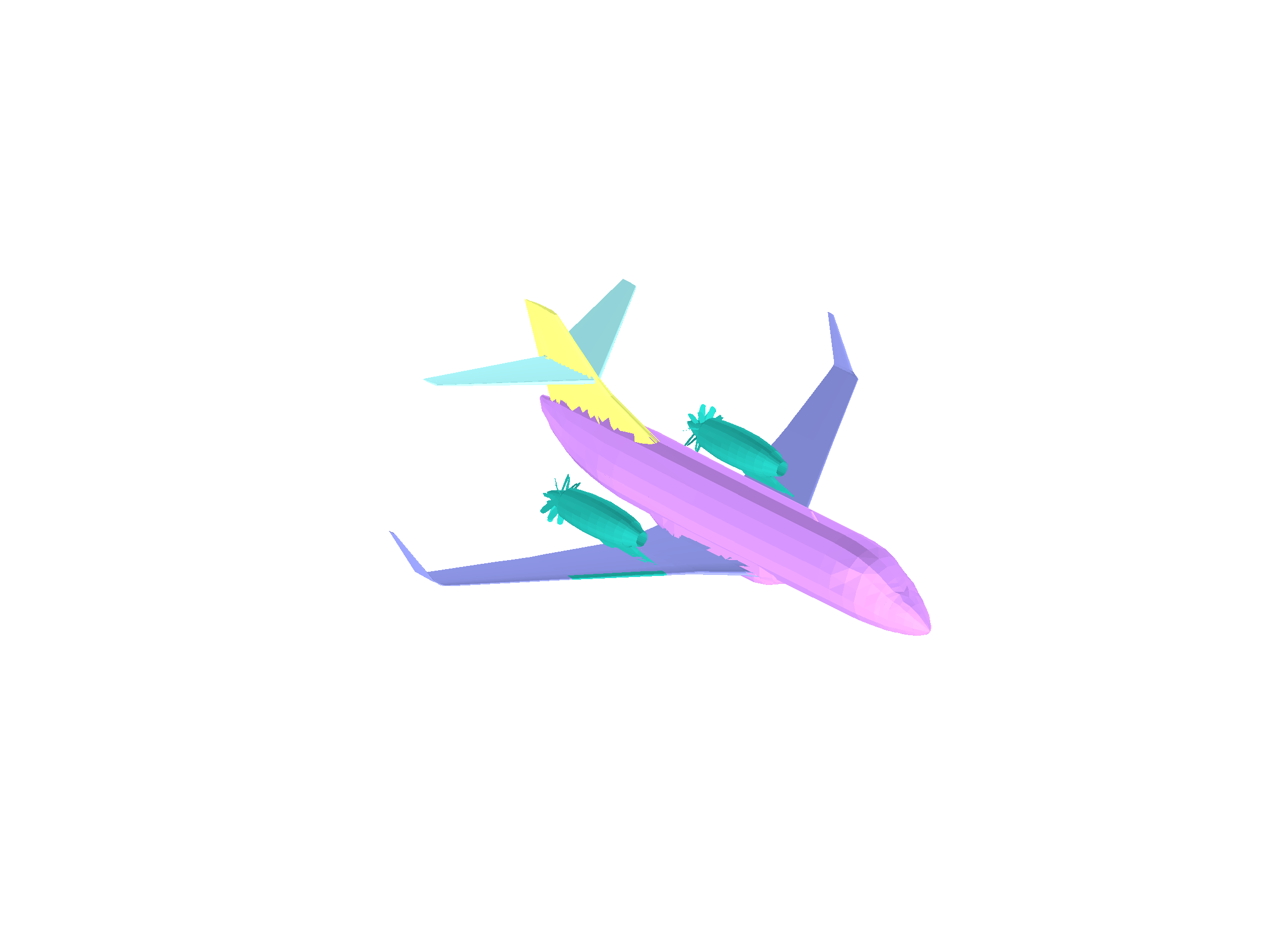} \\

\adjincludegraphics[width=\tfig cm,trim={{.3\width} {.3\height} {.25\width} {.3\height}},clip]{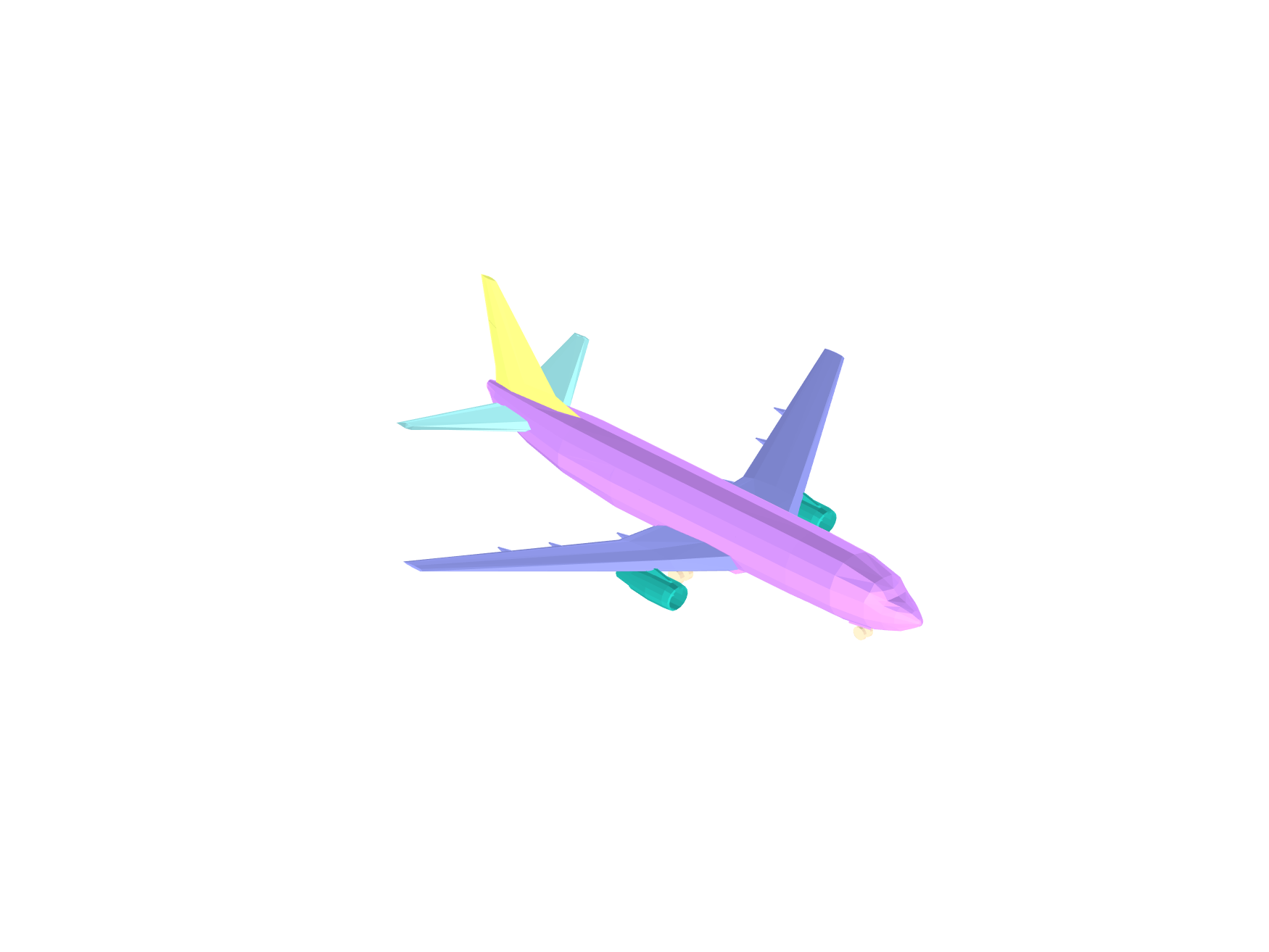} &
\adjincludegraphics[width=\tfig cm,trim={{.3\width} {.3\height} {.25\width} {.3\height}},clip]{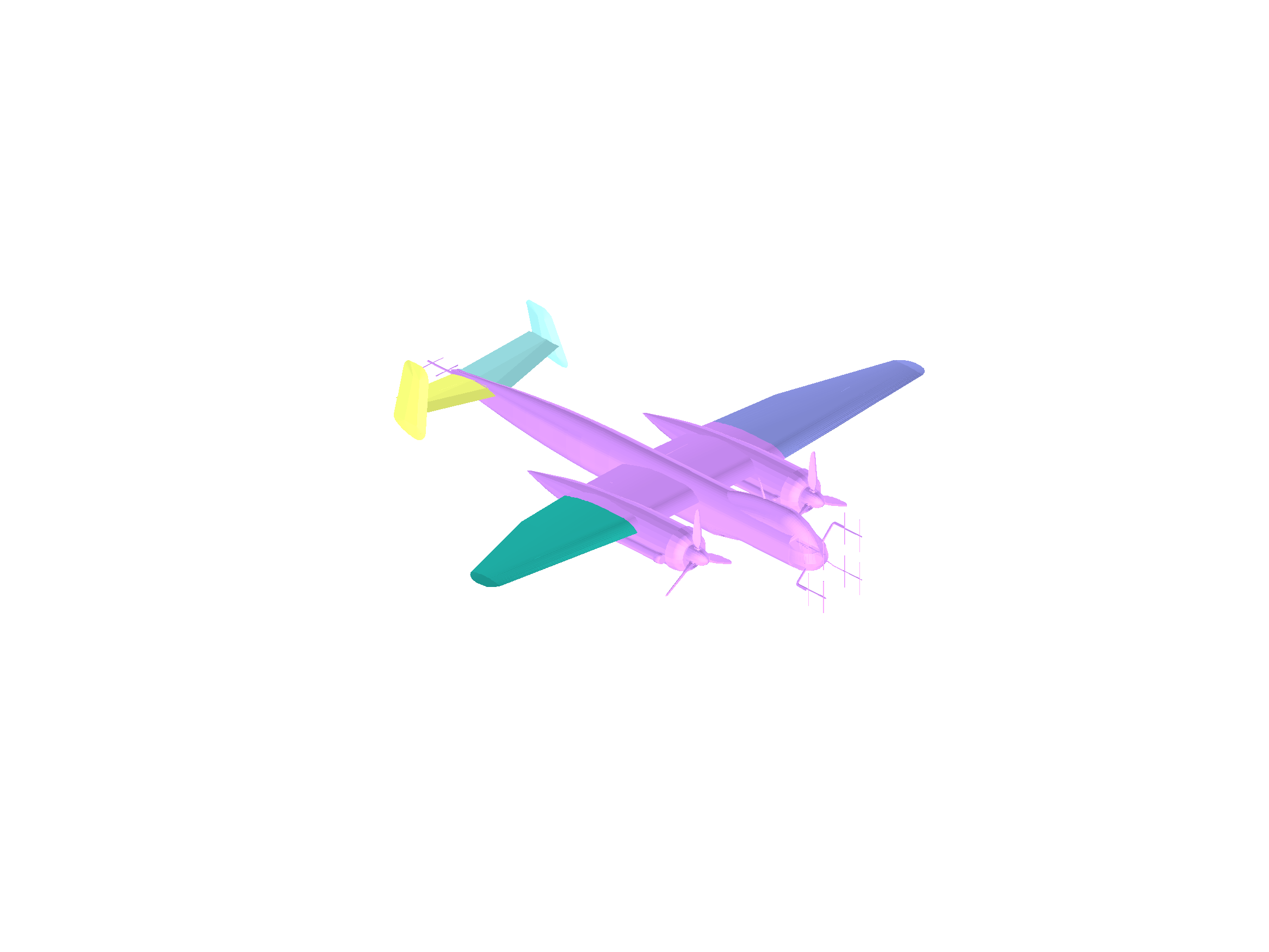} &
\adjincludegraphics[width=\tfig cm,trim={{.3\width} {.3\height} {.25\width} {.3\height}},clip]{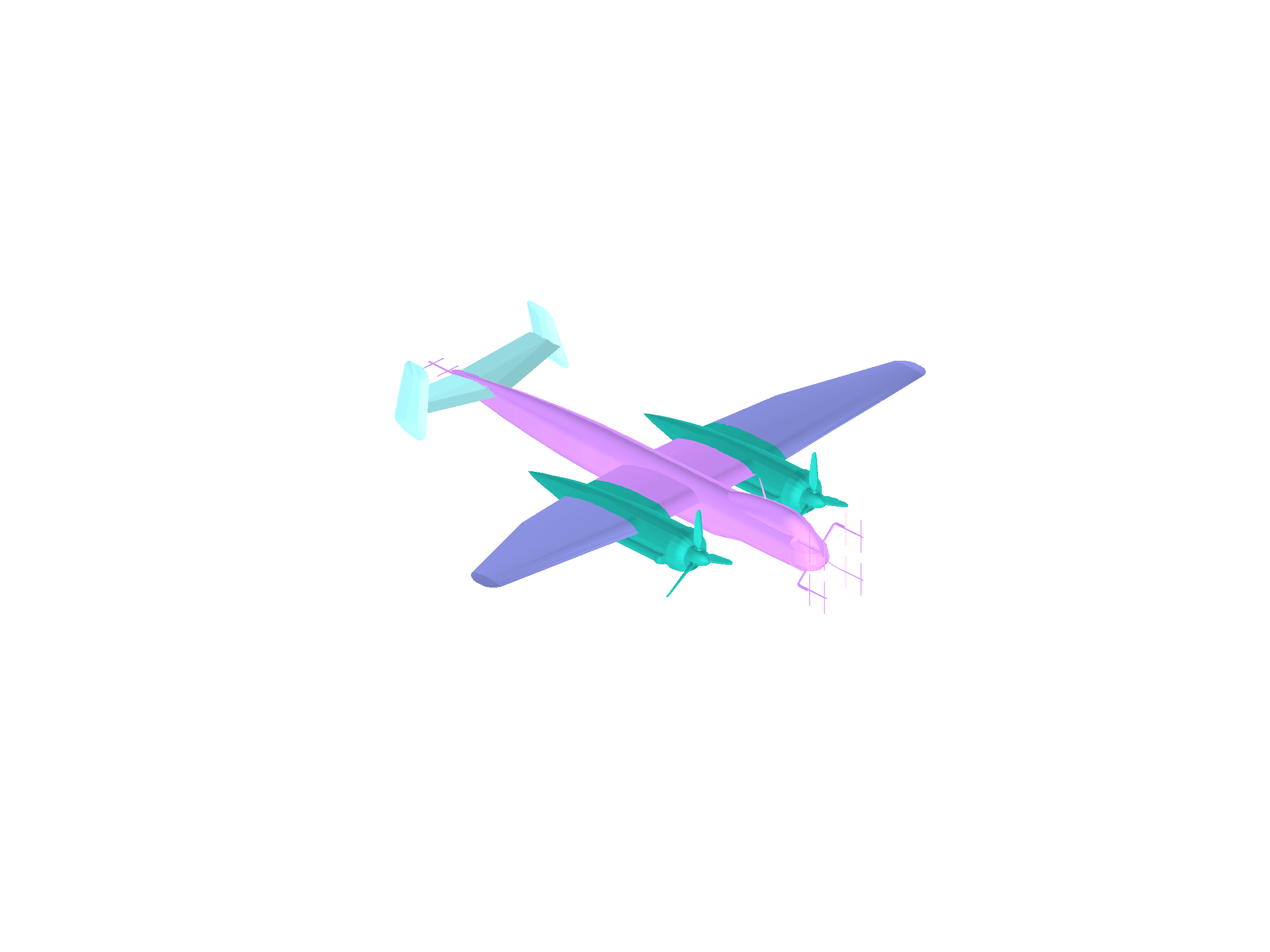} \\

\adjincludegraphics[width=\tfig cm,trim={{.3\width} {.3\height} {.25\width} {.3\height}},clip]{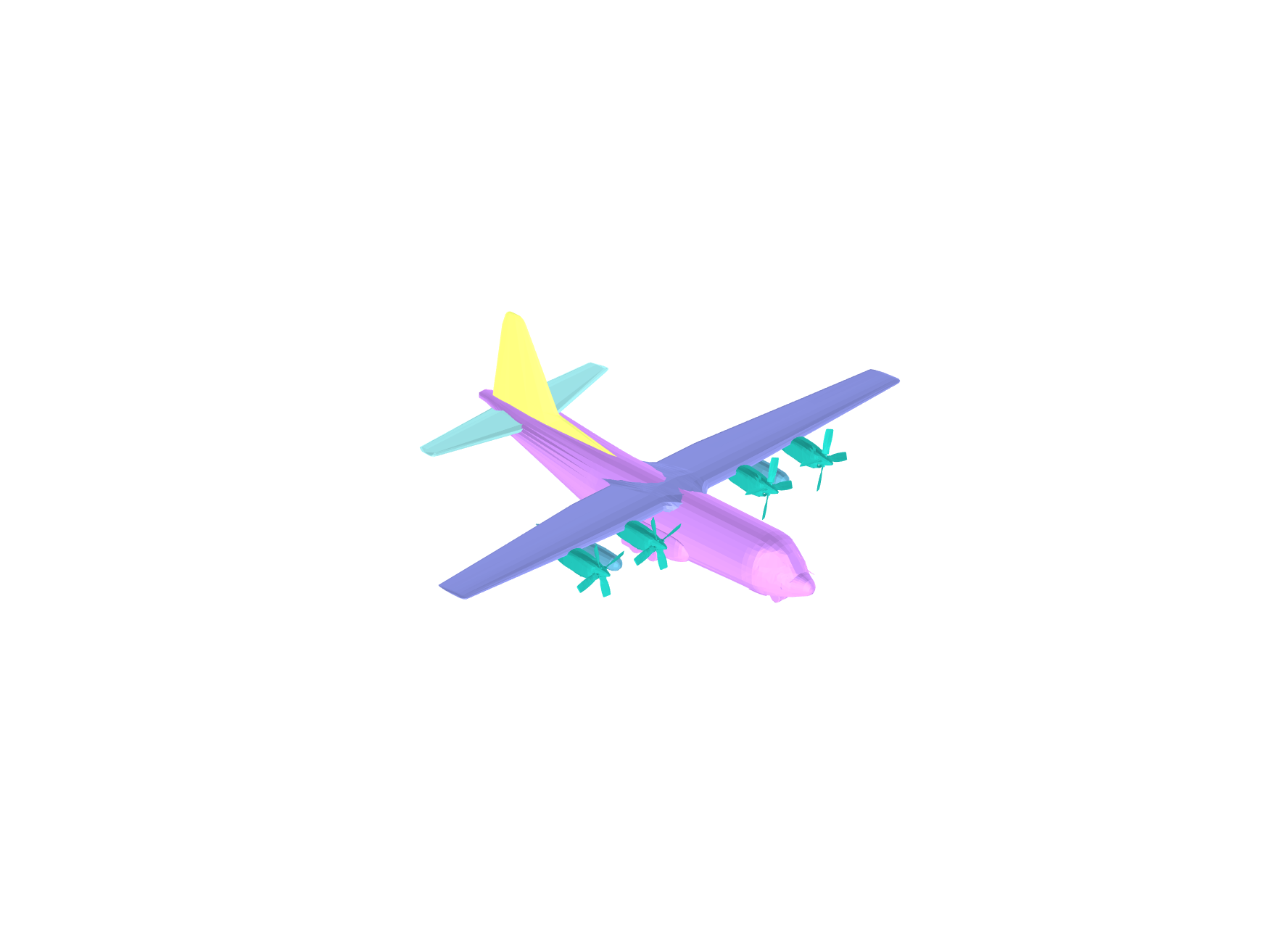} &
\adjincludegraphics[width=\tfig cm,trim={{.3\width} {.3\height} {.25\width} {.3\height}},clip]{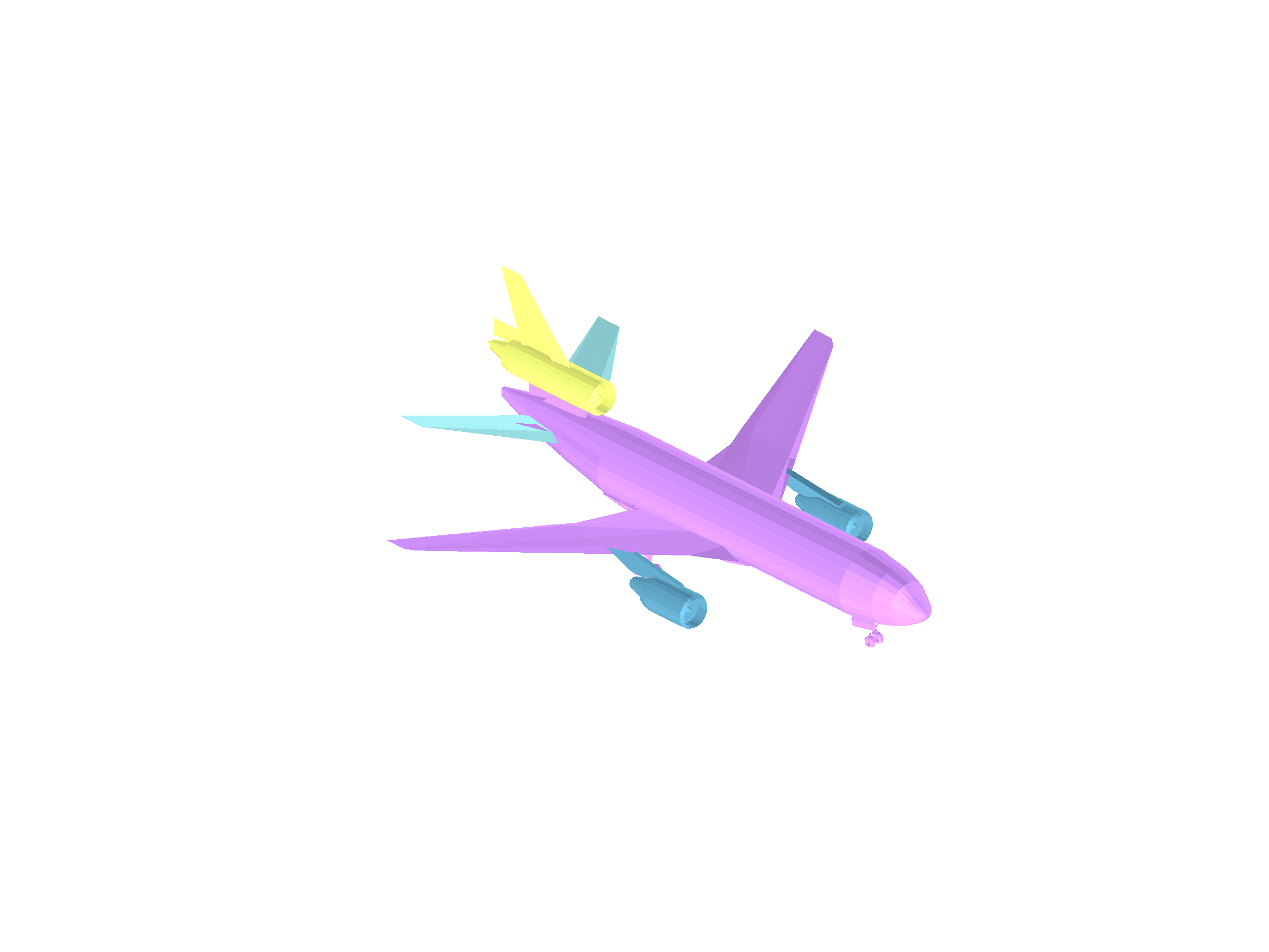} &
\adjincludegraphics[width=\tfig cm,trim={{.3\width} {.3\height} {.25\width} {.3\height}},clip]{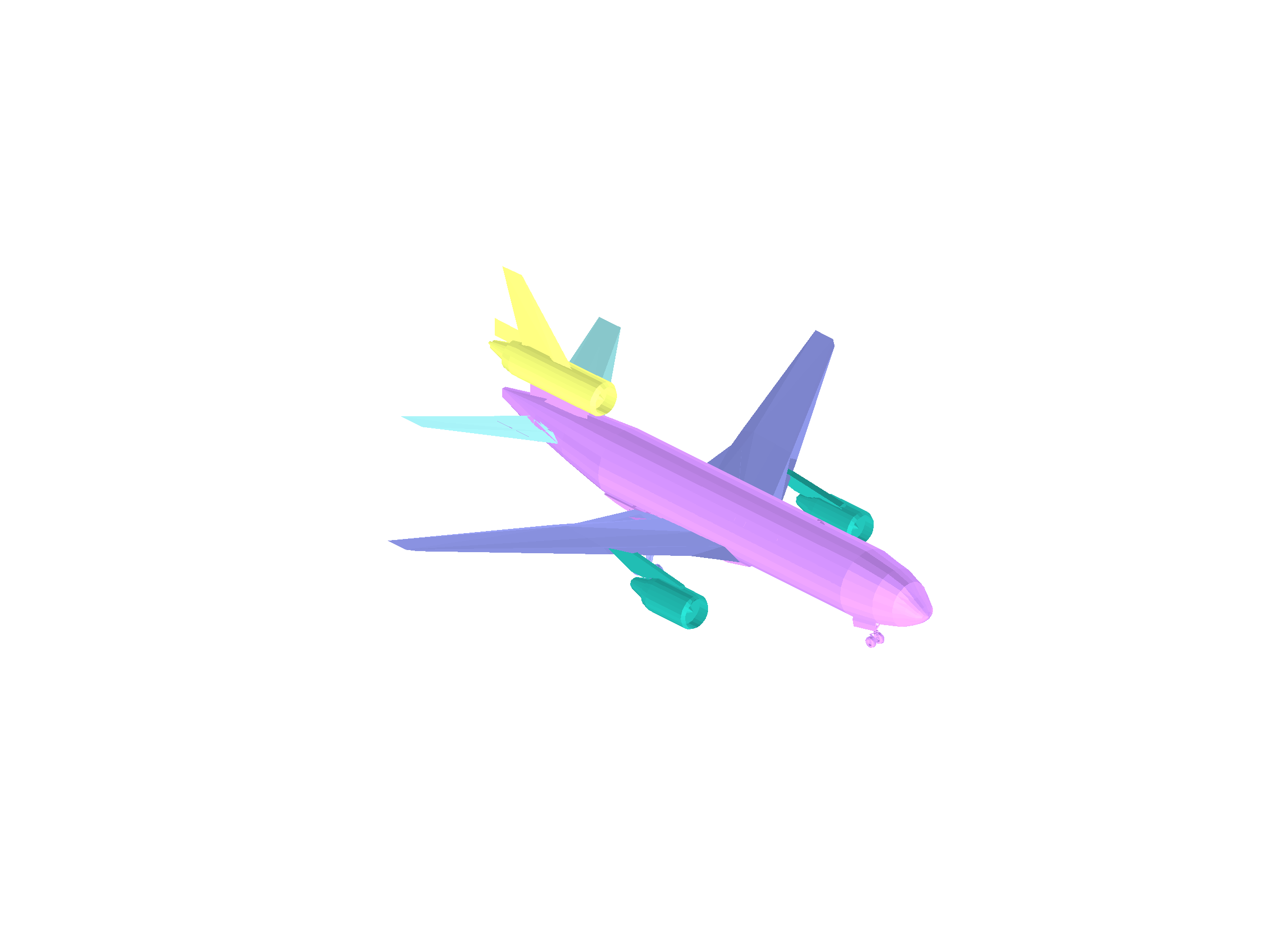} \\


\adjincludegraphics[width=\vfig cm,trim={{.3\width} {.18\height} {.3\width} {.13\height}},clip]{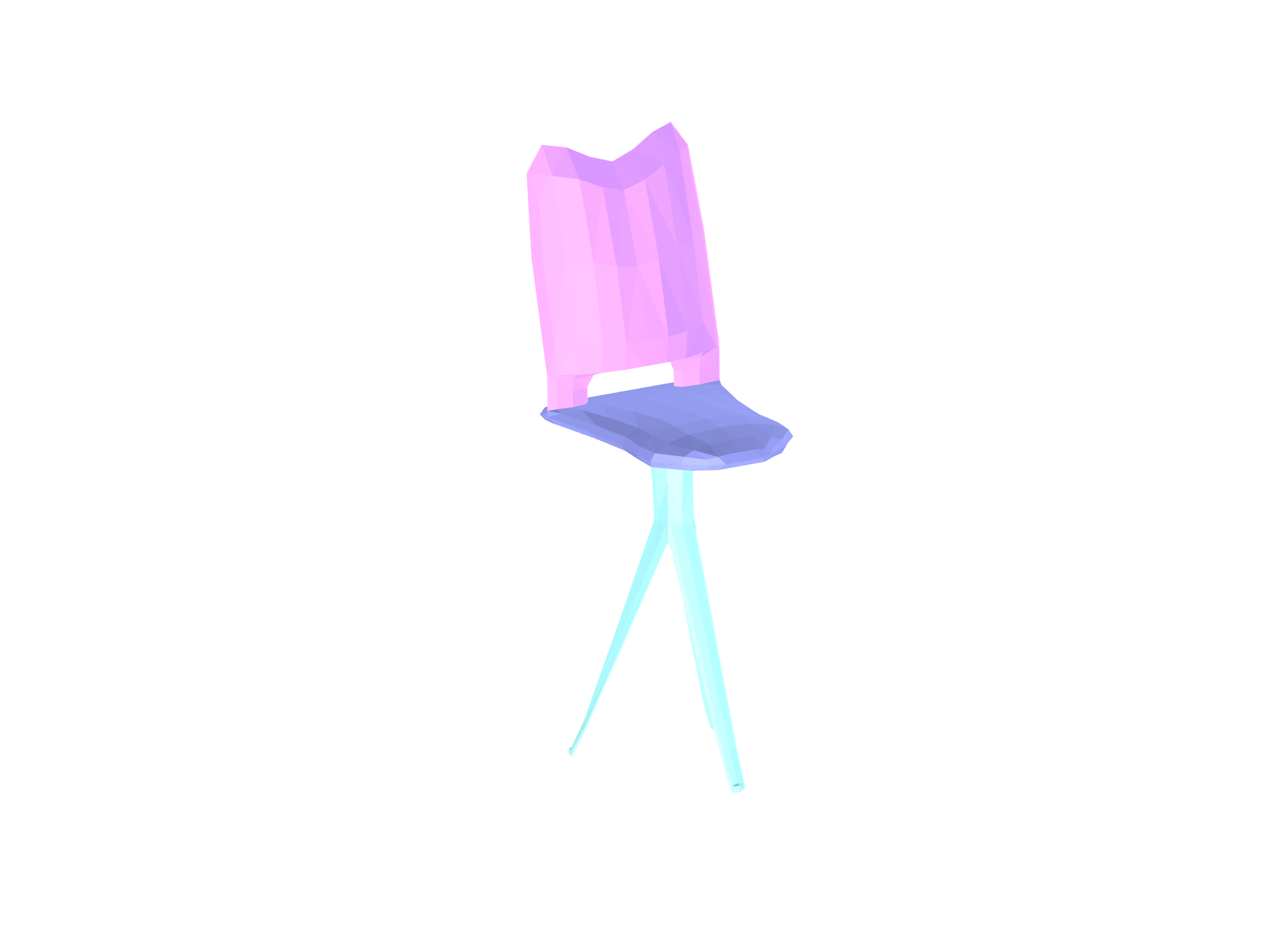} &
\adjincludegraphics[width=\vfig cm,trim={{.3\width} {.18\height} {.3\width} {.12\height}},clip]{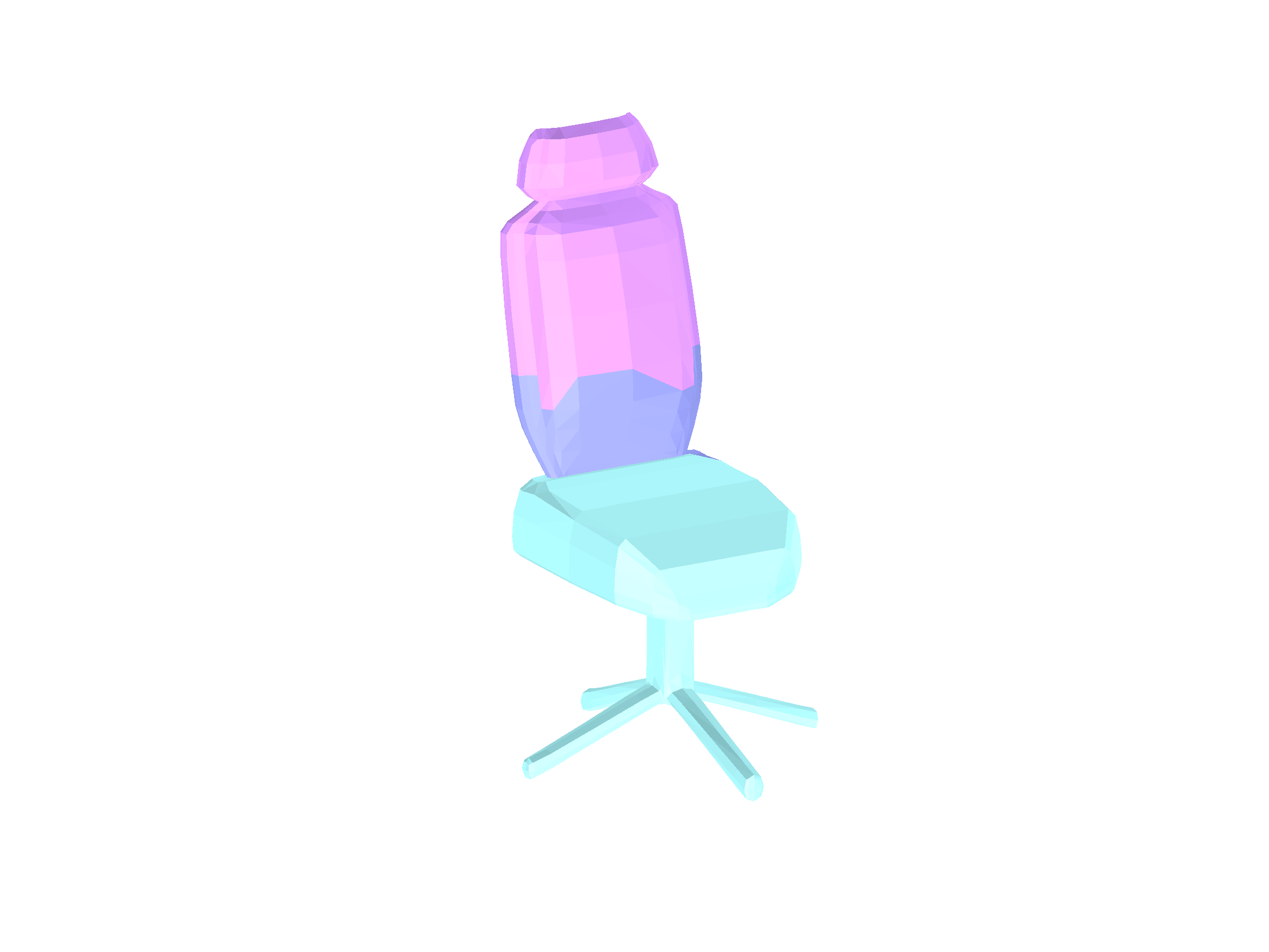} &
\adjincludegraphics[width=\vfig cm,trim={{.3\width} {.18\height} {.3\width} {.12\height}},clip]{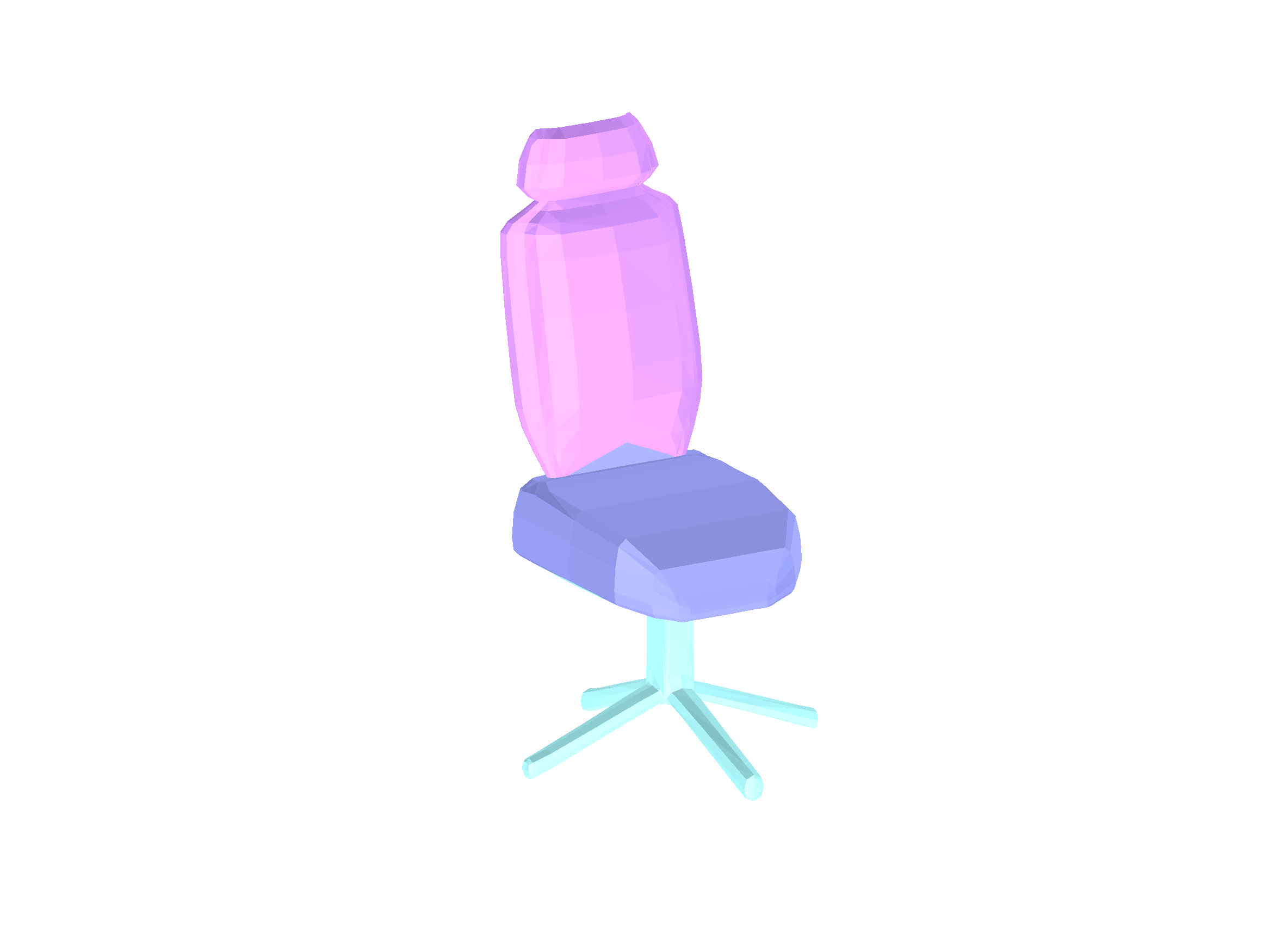} \\

\adjincludegraphics[width=\vfig cm,trim={{.2\width} {.1\height} {.2\width} {.1\height}},clip]{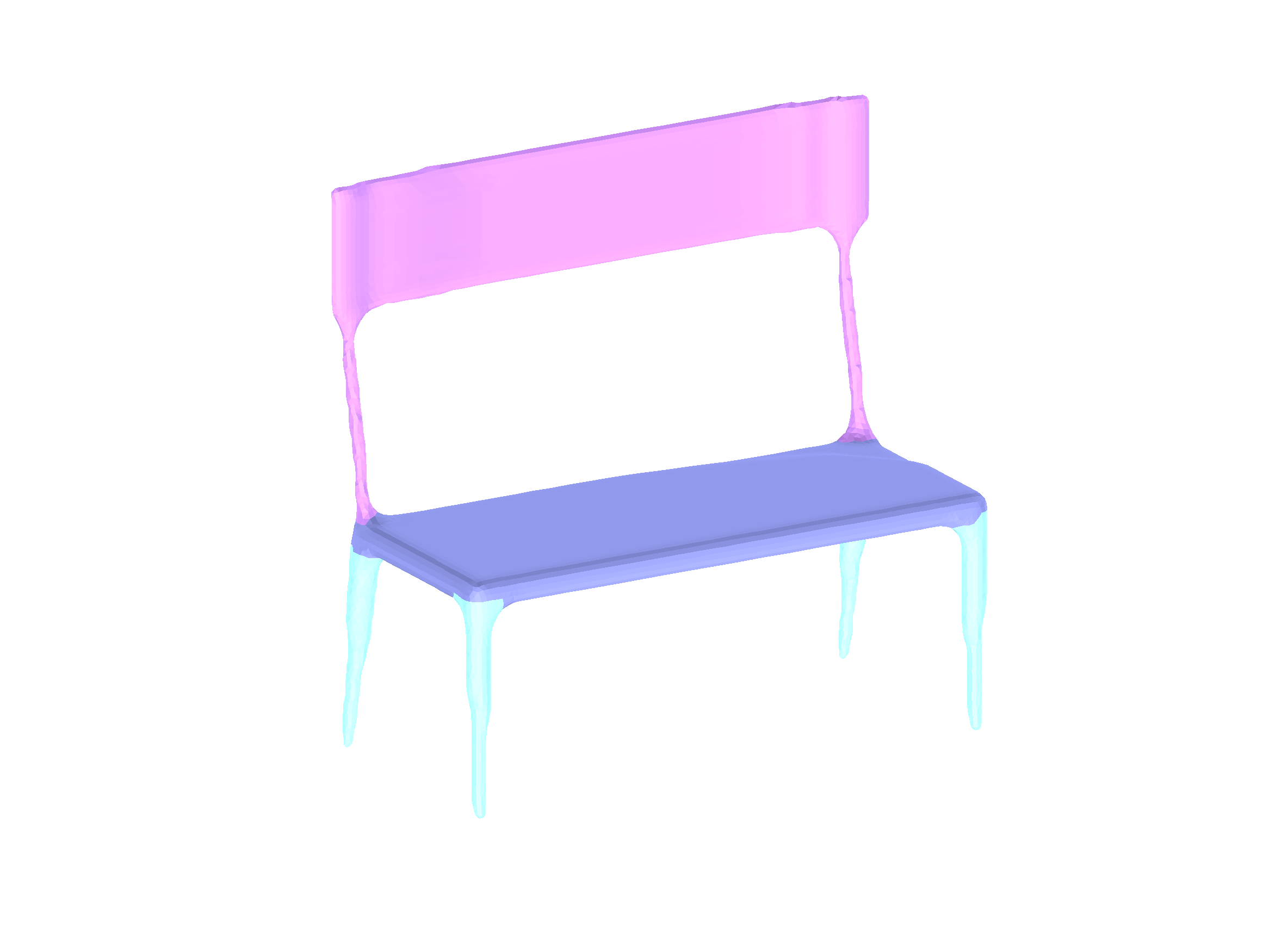} &
\adjincludegraphics[width=\vfig cm,trim={{.2\width} {.1\height} {.2\width} {.1\height}},clip]{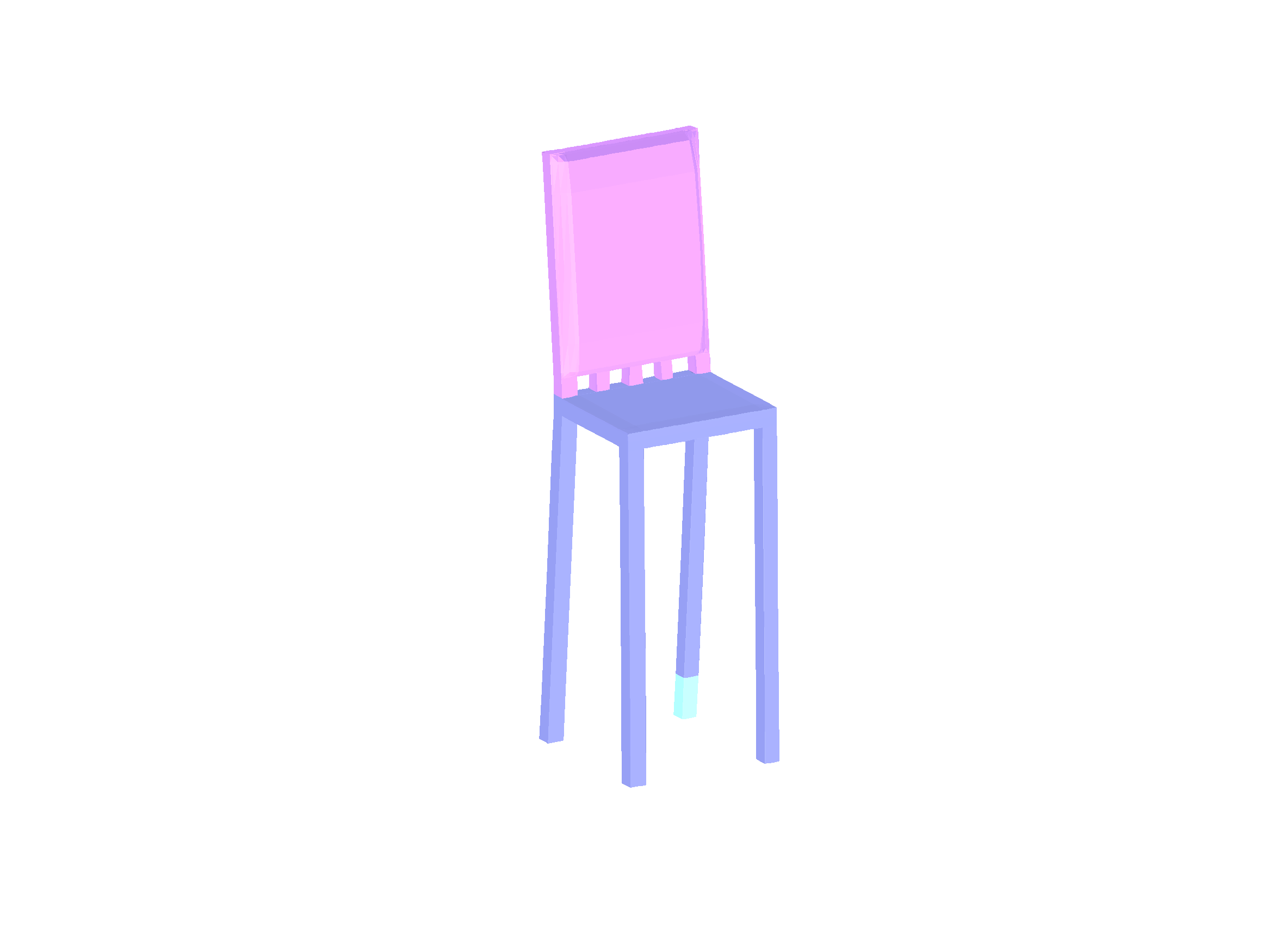} &
\adjincludegraphics[width=\vfig cm,trim={{.2\width} {.1\height} {.2\width} {.1\height}},clip]{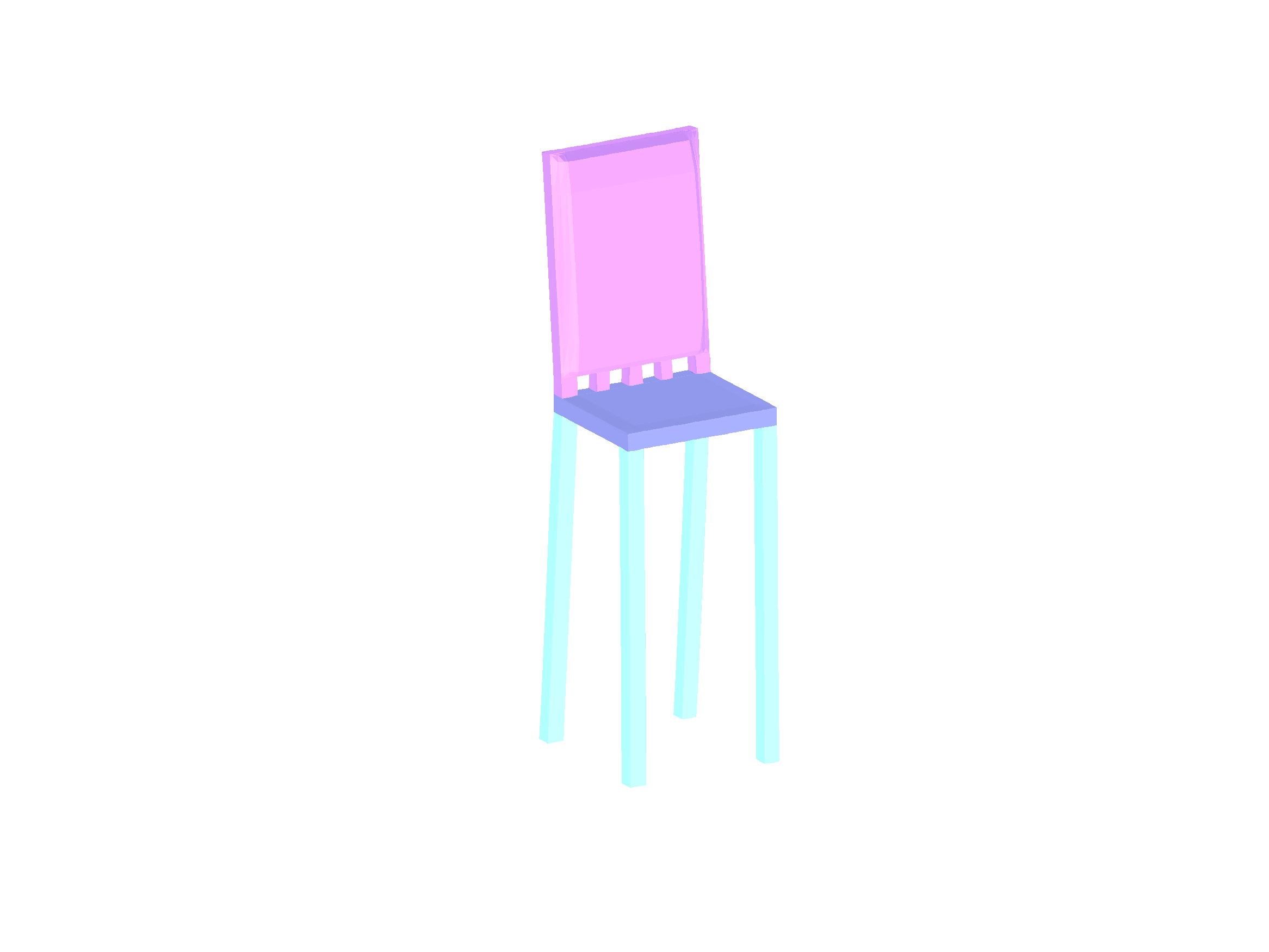} \\

segmented &
naive transfer &
\ourmethod{} transfer \\

\end{tabular}    
\caption{\rev{\ourmethod{} results on 3D segmentation transfer (test set). The labels in the source shape (left) transferred to the target shapes using the correspondence generated with \ourmethod{} (right). The estimated alignment is used in combination with a simple graph cut optimization as a post process. In the middle column, we compare to a baseline transfer, where the identity warp is used to guide the correspondence.}} 
\label{fig:threeD_seg}
\end{figure}

\rev{To further demonstrate the integrity of \ourmethod{} warps, we show results on segmentation transfer in 3D. Using the same trained model that was used in the previous application (Figure~\ref{fig:threeD}), we estimate the deformation between a segmented (source) shape and an unsegmented (target) shape. Given the underlying correspondence determined by the alignment, we transfer the segmentation labels from the source shape faces to their corresponding faces on the target shape, and perform a simple graph cut optimization to obtain a complete and smooth result (see Figure~\ref{fig:threeD_seg}).}

\subsection{Robustness}
\paragraph{\textbf{Symmetry.}}
Bilateral symmetry is characteristic of many man-made as well as natural shapes. Accordingly, most of our datasets are composed either entirely, or almost entirely, of bilaterally symmetric shapes. We have shown that our system is able to recover such symmetries even when a target is missing portions in a pattern that impairs the original symmetry of the shape. However, symmetry is a strong cue that often assists in tasks of reconstruction, and it is possible that \ourmethod{} is more easily trained on a dataset that is inherently symmetric.

\begin{figure}[h]
\newcommand{\asym}{1.7}
\setlength\tabcolsep{1pt}
\begin{tabular}{c c c c}

 &
\includegraphics[width=\asym cm]{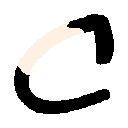} &
\includegraphics[width=\asym cm]{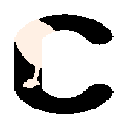} &
\includegraphics[width=\asym cm]{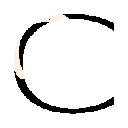} \\

\includegraphics[width=\asym cm]{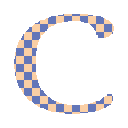} &
\includegraphics[width=\asym cm]{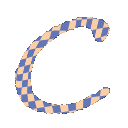} &
\includegraphics[width=\asym cm]{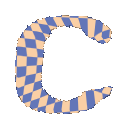} &
\includegraphics[width=\asym cm]{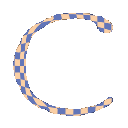} \\


 &
\includegraphics[width=\asym cm]{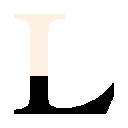} &
\includegraphics[width=\asym cm]{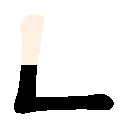} &
\includegraphics[width=\asym cm]{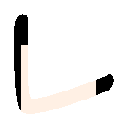} \\

\includegraphics[width=\asym cm]{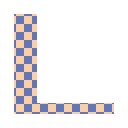} &
\includegraphics[width=\asym cm]{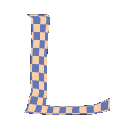} &
\includegraphics[width=\asym cm]{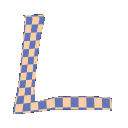} &
\includegraphics[width=\asym cm]{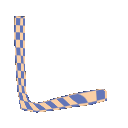} \\

\end{tabular}

\caption{Examples of alignment of asymmetrical shapes.}
\label{fig:asym}
\end{figure}

We are therefore particularly interested in examining the performance of \ourmethod{} on datasets that are asymmetric in nature. Examples for alignments performed on rendered asymmetric letters ("C" and "L") are presented in Figure~\ref{fig:asym}. We observe the satisfactory recovery of the structure of the shapes, suggesting that our approach does not require symmetry as a leading prior to learn to align.

\paragraph{\textbf{Orientation.}}
\addcomment{Our baseline experimental setting assumes that input datasets are consistently pre-oriented, an assumption that holds for many readily available shape datasets. Where this assumption does not hold, consistent orientation methods can be employed (e.g. hierarchical alignment~\cite{shapenet2015}). Even so, we investigate the ability of our system to handle orientation errors.}

\begin{figure}[h]
\newcommand{\rotfig}{1.5}
\setlength\tabcolsep{3pt}
\begin{tabular}{c c c c c}

\includegraphics[width=\rotfig cm]{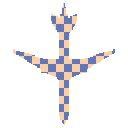} &
\includegraphics[width=\rotfig cm]{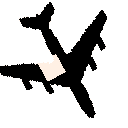} &
\includegraphics[width=\rotfig cm]{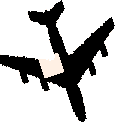} &
\includegraphics[width=\rotfig cm]{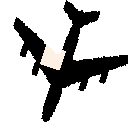} &
\includegraphics[width=\rotfig cm]{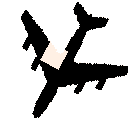} \\

 &
\includegraphics[width=\rotfig cm]{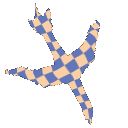} &
\includegraphics[width=\rotfig cm]{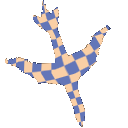} &
\includegraphics[width=\rotfig cm]{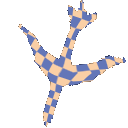} &
\includegraphics[width=\rotfig cm]{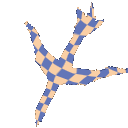} \\

\includegraphics[width=\rotfig cm]{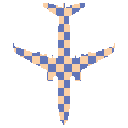} &
\includegraphics[width=\rotfig cm]{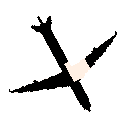} &
\includegraphics[width=\rotfig cm]{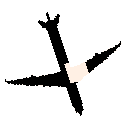} &
\includegraphics[width=\rotfig cm]{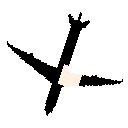} &
\includegraphics[width=\rotfig cm]{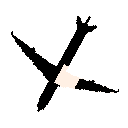} \\

 &
\includegraphics[width=\rotfig cm]{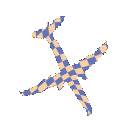} &
\includegraphics[width=\rotfig cm]{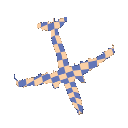} &
\includegraphics[width=\rotfig cm]{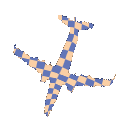} &
\includegraphics[width=\rotfig cm]{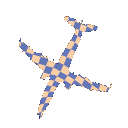} \\

source &
-29$^{\circ}$ &
-19$^{\circ}$ &
19$^{\circ}$ &
29$^{\circ}$ \\

\end{tabular}
\caption{\revb{Robustness to orientation deviations by adding data augmentation: training on offsets within the range $[-30^{\circ},30^{\circ}]$. Test results shown for orientation offsets of -29$^{\circ}$, -19$^{\circ}$, 19$^{\circ}$ and 29$^{\circ}$ for the same source and target pairs.}}
\label{fig:grot}
\end{figure}

\revb{We augment our dataset by introducing rotations uniformly sampled from the range 
$\lbrack -30^{\circ}, 30^{\circ}\rbrack$, and train \ourmethod{} as previously described. Figure~\ref{fig:grot}
shows the qualitative alignment results of our augmented network trained on shapes with arbitrary rotations.}



\begin{figure}[h!]
\newcommand{\tfig}{1.9}
\setlength\tabcolsep{1pt}
\begin{tabular}{c c c}

\includegraphics[width=\tfig cm]{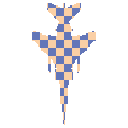} &
\includegraphics[width=\tfig cm]{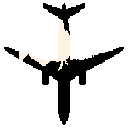} &
\includegraphics[width=\tfig cm]{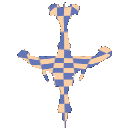} \\

\includegraphics[width=\tfig cm]{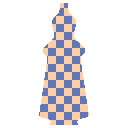} &
\includegraphics[width=\tfig cm]{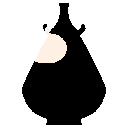} &
\includegraphics[width=\tfig cm]{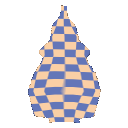} \\

\includegraphics[width=\tfig cm]{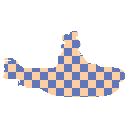} &
\includegraphics[width=\tfig cm]{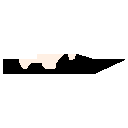} &
\includegraphics[width=\tfig cm]{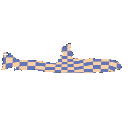} \\


\end{tabular}    
\caption{\rev{Results of a single \ourmethod{} network trained on three classes.}}
\label{fig:multiclass}
\end{figure}

\paragraph{\textbf{\rev{Multi Class.}}}
\rev{A core strength of \ourmethod{} lies in the higher level understanding of a particular shape category, enabling it to preserve geometric features that are typical to the class of objects at hand. However, we investigate the robustness of \ourmethod{} to handle multiple classes simultaneously. By sampling three different classes during training, using the same architecture, we observe that \ourmethod{} can jointly handle several classes simultaneously (see Figure~\ref{fig:multiclass}).}

\begin{figure*}[h]
\newcommand{\tfig}{1.9}
\setlength\tabcolsep{1pt}
\begin{tabular}{c c c c c c c c}

\includegraphics[width=\tfig cm]{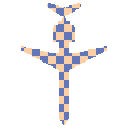} &
\includegraphics[width=\tfig cm]{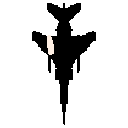} &
\includegraphics[width=\tfig cm]{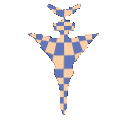} &
\includegraphics[width=\tfig cm]{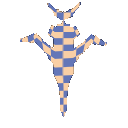} &
\includegraphics[width=\tfig cm]{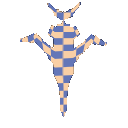} &
\includegraphics[width=\tfig cm]{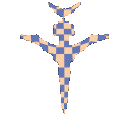} &
\includegraphics[width=\tfig cm]{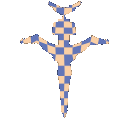} &
\includegraphics[width=\tfig cm]{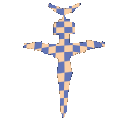} \\

\includegraphics[width=\tfig cm]{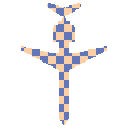} &
\includegraphics[width=\tfig cm]{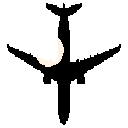} &
\includegraphics[width=\tfig cm]{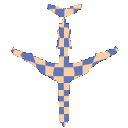} &
\includegraphics[width=\tfig cm]{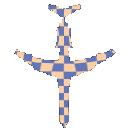} &
\includegraphics[width=\tfig cm]{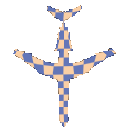} &
\includegraphics[width=\tfig cm]{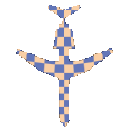} &
\includegraphics[width=\tfig cm]{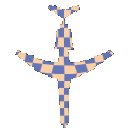} &
\includegraphics[width=\tfig cm]{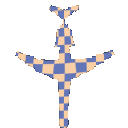} \\

source &
target &
1 class &
2 classes &
3 classes &
4 classes &
5 classes &
6 classes \\

\end{tabular}    
\caption{\rev{Robustness of a single \ourmethod{} network to handle a large number of classes. Given the same network architecture, we incrementally increase the number of trained classes from one class to six (\emph{airplane, vase, vessel, guitar, uppercase L} and \emph{bottle}). Observe that while there is little to no degradation for some examples (bottom row), there is a more pronounced performance degradation in others (top row).}}
\label{fig:multiclass_robust}
\end{figure*}

\rev{To further assess the scalability of \ourmethod{}, we train a single network on an increasing number of classes. We begin with our baseline of single-class training, and add a new class at each step, up to a total of six classes. 
In Figure~\ref{fig:multiclass_robust} we deform a source shape to two different targets from the airplane class, and track the results as the number of trained classes increases. We note the apparent degradation exhibited in the top row, versus the very slight degradation in the bottom row. These results align with our expectations for the general robustness and scalability of the network, accompanied by a little degradation that is a natural outcome of the strain placed on the network as we load more and more classes. To mitigate the effects of degradation, one can consider deepening and widening the network to allow for a larger capacity of shape priors.}

\subsection{Comparisons}
\begin{figure}[h]
\newcommand{\compfig}{1}
\setlength\tabcolsep{0.1pt} 
\begin{tabular}{ c c c c c c c c }
\includegraphics[width=\compfig cm]{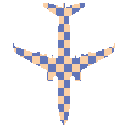} &
\includegraphics[width=\compfig cm]{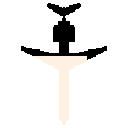} &
\includegraphics[width=\compfig cm]{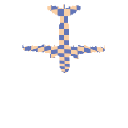} &
\includegraphics[width=\compfig cm]{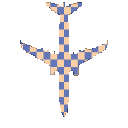} &
\includegraphics[width=\compfig cm]{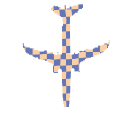} &
\includegraphics[width=\compfig cm]{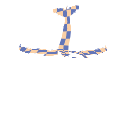} &
\includegraphics[width=\compfig cm]{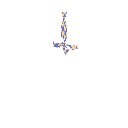} &
\includegraphics[width=\compfig cm]{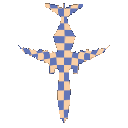} \\

\includegraphics[width=\compfig cm]{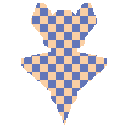} &
\includegraphics[width=\compfig cm]{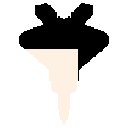} &
\includegraphics[width=\compfig cm]{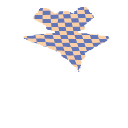} &
\includegraphics[width=\compfig cm]{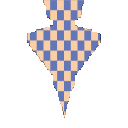} &
\includegraphics[width=\compfig cm]{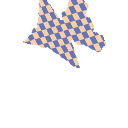} &
\includegraphics[width=\compfig cm]{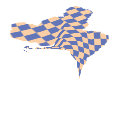} &
\includegraphics[width=\compfig cm]{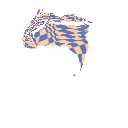} &
\includegraphics[width=\compfig cm]{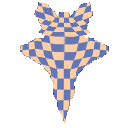} \\

\includegraphics[width=\compfig cm]{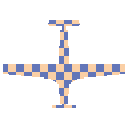} &
\includegraphics[width=\compfig cm]{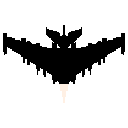} &
\includegraphics[width=\compfig cm]{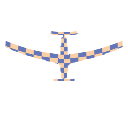} &
\includegraphics[width=\compfig cm]{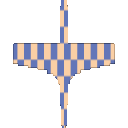} &
\includegraphics[width=\compfig cm]{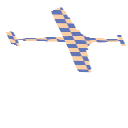} &
\includegraphics[width=\compfig cm]{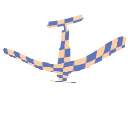} &
\includegraphics[width=\compfig cm]{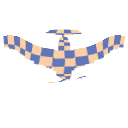} &
\includegraphics[width=\compfig cm]{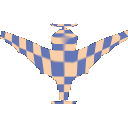} \\

\hline

\includegraphics[width=\compfig cm]{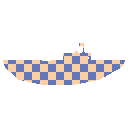} &
\includegraphics[width=\compfig cm]{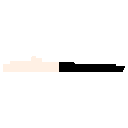} &
\includegraphics[width=\compfig cm]{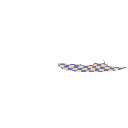} &
\includegraphics[width=\compfig cm]{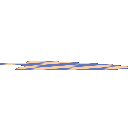} &
\includegraphics[width=\compfig cm]{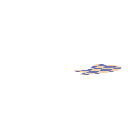} &
\includegraphics[width=\compfig cm]{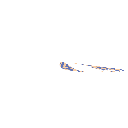} &
\includegraphics[width=\compfig cm]{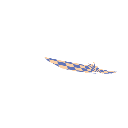} &
\includegraphics[width=\compfig cm]{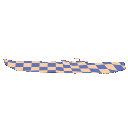} \\

\includegraphics[width=\compfig cm]{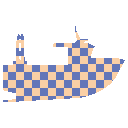} &
\includegraphics[width=\compfig cm]{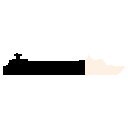} &
\includegraphics[width=\compfig cm]{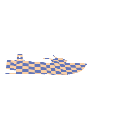} &
\includegraphics[width=\compfig cm]{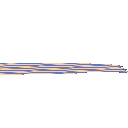} &
\includegraphics[width=\compfig cm]{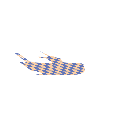} &
\includegraphics[width=\compfig cm]{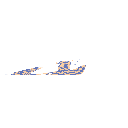} &
\includegraphics[width=\compfig cm]{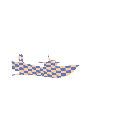} &
\includegraphics[width=\compfig cm]{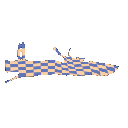} \\

\hline

\includegraphics[width=\compfig cm]{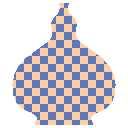} &
\includegraphics[width=\compfig cm]{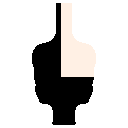} &
\includegraphics[width=\compfig cm]{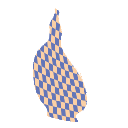} &
\includegraphics[width=\compfig cm]{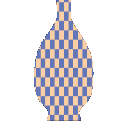} &
\includegraphics[width=\compfig cm]{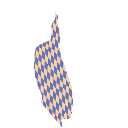} &
\includegraphics[width=\compfig cm]{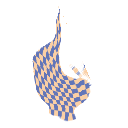} &
\includegraphics[width=\compfig cm]{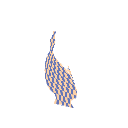} &
\includegraphics[width=\compfig cm]{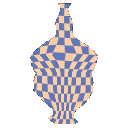} \\

\includegraphics[width=\compfig cm]{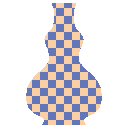} &
\includegraphics[width=\compfig cm]{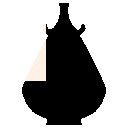} &
\includegraphics[width=\compfig cm]{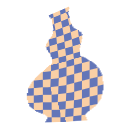} &
\includegraphics[width=\compfig cm]{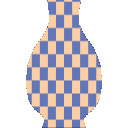} &
\includegraphics[width=\compfig cm]{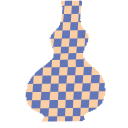} &
\includegraphics[width=\compfig cm]{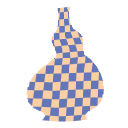} &
\includegraphics[width=\compfig cm]{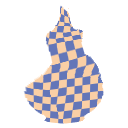} &
\includegraphics[width=\compfig cm]{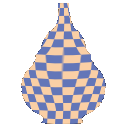} \\

\includegraphics[width=\compfig cm]{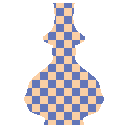} &
\includegraphics[width=\compfig cm]{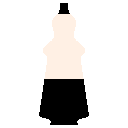} &
\includegraphics[width=\compfig cm]{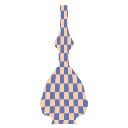} &
\includegraphics[width=\compfig cm]{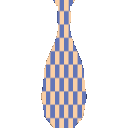} &
\includegraphics[width=\compfig cm]{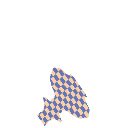} &
\includegraphics[width=\compfig cm]{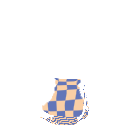} &
\includegraphics[width=\compfig cm]{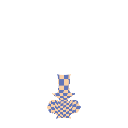} &
\includegraphics[width=\compfig cm]{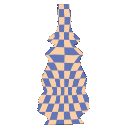} \\

source &
target &
CPD &
\rev{STN} &
RANSAC &
SC &
PRGLS &
\ourmethod{} \\
\end{tabular}    
\caption{Visual examples of test set results for: CPD~\cite{CPD}, STN~\cite{spatialtransformer} (in \ourmethod{} framework), affine RANSAC, SC~\cite{Belongie2002}, PRGLS~\cite{ma2016non} and our method. The same set of parameters is used across all examples for \ourmethod{} and existing approaches. Observe that \ourmethod{} is able to faithfully estimate the scale of missing components and infer plausible mappings on the missing pieces.}
\label{fig:comparisons_qual}
\end{figure}

In this section we compare the results of applying \ourmethod{}, versus existing approaches, on the $870$
reserved test set pairs for each class (as described in Section~\ref{sec:un_data}). The quantitative results are summarized in Table~\ref{table:eval_iou}, with visual examples shown in Figure~\ref{fig:comparisons_qual}.

Since \addcomment{existing} approaches usually operate on point-sets, we compute the deformations on the contour of the shapes. For comparable results, we apply the contour deformation on the entire image. The high order deformation computed using CPD~\cite{CPD} contains two regularization parameters: first to control deformation field smoothness, and second to control the distortion allowed on the source shape. Setting the CPD deformation field smoothness weight too high restricts the expressiveness allowed in the warp field (similar phenomenon as in \ourmethod{}). However, the parameter which controls the source shape distortion is trickier; a large weight restricts the source shape from conforming to the partial components while also restricting the source shape to conform to the target. We also show results of the shape context descriptor (SC), matching and alignment of~\cite{Belongie2002}. In addition to the global and local structure preserving point set registration (PRGLS) of~\cite{ma2016non}.

To demonstrate the restrictiveness of lower order deformations, we present two different affine transformation results. The first, a RANSAC-based approach, samples three points on the source and target to compute an affine transformation, and selects the best transformation based on source and target voting. \addcomment{Since the aforementioned} approaches do not estimate alignment parameters cognizant of the attributes specific to a particular class, estimating simple parameters such as scale becomes difficult (see last row of Figure~\ref{fig:comparisons_qual}).
\rev{To this end, we create an additional data-driven learning-based approach against which to compare. Using the implementation of Spatial Transformer Networks (STN)~\shortcite{spatialtransformer} (available in the form of an affine transformation), we train the STN network \emph{vis-\`a-vis} \ourmethod{}. We train STN using the same architecture, data and learning parameters as \ourmethod{}, but without the regularization components (superfluous for low DoF deformations).}

\rev{Observe that both learning approaches, STN and \ourmethod{}, are better suited to preserve features characteristic to the class of objects being warped (Figure~\ref{fig:comparisons_qual}).
However, the limiting DoF in STN is often over-restrictive, which prevents the scale from being accurately recovered (see rows 2 \& 3 Figure~\ref{fig:comparisons_qual}).}

\begin{table}
\center
\begin{filecontents*}{figures/comparison/vase/eval.csv}
vase,IOU
\ourmethod{},\textbf{81.3}
CPD,62.8
CPD affine,58.9
PRGLS,47.8
SC,48.1
STN,74.1
\end{filecontents*}
\csvautotabular{figures/comparison/vase/eval.csv}
\begin{filecontents*}{figures/comparison/airplane/eval.csv}
airplane,IOU
\ourmethod{},\textbf{69.0}
CPD,44.2
CPD affine,37.2
PRGLS,18.5
SC,36.8
STN, 57.2
\end{filecontents*}
\csvautotabular{figures/comparison/airplane/eval.csv}
\begin{filecontents*}{figures/comparison/vessel/eval.csv}
vessel,IOU
\ourmethod{},\textbf{60.2}
CPD,40.7
CPD affine,37.9
PRGLS,25.7
SC,39.8
STN,59.6
\end{filecontents*}
\csvautotabular{figures/comparison/vessel/eval.csv}
\caption{Quantitative results of average IOU on the entire test set of the \emph{vase}, \emph{airplane} and \emph{vessel} classes. All approaches use a fixed set of parameters for all classes. We compute \ourmethod{} results using TV \& M regularization.}
\label{table:eval_iou}
\end{table}

\begin{figure}[h]
\newcommand{\compfig}{1.2}
\setlength\tabcolsep{0.1pt} 
\begin{tabular}{ c c c c c c c c c }

a &
\includegraphics[width=\compfig cm]{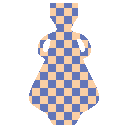} &
\includegraphics[width=\compfig cm]{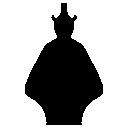} &
\includegraphics[width=\compfig cm]{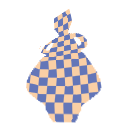} &
\includegraphics[width=\compfig cm]{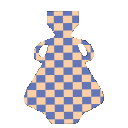} &
\includegraphics[width=\compfig cm]{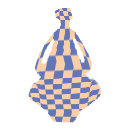} &
\includegraphics[width=\compfig cm]{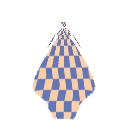} &
\includegraphics[width=\compfig cm]{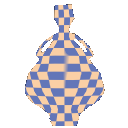} \\

b &
\includegraphics[width=\compfig cm]{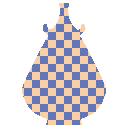} &
\includegraphics[width=\compfig cm]{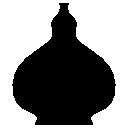} &
\includegraphics[width=\compfig cm]{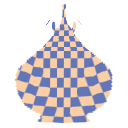} &
\includegraphics[width=\compfig cm]{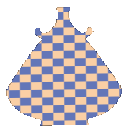} &
\includegraphics[width=\compfig cm]{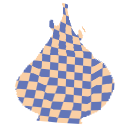} &
\includegraphics[width=\compfig cm]{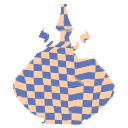} &
\includegraphics[width=\compfig cm]{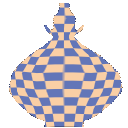} \\

c &
\includegraphics[width=\compfig cm]{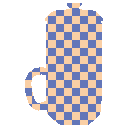} &
\includegraphics[width=\compfig cm]{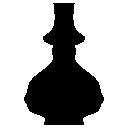} &
\includegraphics[width=\compfig cm]{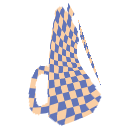} &
\includegraphics[width=\compfig cm]{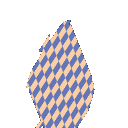} &
\includegraphics[width=\compfig cm]{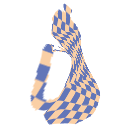} &
\includegraphics[width=\compfig cm]{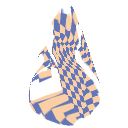} &
\includegraphics[width=\compfig cm]{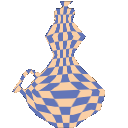} \\

d &
\includegraphics[width=\compfig cm]{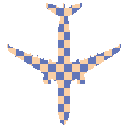} &
\includegraphics[width=\compfig cm]{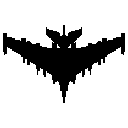} &
\includegraphics[width=\compfig cm]{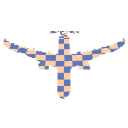} &
\includegraphics[width=\compfig cm]{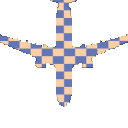} &
\includegraphics[width=\compfig cm]{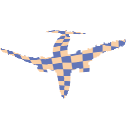} &
\includegraphics[width=\compfig cm]{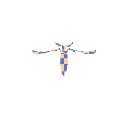} &
\includegraphics[width=\compfig cm]{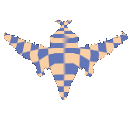} \\

e &
\includegraphics[width=\compfig cm]{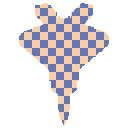} &
\includegraphics[width=\compfig cm]{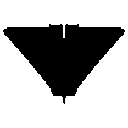} &
\includegraphics[width=\compfig cm]{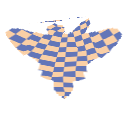} &
\includegraphics[width=\compfig cm]{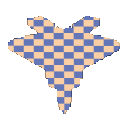} &
\includegraphics[width=\compfig cm]{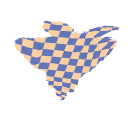} &
\includegraphics[width=\compfig cm]{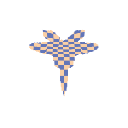} &
\includegraphics[width=\compfig cm]{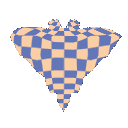} \\

f &
\includegraphics[width=\compfig cm]{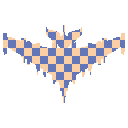} &
\includegraphics[width=\compfig cm]{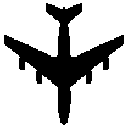} &
\includegraphics[width=\compfig cm]{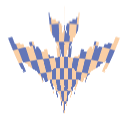} &
\includegraphics[width=\compfig cm]{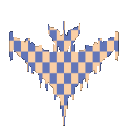} &
\includegraphics[width=\compfig cm]{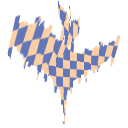} &
\includegraphics[width=\compfig cm]{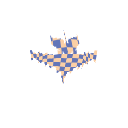} &
\includegraphics[width=\compfig cm]{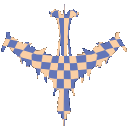} \\

 &
source &
target &
CPD &
STN &
SC &
PRGLS &
\ourmethod{} \\
\end{tabular}    
\caption{\rev{Visual examples of complete shapes (no missing shape data). Test set results for: CPD~\shortcite{CPD}, STN~\shortcite{spatialtransformer} (in \ourmethod{} framework), SC~\shortcite{Belongie2002}, PRGLS~\shortcite{ma2016non} and our method. The same set of parameters is used across all examples for \ourmethod{} and existing approaches. }
\label{fig:comparisons_qual_full}}
\end{figure}

\paragraph{\textbf{\rev{Full Shapes.}}}
\rev{While a core asset of \ourmethod{} is its ability to handle partial shapes, we are interested in the performance of our system in a simpler setting of complete shapes, and extend our evaluation accordingly. The quantitative results are summarized in Table~\ref{table:eval_iou_full}, with visual examples shown in Figure~\ref{fig:comparisons_qual_full}.}

\rev{\ourmethod{} uses the learned prior to compute full-shape warps that maintain shape semantics post-alignment (see  Figure~\ref{fig:comparisons_qual_full}). Observe that non-data-driven approaches often struggle to preserve shape features, for instance the symmetry of vase handles in rows (a) and (b), or the airplane tail in (e). Lastly, many of the non-data-driven approaches use a regularization term to encourage smooth contour-to-contour deformations which may result in a less accurate alignment (see row (d) and quantitatively in Table~\ref{table:eval_iou_full}).}


\begin{table}
\center
\begin{filecontents*}{figures/fullshape/vase/eval_full.csv}
vase,IOU
\ourmethod{},\textbf{81.3}
CPD,75.2
CPD affine,69.9
PRGLS,55.7
SC,70.3
STN,74.1
\end{filecontents*}
\csvautotabular{figures/fullshape/vase/eval_full.csv}
\begin{filecontents*}{figures/fullshape/airplane/eval_full.csv}
airplane,IOU
\ourmethod{},\textbf{70.3}
CPD,57.6
CPD affine,49.3
PRGLS,26.9
SC,59.1
STN,57.3
\end{filecontents*}
\csvautotabular{figures/fullshape/airplane/eval_full.csv}
\begin{filecontents*}{figures/fullshape/vessel/eval_full.csv}
vessel,IOU
\ourmethod{},\textbf{63.7}
CPD,55.9
CPD affine,48.0
PRGLS,33.4
SC,50.2
STN,61.5
\end{filecontents*}
\csvautotabular{figures/fullshape/vessel/eval_full.csv}
\caption{\rev{Quantitative results on complete (not partial) shapes. Average IOU on the entire test set of the \emph{vase}, \emph{airplane} and \emph{vessel} classes. All approaches use a fixed set of parameters for all classes. We compute \ourmethod{} results using TV \& M regularization.}}
\label{table:eval_iou_full}
\end{table}


%



\subsection{Limitations}

We note that a grid-based deformation limits the space of transformations that the shape can undergo for alignment.
A finer resolution increases the degrees of freedom and becomes more susceptible to implausible deformations, which we have shown can be mitigated with an appropriate regularization mechanism. 
\revb{However, as can be seen in Figure~\ref{fig:grid_res}, when the degree of deformation is too large compared to the network size, even our regularization term will not suffice.} 
While the proposed approach produces complex deformations even with regularization, we recognize that such a step in itself results in a confinement of the deformation expressiveness.
\rev{The proposed approach cannot handle large differences in topologies.}
Furthermore, our grid-based deformation would not be an appropriate method of transformation for shapes with articulations.
\revb{We have shown how to incorporate an unsupervised global orientation estimation into \ourmethod{} in Figure~\ref{fig:grot}. However, these results suggest that a more comprehensive solution may be necessary to achieve higher orientation and shape alignment quality.}




\section{Conclusion}

Aligning one shape to another with a high order deformation is a fundamentally ill-posed problem, but when missing parts are thrown into the mix, the ambiguity only deepens. Classic alignment techniques are often at a loss under such circumstances, as the missing information can be highly misleading.
To alleviate this, we take advantage of large collections of shapes, and train a system to align pairs of source and potentially partial target shapes. The system learns the space of plausible deformations associated with the dataset, and effectively becomes agnostic to missing parts, without memorization or over-fitting.

\rev{A notable advantage of our approach lies in its regressive nature. Instead of synthesizing completely new shapes from scratch, we learn to align one shape to another, resulting in a tool that enables us to deform a high quality source shape to a low quality target, thereby forming novel instances. By taking this route, we simplify the learning process, requiring a smaller network that is less sensitive and able to converge faster.}

Our technique introduces a key contribution in the form of a regularization mechanism that promotes smoothness. We have shown (Figure~\ref{fig:visual_reg_ablation}) that allowing the system to run wild with its free form deformation computations is not only detrimental to structure preservation, which is crucial for post processing (\emph{e.g.}, texture \& segmentation transfer and preserving 3D geometry), but also hinders partial agnosticism. Our regularization relies on a differential grid deformation that naturally supports a total variation penalty encouraging piecewise smooth warps.

The benefits associated with data-driven paradigms are central to the performance of our system. The information contained in the training data assists the network to learn a shape prior that helps preserve the characteristic features of the input set of shapes. Our experimental results show that important attributes  such as part symmetry and scale are well recovered by our system, suggesting that the network has learned the expected shape and form of the underlying class. 
We have shown (Figures~\ref{fig:teaser2},~\ref{fig:teaser},~\ref{fig:novel_source_class}, and~\ref{fig:novel_novel_class})
that our trained system generalizes well to unseen examples from the learned classes or from untrained but similar ones, and that the incorporation of prior knowledge benefits not only a complex setting where the target shape may be substantially partial, but also a simpler setting where the target shape is complete (Figure~\ref{fig:comparisons_qual_full}).



Looking forward, we suggest several directions for further research. 
\new{While our current setting uses a simple loss function, it would be interesting to use more complex loss functions such as the perceptual loss~\cite{johnson2016perceptual} or the earth-mover distance. Another possibility is to incorporate the \textit{curriculum-learning} technique: first start with "easy" training examples (similar topologies and no missing parts), and as the network adapts to handle them, gradually include increasingly harder examples.
Furthermore, extending the network scope to handle both partial source and target shapes \newa{is a challenging problem that may lead to a more generic solution}.}
\rev{Our warping grid is set to $8 \times 8$ for 2D shapes, and $7 \times 7 \times 7$ for 3D shapes. We deliberately opt for a milder degree of deformation, which inherently fosters smooth deformations. 
We note that to achieve a tighter alignment, it is possible to apply an ICP-based approach as a post-process.}
Finally, the incorporation of part awareness may potentially support a more semantic alignment, where advanced operations such as part addition and removal could help to maintain the structural integrity of the target shape.

\section{Acknowledgements}
We thank the anonymous reviewers for their helpful comments. This research was supported by the Israel Science Foundation as part of the ISF-NSFC joint program grant number (2217/15, 2472/17), and partially supported by ISF grant 2366/16. This research was partially supported by ERC-StG grant no. 757497 (SPADE).

\bibliographystyle{ACM-Reference-Format}
\bibliography{correspondence}
\end{document}